%% file: prespower.tex
\documentclass[a4paper,UKenglish,cleveref, autoref, thm-restate]{lipics-v2021}

\hideLIPIcs

\usepackage{settigs}
\usepackage{macros}

\title{The complexity of Presburger arithmetic\\ with power or powers}
\titlerunning{The complexity of Presburger arithmetic with power or powers}

\author{Michael Benedikt}{Department of Computer Science, University of Oxford, UK}{michael.benedikt@cs.ox.ac.uk}{https://orcid.org/0000-0003-2964-0880}{funded in part by EPSRC grant EP/T022124/1;}

\author{Dmitry Chistikov}{Centre for Discrete Mathematics and its Applications (DIMAP) \&\\ Department of Computer Science, University of Warwick, Coventry, UK}{d.chistikov@warwick.ac.uk}{0000-0001-9055-918X}{acknowledges the support of IMDEA Software Institute;} 

\author{Alessio Mansutti}{IMDEA Software Institute, Spain}{alessio.mansutti@imdea.org}{https://orcid.org/0000-0002-1104-7299}{funded in part by ERC grant No.\ 852769 (ARiAT).}

\authorrunning{M.~Benedikt, D.~Chistikov, A.~Mansutti} 

\Copyright{Michael Benedikt, Dmitry Chistikov, Alessio Mansutti} 

\ccsdesc[100]{Theory of computation~Logic and verification}

\keywords{arithmetic theories, exponentiation, decision procedures} 

\input{ack}

\nolinenumbers 

\EventEditors{Kousha Etessami, Uriel Feige, and Gabriele Puppis}
\EventNoEds{3}
\EventLongTitle{50th International Colloquium on Automata, Languages, and Programming (ICALP 2023)}
\EventShortTitle{ICALP 2023}
\EventAcronym{ICALP}
\EventYear{2023}
\EventDate{July 10--14, 2023}
\EventLocation{Paderborn, Germany}
\EventLogo{}
\SeriesVolume{261}
\ArticleNo{131}

\begin{document}

\maketitle

\begin{abstract}
\input{abstract}

\end{abstract}

\input{intro}

\input{preliminaries}

\input{algos}

\input{complexity}

\input{conc}

\bibliography{bibliography}

\clearpage

\appendix 

\input{correctness}

\input{appendix-complexity}

\input{appendix-differences}

\end{document}

%% file: ack.tex
\acknowledgements{%
We are grateful to
Hera Brown,
Florian Frohn,
Leon M\"uller, and
Joy Olasunkanmi 
for pointing out mistakes in the proceedings version
and for constructive discussions.
}

%% file: abstract.tex
We investigate expansions of Presburger arithmetic ($\PA$), i.e., the theory of the integers with addition and order, with additional structure related to exponentiation:
either  a function that takes  a number to
the power of $2$, or
a predicate $\powerpred$ for the powers of $2$.
The latter theory, denoted $\prespower$, was introduced by B\"uchi as a first attempt at characterizing the sets of tuples of numbers that  can be expressed
using  finite automata; B\"uchi's method does not give an elementary upper bound, and the complexity of this theory has been open.
The former theory, denoted as $\presexp$, was shown decidable by Semenov;
while the decision procedure for this theory  differs radically from the automata-based method proposed by
B\"uchi,
Semenov's method is also non-elementary. And in fact, the theory with the power function has a non-elementary lower bound.
In this paper, we show that while Semenov's and B\"uchi's approaches yield non-elementary blow-ups for $\prespower$, the theory is in fact decidable in triply exponential time, 
similarly to the best known quantifier-elimination algorithm for $\PA$.
We also provide a $\nexptime$ upper bound for the existential fragment of $\presexp$, a step towards a finer-grained analysis of
its complexity.
Both these results are established by analyzing a single parameterized satisfiability algorithm for $\presexp$, which 
can be specialized to either the setting of $\prespower$ or the existential theory of $\presexp$.
Besides the new upper bounds for the existential theory of $\presexp$ and $\prespower$, we believe our algorithm provides new intuition for the decidability of these theories, and for 
the features that lead to non-elementary blow-ups.

%% file: intro.tex
\section{Introduction} \label{sec:intro} This paper concerns
decision  problems involving first-order logic sentences
over the integers. We are given a sentence in the logic, and
want to know if it holds in a certain infinite structure
over the integers --- we refer to these as ``satisfaction
problems'' below. 
If the sentence can mention ``full arithmetic'' --- both
 addition and multiplication on the integers --- then it is
 well-known that the satisfaction problem is undecidable
 \cite{classicaldecision}. On the other hand, if the
 sentence mentions only addition, inequality, and integer
 constants --- \emph{Presburger arithmetic} ($\PA$) --- then
 the satisfaction problem is decidable \cite{Pre29}.
 Presburger arithmetic  is by no means the maximal decidable
 arithmetic theory. For instance, adding a ``bit predicate''
 to Presburger arithmetic --- a binary predicate holding on
 $(m, n)$ if $m$ is the largest power of $2$ dividing $n$
 --- does not undermine decidability. This extension is
 known as \emph{{\buchi} arithmetic}. A decision procedure
 for the satisfaction problem of this theory is based on
 translating each formula into a finite automaton over
 strings, representing the binary expansions of possible
 solutions to the formula~\cite{bes}. Although both are
 decidable, there is a big difference between Presburger
 arithmetic and {\buchi} arithmetic: the satisfaction
 problem of the former can be decided in triply exponential
 time~\cite{Opp78} and even in doubly exponential
 space~\cite{FerranteR75}, whereas the latter is known to
 have no elementary bound. See~\cite{Berman80}
 and~\cite{Weispfenning90} for a finer-grained analysis of
 the complexity of Presburger arithmetic, in terms of
 alternating Turing machines.

Sitting in between Presburger arithmetic and {\buchi}
arithmetic is the extension of~$\PA$ with a predicate for
the powers of $2$: we refer to this set of numbers as
$2^{\Nat}$ and to the theory as $\prespower$. This predicate
is clearly definable in {\buchi} arithmetic, so the
first-order theory of~$\prespower$ is again decidable with
automata. An alternative decision procedure, avoiding
automata, was developed by Semenov
in~\cite{semenovprespower}. It proceeds by eliminating
quantifiers, arriving at a quantifier-free formula in an
enhanced signature --- including, for example, a predicate
for the highest power of $2$ below a given integer.
Semenov's procedure applies more broadly to extensions of
Presburger arithmetic by a unary predicate satisfying a
condition ``effective sparseness'': thus it isolates
combinatorial properties of $2^{\Nat}$ that underlie
decidability, rather than automata-theoretic constructions.
Semenov's procedure has been refined and extended by Point;
see, e.g.,~\cite{pointexpansion}. The complexity of the
procedure and the complexity of $\prespower$ has, to our
knowledge,  received no attention.

Instead of adding  a unary predicate, one can add to~$\PA$
the function taking a number $n$ to $2^n$: the power
function for short, rather than the powers predicate above.
The theory, which we denote $\presexp$,  subsumes
$\prespower$. Semenov proved decidability of this theory as
well \cite{semenovpresexp}. But in this case a
non-elementary lower bound follows from
\cite{comptonhenson}, see  \cite{pointcherlin}. We are not
aware of any finer-grained analysis of the complexity of the
theory. Note that $\presexp$ is incomparable to {\buchi}
arithmetic in expressiveness: in  fact the union of the two
is undecidable \cite{pointcherlin}. 

In this paper we show that the complexity of $\prespower$ is
elementary, and is in fact contained in $\threeexptime$. In
this sense $\prespower$ is quite similar to Presburger
arithmetic in complexity. We also show that the existential
fragment of the theory $\presexp$ has elementary complexity: its
satisfaction problem is in $\nexptime$.

We show our results on extending~$\PA$ with powers or the
power function using a single parameterized algorithm. The
algorithm can be applied to decide satisfaction of a
$\presexp$ sentence $\phi$ in time tower of $|\phi|$,
matching the prior non-elementary complexity. But it can be
specialized to the context of either a  $\prespower$
sentence or an existential $\presexp$ sentence, giving in
each case an elementary bound. The algorithm is based on
eliminating quantifiers: it makes heavy use both of existing
Presburger quantifier elimination algorithms
\cite{Weispfenning90} and the core of the method of Semenov,
which involves removing  ``problematic occurrences'' of a
variable within a formula. Intuitively, an occurrence of a
variable in an atomic formula is unproblematic if it occurs
only outside of power functions, or if the atomic formula is
just a comparison between two power terms. In the latter
case, the exponentiation of the variable can be eliminated
by taking logarithms. We factor this core Semenovian idea
out into a self-contained subroutine. We give a short
top-level procedure that interleaves calls to this
subroutine with calls to a variant of Presburger quantifier
elimination. The latter enables us to remove quantified
variables completely. A meticulous complexity analysis that
tracks several parameters of the input formula shows that
the procedure achieves the desired bounds in the special
cases of $\prespower$ and existential $\presexp$.

Our work brings the following ideas:
\begin{itemize}
\item
For \prespower, we rewrite the formula by processing
quantifier blocks inside out but do not eliminate quantifier
alternation. The resulting formula is put in a nontrivial
fragment of \PA: integer octagon arithmetic (\Oct), which we
observe to be decidable in \pspace.
\item
In eliminating problematic variables from formulae, our algorithm exploits a new
substitution strategy:
it tailors
substitutions to individual inequalities, rather than
applying them in the entire formula globally or in an
individual disjunct in the DNF (\`a la Reddy and
Loveland~\cite{ReddyLoveland78}). 
For both \prespower and $\exists\presexp$, this will be  key to
obtaining strong bounds on the number of homogeneous terms produced during the transformation, and bounds on these
terms will give us bounds on the running times of the algorithms.
\end{itemize}
We believe that our procedure, in addition to providing the
desired bounds, gives a good intuition for the decidability
of $\presexp$ and the sources of non-elementary blow-up
within~it. Note that in this work we deal for simplicity
with powers of $2$, but  the same complexity results  apply
to any other base $k \in \Nat$, $k > 2$. 
Expansion with two bases is undecidable~\cite{phstrongcobham}.

%% file: preliminaries.tex
\section{Preliminaries}
\label{section:preliminaries}

The symbols~$\Zed$, $\Nat$ and~$\PNat$ denote the set
of integers, natural numbers (including zero), and
positive integers, respectively.
We write $\card{\aset}$
for the cardinality of a finite set~$\aset$. 
Given $n,m \in \Zed$, we define $[n,m] \egdef \{n,n+1,\dots,m\}$
and, if $n \in \PNat$, $[n] \egdef [0,n-1]$. For two sets $D$ and $C$, $[D
\to C]$ stands for the set of all functions from~$D$
to~$C$.
We write $\floor{\cdot}$ and $\ceil{.}$ to denote the \emph{floor} and \emph{ceiling} functions, respectively, 
$\abs{\cdot}$ to denote the \emph{absolute value} function, 
and $\log(\cdot)$ to denote the \emph{binary logarithm} function.
Note that $n \in \Nat$ can be represented in binary using $\ceil{\log (n +1)}$ many bits.

We sometimes apply standard set operations and
predicates, such as for instance~$\in$, $\subseteq$
and $\setminus$, to vectors $\vec v = (v_1,\dots,v_d)$.
In these cases, there is an implicit conversion of
$\vec v$ into the set $V = \{v_1,\dots,v_d\}$. As an example, $v \in \vec {v}$ and $\vec {v} \setminus A$
stand for $v \in V$ and $V \setminus A$, respectively,
where $\aset$ is a set (or another vector).

\subparagraph*{Presburger arithmetic with a power function.}
We consider the structure $\presexp \coloneqq \langle
\Zed, 1, +, (a \cdot x)_{a \in \Zed}, \pow{x}, (q \divides
{x})_{q \in \PNat}, < \rangle$, in which the classical
signature of Presburger arithmetic ($\PA$) is enriched with the
unary \emph{power of the absolute value} function $x
\mapsto \pow{x}$. As usual, $1$ is the constant
(interpreted as) $1 \in \Zed$, $+$ and $<$ stand for
addition and strict ordering over~$\Zed$,
respectively, $x \mapsto a \cdot x$ is the unary
function multiplying its input by the constant $a \in
\Zed$, and $x \mapsto q \mid x$ is the unary relation
that is true for integers divisible by~$q \in \PNat$.

The \emph{first-order formulae} $\Phi,\Psi,\phi,\psi,\dots$ \text{of} $\presexp$ are generated from the grammar 
\begin{align*} 
  \Phi,\Psi &\coloneqq  \alpha \,\mid\, \top \,\mid\, \bot \,\mid\, \lnot \Phi \,\mid\, \Phi \land \Psi \,\mid\, \exists x \, \Phi \,\mid\, \forall x\, \Phi
  \qquad\qquad 
  \alpha \coloneqq t_1 < t_2 \,\mid\, (q \mid t_1),
\end{align*}
where $x$ is a first-order variable from an
infinite countable set $\V$. 
Each variable in a formula is either free or bound,
and in the latter case there is no nested quantification of the same variable.
The elements of $\alpha$
are the \emph{atomic formulae} of the logic, i.e., they
are \emph{inequalities} $t_1 < t_2$ between
terms $t_1$ and $t_2$, or \emph{divisibility
constraints} $q \mid t_1$, where $q \in \PNat$.
Instead of allowing arbitrary terms of the signature,
we will deal with a simpler  language where \emph{terms} are expressions of the form $\sum_{i \in
I} a_i \cdot \pow{{x}_i} + \sum_{j \in J} b_j \cdot {x}_j
+ c$ where $c \in \Zed$ is the \emph{constant} of the
term, $a_i,b_j \in \Zed \setminus \{0\}$ are the
\emph{coefficients} of the \emph{power terms}
$(\pow{x_i})_{i \in I}$ and of the \emph{linear} variables $(x_j)_{j
\in J}$, and $I,J$ are finite sets of indices (which
may overlap). We also restrict
to terms where no variable occurs linearly twice, or occurs
exponentiated twice. It is easy to see that formulae of the full language can be converted to use this restricted term language, and
our algorithms will not take us out of this fragment.
When we talk about equality of terms, we mean modulo
associativity and commutativity of $+$. 
A term is said
to be \emph{homogeneous} if its constant is $0$.

The Boolean connectives $\lor$, $\implies$ and $\iff$, 
and the standard (in)equalities between terms $\leq$, $=$, $\geq$ and $>$ are defined from $\land$, $\lnot$ and $<$,
as usual. We use the absolute value of variables occurring linearly as a shortcut: e.g., $a \cdot \abs{x} < t$ is equivalent to the formula $(x \geq 0 \implies a \cdot x < t) \land (x < 0 \implies -a \cdot x < t)$.
We write $\aformula({x}_1,\dots,{x}_d)$ or $\aformula(\vec x)$ 
to highlight the fact that all free variables of the formula~$\aformula$
are in $\vec x$ and are distinct. A formula without free variables is said to be a \emph{sentence}.

The \emph{satisfaction
problem} asks whether a given
sentence is true.

\subparagraph*{Covers.}

We write
$\aformula \fequiv
\aformulabis$ whenever $\aformula$ and $\aformulabis$
are equivalent.
A finite set of formulae $\{\aformula_i : i
\in I \}$ is said to be a \emph{cover for} (or \emph{to cover}) a formula
$\aformulabis$ whenever $\aformulabis \fequiv
\bigvee_{i \in I} \aformula_i$.
Analogously, $S$ 
is said to be a \emph{cover for} (or \emph{to cover}) a formula
$\aformulabis$ \emph{under} a sequence of quantifiers $\quant_1 w_1 \ldots \quant_k w_k$
whenever $\aformulabis \fequiv
\quant_1 w_1 \ldots \quant_k w_k .\bigvee_{i \in I} \aformula_i$.

\subparagraph*{Term and formula normalization.}

To simplify the exposition, we often bring terms and
formulae to convenient (normal) form, without mentioning this explicitly every time.
This normalization
does not change our bounds on the asymptotic running time of the algorithms,
nor their correctness. 
\begin{itemize}
\item
We assume inequalities to have the form $t < 0$, where $t$ is a term.
Thus, 
we convert inequalities of the form $t_1 < t_2$ into $t_1 - t_2 < 0$.
The construction of $t_1 - t_2$ follows the convention on terms described above.
We will still sometimes
refer to more general inequalities $t_1 < t_2$ for brevity, but these should be taken as abbreviations for inequalities of the
above form.
\item
We rearrange terms following associativity and commutativity of~$+$.
We also evaluate arithmetic expressions, including, e.g., $\pow a$ and $\abs{a}$,
where $a \in \Zed$.
\item
In divisibility constraints of the form ${q \mid \sum_{i \in
I} a_i \cdot \pow{{x}_i} + \sum_{j \in J} b_j \cdot {x}_j
+ c}$, we always assume $a_i,b_j,c \in [q]$.
\item
Inequalities $a < b$ and divisibility constraints $q \mid a$ on integers $a,b \in \Zed$ and $q \in \PNat$ are
evaluated to $\true$ or $\false$. So are divisibility constraints $1 \mid t$ or $q \mid 0$, where $q \in \PNat$ and $t$ is a term (these are $\true$).
\item
Inequalities of the form $a \cdot x < b$ with $a,b \in \Zed$ and $\abs{a} \geq 2$ are rewritten into $x \leq \floor{\frac{b-1}{a}}$ if $a > 0$, and to 
$x \geq \ceil{\frac{b-1}{a}}$ if $a < 0$.
This normalization is required in the context of the quantifier elimination (q.e.) procedure for Presburger arithmetic applied to octagons (see~\emph{integer octagon arithmetic} below).
\item
Trivial inequalities involving power terms, that is those that can be evaluated by a simple sign analysis that relies on the fact that $\pow{x}$ is positive for all values of $x$
are also rewritten as $\top$ or $\bot$.
For instance, $a \cdot \pow{x} < c$ when $a$ and $c$ have different signs or when $c = 0$;
or $a \cdot \pow{x} < b \cdot \pow{y}$ where $a$ and $b$ have different signs.
\end{itemize}

Beyond normalization, we need the following operations and notation
for terms.
We write $t(\vec x)$ if all variables appearing in the term $t$
are  in $\vec x$.
Let $\alpha_1$ be a formula (resp.,~a term of the form $x$ or $\pow{x}$).
Given a second formula (resp.,~term) $\alpha_2$,
$\aformula\substitute{\alpha_2}{\alpha_1}$ stands for the
formula obtained from $\aformula$ by replacing every
occurrence of $\alpha_1$ by $\alpha_2$. Additionally, when $\alpha_1$ and $\alpha_2$ are two terms $t_1$ and $t_2$, and given $n \in \PNat$, we write
$\aformula\substitute{\frac{t_2}{n}}{t_1}$ for the formula
obtained from $\aformula$ by replacing each
inequality $a \cdot t_1 + t' < t''$ by $a \cdot t_2 + n \cdot
t' < n \cdot t''$ and each divisibility constraint $q \mid a
\cdot t_1 + t'$ by $n \cdot q \mid a \cdot t_2 + n \cdot
t'$. This operation can be seen as
scaling by $n$ the atomic formulae where $t_1$
occurs linearly, relying on the equivalences $s_1 < s_2 \fequiv n
\cdot s_1 < n \cdot s_2$ and ${q \mid s \fequiv n \cdot q \mid
n \cdot s}$, followed by the replacement of each $n
\cdot t_1$ by $t_2$.
We will restrict the use of term substitutions to the following cases:
$\Phi\substitute{t}{x}$ and $\Phi\substitute{\frac{t}{n}}{x}$, where $x$
is a variable only occurring linearly in $\Phi$;
and $\aformula\substitute{t}{\pow x}$ and $\aformula\substitute{\frac{t}{n}}{\pow x}$. Note that in the last two cases,
all linear occurrences of the variable $x$ are left untouched.
We extend the notion of substitution
to multiple terms or formulae: $\aformula\substitute{\beta_i}{\alpha_i : i \in [1,k]} \coloneqq (\ldots(\aformula\substitute{\beta_1}{\alpha_1})\substitute{\beta_2}{\alpha_2}\dots)\substitute{\beta_k}{\alpha_k}$.

\subparagraph*{Parameters of formulae.}
As often done for~$\PA$, the complexity analysis of our
procedure requires the introduction of several
parameters for a formula. We define functions
$\linterms(\cdot)$, $\homterms(\cdot)$, $\maxvars(\cdot)$,
$\fmod(\cdot)$, $\boolnum(\cdot)$, and $\alt(\cdot)$,
to track various features of a formula $\aformula$
from~$\presexp$:

\begin{itemize}
  \item
  $\linterms(\aformula)$ is the set containing the
  terms $0$ and $2$ as well as all the terms $\aterm$ that
  appear in inequalities~$\aterm < 0$
  of~$\aformula$ 
  (recall that $\aterm_1 < \aterm_2$ is a shorthand for~$\aterm_1 - \aterm_2 < 0$);
  \item
    $\homterms(\aformula)$ is the set of
    \emph{homogeneous terms} obtained from the
    terms in $\linterms(\aformula)$ by eliminating
    their constant term~$c$ (alternatively, updating $c$ to $0$);
  \item $\maxvars(\aformula)$ is the maximum number of
  variables in a term of $\aformula$;
  \item
  $\fmod(\aformula)$ is the least common multiple of all~$q
  \in \PNat$ appearing in constraints $q \mid \aterm$ of
  $\aformula$ (if the formula $\aformula$ has no divisibility constraints, then we postulate~$\fmod(\aformula)=1$);
  \item $\boolnum(\aformula)$ denotes the number of occurrences of
  negations~$\lnot$ and conjunctions~$\land$ in $\aformula$
  (note that the syntax permits binary conjunction only);
  \item $\alt(\aformula)$ is the \emph{quantifier
  alternation rank} (number of quantifier blocks) of a formula $\aformula$
  in prenex normal form.
  \end{itemize}

Throughout the paper, we assume an encoding of terms
where constants and coefficients are given in binary
representation.
By
$\len{\aformula}$ we denote the \emph{length} of the
formula $\aformula$: the number of bits
required to write it down. For simplicity, we assume it is always at least $2$.
We extend the notion of infinity norm to terms. The
\emph{infinity norm} $\norminf{t}$ of a
term~$\aterm$ is the maximum absolute value of a
coefficient or constant appearing in $t$. For a finite
set of terms~$T$, we define $\norminf{T} \egdef
\max\{\norminf{t} :  t \in T\}$.
The \emph{$1$-norm} of~$t$, denoted $\normone{t}$, is the sum of
absolute values of all its coefficients and of its constant;
this is always non-negative. 

\section{Summary of main results}
\label{s:summary}

This paper  focuses on two fragments of
$\presexp$:
\begin{itemize}
\item
The 
first-order theory of ${\prespower
\coloneqq \langle \Zed, 1, +, (a \cdot x)_{a \in \Zed},
\pp(x), (q \mid {x})_{q \in \PNat}, < \rangle}$, 
that is, the structure which
enriches Presburger arithmetic with the unary relation
$x \mapsto \pp(x)$ that is true for the powers of $2$, i.e.,
$\pp(x) = \top$ iff $x \in \{1, 2, 4, \ldots \}$.
Note that the relation $\pp(x)$ can be expressed in $\presexp$, with the formula
$\exists y .\, x = \pow{y}$.
\item
The existential fragment of $\presexp$,
denoted by $\exists\presexp$. Formulae of this fragment
are of the form $\exists \vec x. \Phi$, where $\Phi$ is
quantifier-free (q.f., in short). 
\end{itemize}
The main results of this paper are summarized
below:
\begin{theorem}
  \label{theo:exexp-nexp}
  The satisfaction problem for $\exists\presexp$ is
   in $\nexptime$.
\end{theorem}
\begin{theorem}
  \label{theo:pow-3exp}
  The satisfaction problem for $\prespower$ is
   in $\threeexptime$.
\end{theorem}
\Cref{theo:pow-3exp,theo:exexp-nexp} are based
on a common core procedure for $\presexp$ that we
introduce in~\Cref{sec:procedure}.
The procedure manipulates the subformulae of an input
formula so that they (eventually) enter the following  fragments
of $\presexp$:
\begin{itemize}
  \item The \emph{power comparisons fragment}, denoted
  by~$\PT$. In this fragment, inequalities are
  restricted to the form $a \cdot \pow{x} < b \cdot
  \pow{y}$ or $a \cdot \pow{x} < b$, where $a,b \in
  \Zed$, and divisibility constraints 
  are of the form~$q \mid \pow{x} -
  r$, where $q \in \PNat$ and $r \in [q]$.
  
  \item \emph{Integer octagon arithmetic}, denoted by $\Oct$ (see e.g.~\cite{KapurZHZLN13,Mine06}), 
  that is, the fragment of~$\PA$ in which inequalities
  are restricted to the forms $\pm x \pm y < c$ and $\pm x < c$, where
  $c \in \Zed$, and divisibility constraints are of the form $q \mid x - r$, where $q \in \PNat$ and $r \in [q]$.

  \item The~fragment $\QFPC$ (short for \emph{Semenov}, as this fragment is related to the one used 
  in~\cite{semenovprespower}).
  In formulae $\Phi$ of this fragment,
  each variable appears either always linearly or always in a power, and every \emph{bound} variable $x$
  appears only in atomic formulae from $\PT$ (hence, $x$ is always in a power). Moreover, divisibility constraints in $\Phi$ are \emph{simple}, i.e., 
  they are of the form~$q \mid \pow{x} -
  r$ or of the form~$q \mid x - r$, where $q \in \PNat$ and $r \in [q]$.
Notice that a sentence in this fragment must be in $\PT$.

  \item The quantifier-free fragment of $\presexp$, denoted $\QF$, consisting of all q.f.\ formulae.
\end{itemize}

%% file: algos.tex
\section{The core procedure} \label{sec:procedure}

\subparagraph*{Overall organization.}
Our final decision procedures, which will be presented  in Section \ref{sec:complexity}, rely on a ``Master procedure'' (Algorithm~\ref{algo:top-level}) that
interleaves calls to
what are essentially quantifier elimination subroutines \`a la Presburger~\cite{Pre29} and Semenov~\cite{semenovprespower},
respectively, to be explained further below.
The input of~Algorithm~\ref{algo:top-level} is a formula $\Phi$ of \presexp in prenex normal form.
The procedure can be run in two modes,
taking an additional parameter $\Fragment$~accordingly.
This parameter specifies the ``target'' fragment of the~logic:
\begin{itemize}
\item
For $\Phi$ obtained as a translation of a \prespower formula
into \presexp, set~$\Fragment = \QFPC$.
\item
For general formulae of \presexp, and for
the handling of existential \presexp in non-deterministic exponential time,
set $\Fragment = \QF$. 
\end{itemize}
The output of the algorithm is a simplified formula: more specifically,
it is a formula of the form $\exists \vec x. \phi$
                    or $\lnot\exists \vec x. \phi$, where $\phi$
is in $\Fragment$: ``\emph{alternation-free modulo $\Fragment$}'' below.
If the input is a sentence, then the output has no leading quantifiers
and is thus a sentence of $\Fragment$.

\begin{algorithm}[!pt]
  \caption{Master procedure.}\label{algo:top-level}
  \begin{algorithmic}[1]
    \Require
    fragment $\cT \in \{\QF, \QFPC\}$;
    \newline{}\qquad\hspace*{2em}%
    formula $\Phi(\vec y)$ in $\presexp$ in prenex normal form with quantifier-free part from $\cT$
    \newline{}\qquad\hspace*{2em}%
    in which all divisibility constraints are simple
    \Ensure
    an equivalent formula $\Phi'(\vec y)$, alternation-free modulo $\cT$;
    \newline{}\qquad\hspace*{2.8em}%
    if $\Phi$ is a sentence, $\Phi'$ is a sentence of $\Fragment$
    \medskip
    \While{true} \label{top:outer-loop}
      \State $\Pi$ $\gets$
             the shortest quantifier prefix of $\Phi$, possibly with $\lnot$ in front,\newline{}
             \hspace*{1.25em}
             \phantom{$\Pi$ $\gets$}
             such that $\Phi = \Pi . \exists  \vec u . \Psi$ where $\Psi$ is in $\cT$
             (if necessary, rewrite $\forall \vec u$ as $\lnot \exists \vec u. \lnot$) 
             \label{top:decompose-prefix}
      \State $Q$ $\gets$ $\{(\vec u, \Psi)\}$; \ $D$ $\gets$ $\emptyset$ \label{top:init-q-and-d}
      \State $\Pi'$ $\gets$ empty string of quantifiers \label{top:init-pi-prime}
      \Comment $\Pi'$ is a \textbf{global} variable
      \While{$(\vec x, \phi)$ $\gets$ $\text{pop}(Q)$} \label{top:inner-loop}
        \If{$\vec x$ is empty}
          add $\phi$ to $D$
          \label{top:move-to-d}
        \ElsIf{some $x \in \vec x$ does not appear in $\phi$} 
          push pair $(\vec x \setminus \{x\}, \phi)$ to $Q$ 
          \label{top:remove-variable}
        \ElsIf{$\exists x. \phi$ is in $\cT$ for some $x \in \vec x$} 
          push pair $(\vec x \setminus \{x\}, \exists x. \phi)$ to $Q$
          \label{top:fall-into-fragment}
        \ElsIf{some $x \in \vec x$ occurs only linearly in $\phi$}
          bulk push {\funcPA}($x,\vec x, \phi$) 
          to~$Q$%
          \label{top:call-presburger}
        \Else \
          bulk push {\funcLinearise}({\funcSem}($\vec x, \phi$)) to $Q$
          \label{top:call-semenov}
          \Statex\Comment{note: \funcLinearise is no-op if $\cT = \QFPC$}
        \EndIf
      \EndWhile
      \State $\Phi \gets \Pi.\Pi'.\bigvee\limits_{\phi \in D}\phi$ \label{top:update-phi}
      \If{$\Pi$ contains no quantifiers}
        \textbf{return} $\Phi$ \label{top:return}
      \EndIf
    \EndWhile
  \end{algorithmic}
\end{algorithm}

\begin{algorithm}[!pt]
  \caption{Function $\funcLinearise$.}\label{func:linearise}
  \begin{algorithmic}[1]
    \Require a set $S$ of pairs $(\vec x, \theta)$, with $\vec x$ a vector of variables and $\theta$ a formula
    \Ensure 
    for each $(\vec x, \theta)$, a pair $(\vec x, \theta')$ where $\theta \fequiv \theta'$ and,
    for every $x \in \vec x$,
    \newline{}
    \hspace*{1.5em}%
    if $\pow{x}$ only occurs in constraints from \PC in $\theta$, then
    $x$ only occurs linearly in $\theta'$
    
    \medskip
    \If{$\cT = \QFPC$} return $S$
    \Comment{do nothing unless $\cT = \QF$}
    \EndIf
    \label{linearise:hack}
    \For{$(\vec x, \theta) \in S$}
      \State $\vec x'$ $\gets$ vector of all $x \in \vec x$ s.t.\ $\pow x$ only occurs in constraints from \PC in $\theta$
      \For{$x \in \vec x'$}
          \State\label{linearise:cases} update $\theta$
                 by applying all of the following replacements: 
          \vspace{5pt}
          \State\label{lin:in-1} $a \cdot \pow{x} < b$ \ $\isreplacedby$ \  
            $\begin{cases}
              \abs{x} < \lceil \log_2 (b/a) \rceil
              & \text{if $a > 0$ and $b > 0$}
              \\[-1ex]
              \abs{x} > \lfloor \log_2 (b/a) \rfloor
              & \text{if $a < 0$ and $b < 0$}
            \end{cases}$
          \State\label{lin:in-2} $a \cdot \pow{x} < b \cdot \pow{y}$ \ $\isreplacedby$ \ 
            $\begin{cases}
              \abs{x} < \abs{y} + \lceil \log_2 (b/a) \rceil
              & \text{if $a > 0$ and $b > 0$}
              \\[-1ex]
              \abs{x} > \abs{y} + \lfloor \log_2 (b/a) \rfloor 
              & \text{if $a < 0$ and $b < 0$}
            \end{cases}$
          \State\label{lin:mod} $q \divides \pow{x} - r$ \ $\isreplacedby$ \ 
            $\begin{cases}
              (q' \divides \abs{x} - r') \land (\abs{x} \ge r')
              & \text{if $r' = \min \{ s \ge 0 : q \divides 2^s - r \}$,}\\[-1ex]
              & \text{\phantom{if} $q' = \min \{ \rlap{$t$}\phantom{s} > 0 : q \divides r \cdot (2^t - 1) \}$}
              \\
              \abs{x} = r'
              & \text{if $r' = \min \{ s \ge 0 : q \divides 2^s - r \}$,}\\[-1ex]
               & \text{\phantom{if $r' = \min$}$\{ \rlap{$t$}\phantom{s} > 0 : q \divides r \cdot (2^t - 1) \} = \emptyset$}\\
              \bot 
              & \text{otherwise, i.e., $\{s \ge 0 : q \divides 2^s - r \} = \emptyset$}
            \end{cases}$
          \vspace*{1ex}
          \Statex
          \Comment{in the replacements in line~\ref{lin:mod}, search for $s, t \le q - 1$ only}
      \EndFor
    \EndFor
    \State \textbf{return} $S$
  \end{algorithmic}
\end{algorithm}

\begin{algorithm}[!pt]
  \caption{Function \funcPA.}\label{func:presburger}
  \begin{algorithmic}[1]
    \Require variable $x$; vector of variables $\vec x$, where $x \in \vec x$;
    \newline{}\qquad
    formula $\phi(x, \vec y)$ of $\cT$ where $\vec x \setminus \{x\} \subseteq \vec y$ and $x$ appears only linearly in atomic formulae 
    \Ensure a set of pairs $(\vec x, \psi(\vec y))$ where $\psi \in \cT$, 
     and the set of all $\psi$ is a cover for $\exists x . \phi$
    \medskip
    \State $T \gets$ $\{(a,-t(\vec y)) : a > 0,\ a \cdot x + t \in \linterms(\phi) \} \cup \{(-a,t(\vec y)) : a < 0,\ a \cdot x + t \in \linterms(\phi) \} \cup \{(1,0)\}$\label{pres:line1}
    \State $\Gamma \gets \{ \phi\substitute{\frac{t+k}{a}}{x} \land (a \divides t+k) \,:\, (a,t) \in T, k \in [-r,r] \text{ where } r \coloneqq a \cdot \fmod(\phi) \}$\label{pres:line-sub}
    \State \textbf{return} $\{(\vec x, \psi) : \psi \in \simplify(\gamma), \gamma \in \Gamma \}$ 
  \end{algorithmic}
\end{algorithm}

\begin{algorithm}[!pt]
  \caption{Function \funcSem.} \label{func:semcov}
  \begin{algorithmic}[1]
    \Require vector $\vec x$ of variables; formula $\phi(\vec x, \vec z)$ of $\cT$, containing $\pow{x}$ for each $x \in \vec x$
	  \Ensure a set of pairs $(\vec x, \psi(\vec x, \vec z, \vec w))$, where: $\psi \in \cT$, variables $\vec w$ are fresh,\newline{}
    \hspace*{\parindent+1ex}
          $\exists \vec x. \phi(\vec x, \vec z) \fequiv \forall \vec w. \bigvee \exists \vec x. \psi(\vec x, \vec z, \vec w)$
          (i.e., the set of all $\exists \vec x.\psi$ covers $\exists \vec x.\phi$ under $\forall \vec w$),
    \newline{}
    \hspace*{\parindent+1ex}
    and
    in every $\psi$ 
    some $\pow x$ ($x \in \vec x$) only occurs in constraints from \PowComp
    \newline{}
    \hspace*{-\parindent}
    \textbf{Side effect:} $\forall w$ (for each $w \in \vec w$) is added to global variable $\Pi'$ (string of quantifiers)
    \medskip
    \For{$x \in \vec x$}
    \label{sem:outer-loop}
      \State $I \gets $ set of inequalities in $\phi$ outside $\PT$ in which $x$ appears as a power 
      \label{sem:I}
      \State $H$ $\gets$ $\{(\eta,\sigma) : \text{$\eta(\vec x) + \sigma(\vec z) + c < 0$ in $I$; $\eta$ and $\sigma$ homogeneous}\}$
      \label{sem:H}
      \State $\Gamma_x$ $\gets$ $\{\phi\}$
      \For{$(\eta,\sigma) \in H$}
      \label{sem:inner-loop}
        \State $A$ $\gets$ subset of $I$ with these $\eta$ and $\sigma$ (only $c$ varies)
        \label{sem:A}
	\State $2^g$ $\gets$ $2^7 \cdot \bigl(\lambda(\normone\eta + \max \{ \abs{c} : (\eta+\sigma+c < 0) \in A \})\bigr)^2$
        \Statex
        \Comment factors up to $2^g$ are considered ``small''
        \label{sem:g}
        \State $a$ $\gets$ coefficient at $\pow{x}$ in $\eta$
        \label{sem:a}
        \State $V$ $\gets$ variables in $\eta$ except $x$
        \label{sem:V} 
        \State $\beta$ $\gets$ $\pow{x} > 2^g \land (\bigwedge\limits_{u \in V}\pow{x} > 2^g \cdot \pow{u})$
        \label{sem:beta}
        \State $\Gamma_x$ $\gets$ $\big\{ 
        \pow{x} = 2^j \land \gamma\substitute{\alpha\substitute{2^j}{\pow{x}}}{\alpha : \alpha \in A}\,,$
        \label{sem:case-1}
        \State\phantom{$\Gamma_x$ $\gets$ $\big\{$}%
          $\pow{x} > 2^g \land \pow{x} = 2^j \cdot \pow{v} \land \gamma\substitute{\alpha\substitute{2^j \cdot \pow{v}}{\pow{x}}}{\alpha : \alpha \in A}\,,$
          \label{sem:case-2}
        \State\phantom{$\Gamma_x$ $\gets$ $\big\{$}%
          $\beta \land \lambda(a) \cdot \pow{x} < \lambda(\sigma) \land \sigma < 0 \land \gamma\substitute{\true}{\alpha : \alpha \in A}\,,$
          \label{sem:case-3}
        \State\phantom{$\Gamma_x$ $\gets$ $\big\{$}%
          $\beta\land \lambda(a) \cdot \pow{x} < \lambda(\sigma) \land \sigma \geq 0 \land \gamma\substitute{\false}{\alpha : \alpha \in A}\,,$
          \label{sem:case-4}
        \State\phantom{$\Gamma_x$ $\gets$ $\big\{$}%
          $\beta\land \lambda(a) \cdot \pow{x} = \lambda(\sigma) \land \gamma\substitute{\alpha\substitute{\frac{\lambda(\sigma)}{\lambda(a)}}{\pow{x}}}{\alpha : \alpha \in A}\,,$
          \label{sem:case-5}
        \State\phantom{$\Gamma_x$ $\gets$ $\big\{$}%
          $\beta\land \lambda(a) \cdot \pow{x} = 2 \cdot \lambda(\sigma) \land \gamma\substitute{\alpha\substitute{\frac{2 \cdot \lambda(\sigma)}{\lambda(a)}}{\pow{x}}}{\alpha : \alpha \in A}\,,$
          \label{sem:case-6}
        \State\phantom{$\Gamma_x$ $\gets$ $\big\{$}%
          $\beta \land \lambda(a) \cdot \pow{x} > 2 \cdot \lambda(\sigma) \land a < 0 \land \gamma\substitute{\true}{\alpha : \alpha \in A}\,,$
          \label{sem:case-7}
        \State\phantom{$\Gamma_x$ $\gets$ $\big\{$}%
          $\beta\land \lambda(a) \cdot \pow{x} > 2 \cdot \lambda(\sigma) \land a > 0 \land \gamma\substitute{\false}{\alpha : \alpha \in A}$
          \label{sem:case-8}
        \State\phantom{$\Gamma_x$ $\gets$ $\big\{$}%
        $:  \ \ \gamma \in \Gamma_x,\ 0 \leq j \leq g,\ v \in V \big\}$
          \label{sem:end-case}
      \EndFor
    \EndFor
    \State $\Gamma \gets \bigcup\limits_{x \in \vec x} \{ (\bigwedge\limits_{y \in \vec x} \pow{x} \geq \pow{y}) \land \gamma : \gamma \in \Gamma_x \}$
    \label{sem:union-gamma}
    \Comment we next remove all occurrences of $\lambda$
    \State $\Sigma \gets \{ \sigma : \text{$\lambda(\sigma)$ is a subterm of some $\gamma \in \Gamma$} \} \setminus \{ 0 \}$
    \label{sem:delambda-start}
    \For{$\sigma \in \Sigma$}
      \If{$\forall w_\sigma$ is not in $\Pi'$}
      \label{sem:w-intr}
        \State $w_\sigma$ $\gets$ fresh variable; add $\forall w_\sigma$ to $\Pi'$
      \Comment update \textbf{global} $\Pi'$
      \EndIf
    \EndFor
    \State $\Theta$ $\gets$ $\{ (\sigma \neq 0 \land \lnot (\pow{w_\sigma} \leq \abs{\sigma} < 2 \cdot \pow{w_\sigma}))  : \sigma \in \Sigma \}$
           \label{semenov:guard-for-lambda}
    \For{each $\Sigma' \subseteq \Sigma$ and each $\gamma \in \Gamma$}
    \label{sem:for-sigmaprime}
      \State add to $\Theta$ the following formula:
      \State $(\hspace{-2pt}\bigwedge\limits_{\sigma \in \Sigma'}\hspace{-2pt} \pow{w_\sigma} \leq \abs{\sigma} < 2 \cdot \pow{w_\sigma}) \land
         (\hspace{-8pt}\bigwedge\limits_{\sigma \in \Sigma \setminus \Sigma'}\hspace{-9pt} \sigma = 0) \land 
         \gamma
         \substitute{\pow{w_\sigma}}{\lambda(\sigma) : \sigma \in \Sigma'}
         \substitute{0}{\lambda(\sigma) : \sigma \in \Sigma \setminus \Sigma'}$
         \label{semenov:lambda-sub}
    \EndFor
    \vspace*{-1ex}
    \State \textbf{return} $\{ (\vec x, \theta) : \theta \in \Theta \}$
    \label{sem:return}
  \end{algorithmic}
\end{algorithm}

\begin{algorithm}
  \caption{Function $\simplify$.}\label{func:simplify}
  \begin{algorithmic}[1]
    \Require formula $\phi \in {\Fragment}$, except for $\Fragment= \QFPC$, may contain non-simple divisibility constraints
    \Ensure a cover for $\phi$ of formulae from $\cT$, in which all divisibility constraints are simple
    \medskip
    \State $G$ $\gets$ set of non-simple divisibilities in $\phi$
    \State $d$ $\gets$ least common multiple of all divisors in $G$
    \State $\vec t$ $\gets$ all variables $x$ and powers $\pow{y}$ appearing in $G$
    \State $\Gamma$ $\gets$ $\emptyset$
    \For{$r \in [\vec t \to [d]]$}\label{simp:loop}
      \State add $\left((\bigwedge_{t \in \vec t} d \mid t - r(t)) \land \phi\substitute{r(\alpha)}{\alpha : \alpha \in G}\right)$ to $\Gamma$ 
      \label{simp:add-to-gamma}
      \State \quad \textbf{where} $r(q \mid \sum_{i=1}^n a_i \cdot t_i + c) \coloneqq  q \mid \sum_{i=1}^n a_i \cdot r(t_i) + c$
      \Comment simplifies to $\top$ or $\bot$
    \EndFor%
    \State \textbf{return} $\Gamma$
  \end{algorithmic}
\end{algorithm}

The procedure processes blocks of quantifiers at a time, eliminating them one by one.
Each block corresponds to one iteration of the outer \textbf{while} loop.
In line~\ref{top:decompose-prefix} we split the quantifier prefix at the innermost existential block that takes us out of the fragment \Fragment.
There may be a choice as to whether $\lnot$
appears at the beginning of $\Pi$, but
this introduces no ambiguity to the choice of $\vec u$,
because $\forall v. \Psi$ is in $\cT$ iff $\exists v. \Psi$ is in $\cT$.
This follows because both fragments $\cT = \QFPC$ and $\cT = \QF$
are closed under negation.

The organization of the procedure maintains a DNF-like structure.
The set $Q$ acts as a worklist containing the formulae;
intuitively, they are the conjunctions (although not necessarily of atomic formulae).
The $\funcPA$ and $\funcSem$  subroutines
embed Reddy and Loveland's optimization for \PA~\cite{ReddyLoveland78}:
whenever a pair $(\vec x, \phi_1 \lor \phi_2)$ could be produced,
it is split 
into two pairs $(\vec x, \phi_1)$ and $(\vec x, \phi_2)$
evolving independently for as long as possible.
Thus, the DNF-like structure is maintained within each iteration of the outer
\textbf{while} loop of the Master procedure:
\begin{equation*}
\Phi \fequiv \Pi. \Pi' . \left[ \left( \bigvee_{(\vec x, \phi) \in Q} \exists \vec x. \phi \right) \lor \bigvee_{\phi \in D} \phi\right].
\end{equation*}
For each $\phi \in D \cup \{ \phi : \text{$(\vec x, \phi) \in Q$ for some $\vec x$}\}$, we have $\phi \in \cT$.
Pairs from $Q$ are processed in the inner \textbf{while} loop one at a time.
Formulae from $D$ are ``done'' and will only be picked
up again after leaving the current block:
the algorithm will no longer process them within the current block.
Thanks to the DNF-like structure,
our analysis of the parameter growth for an individual pair $(\vec x, \phi)$
can ignore the complexity of the big disjunction (i.e., other pairs in $Q$ and $D$).

Above we have presented Algorithm \ref{algo:top-level}
deterministically:
any deterministic choice can be made in line~\ref{top:inner-loop} when popping an element from~$Q$,
and in lines~\ref{top:remove-variable}--\ref{top:call-presburger} when choosing an appropriate $x \in \vec x$.
This implementation will be employed to obtain the claimed triply-exponential bound for $\prespower$, but not the
$\nexptime$ bound for the existential fragment of $\presexp$. In the latter case, we will only perform the outer loop once. The prefix
$\Pi$ will always be empty, and thus the formula we are processing can always be considered an existentially quantified DNF, or equivalently
a disjunction of existentials.  It suffices to guess one disjunct, corresponding to one element of $Q$ that is satisfiable. Thus, we will
replace a deterministic inner loop that maintains a set of pairs in $Q$ with a non-deterministic algorithm that maintains a single pair from
$Q$. In the deterministic interpretation, calls to  the subroutines  in  lines \ref{top:call-presburger}
          and \ref{top:call-semenov} replace a single element of $Q$ with a set of pairs.
In the non-deterministic interpretation, we guess one pair in the output of the subroutine as the new
element of $Q$.

\subparagraph*{Subroutines.}
We turn from the Master procedure to its subroutines.
The core of the subroutine $\funcPA$ (Algorithm~\ref{func:presburger})
corresponds to Weispfenning's quantifier elimination for \PA~\cite{Weispfenning90}, while  $\funcLinearise$ given in Algorithm~\ref{func:linearise}
is a simple procedure taking
$\powcomp$ atomic formulae like $\pow{x}< \pow{y}$ and transforming them to Presburger formulae $x<y$ by ``taking logs''.
We remark that all three types of divisibility replacements in line~\ref{lin:mod} of {\funcLinearise} are possible:
e.g.,
$(7 \divides \pow{x} - 4) \mapsto (3 \divides \abs{x} - 2) \land (\abs{x} \ge 2)$, and
$(20 \divides \pow{x} - 2) \mapsto (\abs{x} = 1)$, and
$(6 \divides \pow{x} - 3) \mapsto \bot$.

The $\funcSem$ subroutine (Algorithm~\ref{func:semcov})
is  a variation of procedures dating back to Semenov's \cite{semenovprespower}. This
 will be less familiar to most readers, and so we discuss it in detail here.

The purpose  of subroutine $\funcSem$ is to ensure that, in
each of the pairs $(\vec x, \psi)$ in its output, for some
variable $x \in \vec x$ every occurrence is either linear or
in an atomic formula from $\PT$. Across all outputs, the
identity of the variable $x$ may differ. Thus, the
subroutine is essentially ``$\PC$-ifying'' the formula. The
significance of converting atomic formulae to $\powcomp$ is
that powers can then be eliminated by just ``taking
logarithms'', i.e., by invoking 
$\funcLinearise$.
And once a quantified
variable is so heavily processed that it occurs only linearly (outside powers), then by applying
standard Presburger arithmetic quantifier elimination, we can eliminate the variable completely using
$\funcPA$.

To be more precise about how \funcSem assists the Master procedure, consider what happens
when $(\vec x, \psi)$ from the output of \funcSem
gets popped from $Q$ in line~\ref{top:inner-loop} of the Master procedure.
Our actions depend on the chosen fragment
(unless $x$ is eliminated from $\psi$ entirely, in which case line~\ref{top:remove-variable} takes care of it).
\begin{description}
\item[If $\cT = \QFPC$:] for some $x \in \vec x$, we can move $\exists x$ into $\psi$ while still staying in the fragment, since $x$ occurs only in power comparisons (line~\ref{top:fall-into-fragment}).
\item[If $\cT = \QF$:] all occurrences of $x$ became linear
after the execution of $\funcLinearise$ on the output of \funcSem,
so the variable $x$ can be eliminated by \funcPA (line~\ref{top:call-presburger}).
\end{description}
We flag line~\ref{linearise:hack} of \funcLinearise.
It ensures that, in line~\ref{top:call-semenov} of the Master procedure, the output
of \funcSem is directly pushed (in bulk) to $Q$, without linearization, if $\cT = \QFPC$.

\subparagraph*{A look inside subroutine \funcSem.} 
Intuitively, the overall workflow of the Master procedure
is repeated processing of atomic formulae lying within the scope of a particular block of quantifiers.
The constraints we process will  be those containing ``problematic  quantified variables'': those
that appear in atomic formulae involving powers, but are outside the fragment $\powcomp$.
We exhibit the idea using the following subformula:
\begin{equation}
\label{eq:example}
\exists x. \exists y.\quad 3 \cdot \pow{x}- 5 \cdot \pow{y}-z<0.
\end{equation}
Here both $x$ and $y$ are problematic within the sole atomic subformula of the quantified formula.
A major  component of all prior procedures is to replace such  a formula with a quantified DNF  corresponding to a case analysis on the relative values of the problematic
variables.  These cases correspond to lines~\ref{sem:case-1} to~\ref{sem:end-case} of~Algorithm~\ref{func:semcov} and are a cover for the formula under analysis, hence the name ``Semenov cover'' given to the algorithm.
Each case is defined by a $\powcomp$ ``guard'' and,
under the assumption specified in  a given case, we will be able to eliminate one problematic variable within a  constraint,
without introducing new problematic existentially-quantified variables.
Thus, by applying the procedure repeatedly, we can expunge all problematic quantified variables.

The case analysis includes  a guess as to which existentially quantified variable is the largest. In the example, one such
guess is that $\pow{x}$ is the largest. In all the subcases for this guess, we will make $x$ unproblematic. 
The Semenov cover breaks up this guess into several subcases. One subcase is where $\pow{x}$ is not much bigger than
 one of the other power
terms, say $\pow{x} = 4 \cdot \pow{y}$. In such cases we can replace $\pow{x}$ by a constant  multiple of the other term, where the constant is itself
a power of $2$.
Returning to the subcase  mentioned just above, where $\pow{x} = 4 \cdot \pow{y}$, we can replace $\pow{x}$ by $4 \cdot \pow{y}$.
The remaining case is where $\pow{x}$ is significantly bigger than all other power terms like $\pow{y}$;
the threshold for ``significantly'' is set by line~\ref{sem:g} of Algorithm~\ref{func:semcov}.
In this case we further analyze
the most significant digit of the binary expansion for each term.
\begin{mydefinition}
For any integer $N$, let $\lambda(N)$ denote the \emph{highest
power of $2$ not exceeding $|N|$}~\footnote{We will be mostly concerned with this function on positive integers, but using absolute values
gives us the convenience of avoiding partial functions.};
we have $\lambda(0) = 0$ and $\lambda(N) \le |N| < 2 \lambda(N)$.
\end{mydefinition}
Algorithm~\ref{func:semcov} will make use  of intermediate terms that contain $\lambda$'s --- for example,
$\lambda(\sigma)$ for $\sigma$ an ordinary $\presexp$ term. The semantics of such terms is the obvious one, which could be formalized by
translation into $\presexp$, where the function $\lambda$ is definable.

Returning to the example, our ``significantly bigger'' hypothesis implies that 
\[
\lambda(3 \cdot \pow{x}-5 \cdot \pow{y})= \lambda(3 \cdot \pow{x})= 2^{x+1}.
\]
This equality in turn implies that, when  $\lambda(3 \cdot \pow{x})$ is strictly below $\lambda(z)$, the corresponding
inequality in Equation~\eqref{eq:example} is true: 
$\lambda(3\cdot \pow{x}-5 \cdot \pow{y})$ is strictly below $\lambda(z)$, and while each term can differ from the corresponding $\lambda$,
the difference cannot be large enough to make the inequality go the other way.
By a similar argument, in the subcase where $\lambda(3 \cdot \pow{x})$ is at least four times greater than $\lambda(z)$, the inequality
must be false. Here we reason that if  $\lambda(3 \cdot \pow{x} -5 \cdot \pow{y})$ is at least four times greater than $\lambda(z)$, then the offset of each term from its $\lambda$
value cannot change the inequality from true to false.

This leaves some subcases where $\lambda(3 \cdot \pow{x})$ is close to $\lambda(z)$, and in these cases we can substitute away $\pow{x}$ as well.
For example, in the subcase where $\lambda(3 \cdot \pow{x})=\lambda(z)$, we note that $\lambda(3 \cdot \pow{x})=2 \cdot \pow{x}$, and thus
we could replace $\pow{x}$ with $\lambda(z)/2$.  By multiplying through the inequality by $2$, we can eliminate the division by $2$.

\subparagraph*{Using the output of \funcSem.}
The procedure above removed $x$, but there are several caveats. Firstly, each case was associated  with a condition, where the problematic
variable $x$ still appears!
However, these conditions are in $\powcomp$, and therefore all occurrences of $x$ are now unproblematic.
Secondly, in some of our substitutions to eliminate $x$, we introduced $\lambda$ terms, which appear both in the condition describing the
case and in the formula obtained by substituting (assuming the condition).  One solution to this problem, applied
in earlier procedures such as Point's 
\cite{pointexpansion}, is to  extend the signature
with several functions such as $\lambda$, and declare that such conditions are acceptable. In this way one can  obtain quantifier elimination in the
extended signature.
In our $\funcSem$ subroutine, we proceed slightly differently, eliminating $\lambda$ terms in favor of new variables that are bound by \emph{definitional quantifiers}. That is, the new variables
are associated with additional conditions which  define them  from the free variables.
For example,  $\lambda(z)$ can be replaced by $2^w$, with  additional
conditions $2^w  \leq |z| < 2 \cdot 2^w$. There is a unique such $w$ for a given $z$, so the quantification over $w$ can be thought of simultaneously
as an existential conjoined with this condition and as a universal relativized to this condition.
Such quantifications take us
out of  $\powcomp$. But when considered as leading universal quantifiers, they will not increase the quantifier alternation of the global formula --- they will
add on variables to the next quantifier block considered in the Master procedure.
 And since Algorithm~\ref{algo:top-level}
  will process from inner  quantifier blocks outward, the fattening of outer quantifier blocks
does not jeopardize termination of our procedure. 
Note that at the end of processing quantifier blocks outward with  Algorithm~\ref{algo:top-level}, we will have only an outermost block of definitional quantifiers.

\begin{remark*}
If
there are no free variables in the top-level input formula~$\Phi$, the quantified variables will depend only on constants, and thus can be replaced by numbers, leading to a 
sentence that is in the fragment. If the input formula $\Phi$ contained free variables,  these quantifiers can be converted to either existential  or universal quantifiers.
Thus the output of  Algorithm~\ref{algo:top-level} is not just alternation-free modulo  $\Fragment$ but
actually  ``definitional modulo  $\Fragment$'': the only quantifiers outside of $\Fragment$ will be definitional, and thus
the formula can be rewritten to be either universal or existential modulo $\Fragment$.
\end{remark*}

\subparagraph*{Analysis of the core procedure.}
We analyze the procedure, showing in particular
that each of Algorithms~\ref{algo:top-level}--\ref{func:simplify}
correctly implements its specification.
In the sequel we will also need the following facts.

\begin{restatable}{lemma}{OnlySimpleDiv}
\label{l:only-simple-div}
All divisibility constraints
in $D \cup \{ \phi : \text{$(\vec x, \phi) \in Q$ for some $\vec x$}\}$
are simple.
\end{restatable}

Thanks to \Cref{l:only-simple-div}, in the forthcoming complexity analysis we
only consider formulae whose divisibility constraints are simple.

\begin{restatable}{lemma}{TerminationAndCorrectness}
\label{l:termination-and-correctness}
The Master procedure always terminates and, on a formula $\Phi(\vec y)$,
returns an equivalent formula $\Phi'(\vec y)$ such that:
\begin{itemize}
\item $\Phi'(\vec y)$ is equal to either $\exists \vec w. \phi'(\vec w, \vec y)$ or
      $\lnot \exists \vec w. \phi'(\vec w, \vec y)$,
      where $\phi' \in \cT$,
\item if $\Phi$ is a sentence, then $\Phi' \in \cT$
\textup{(}in other words, if $\vec y$ is empty, then $\vec w$ is empty\textup{)}.
\end{itemize}
\end{restatable}

In fact, $\Phi'$ 
starts with $\lnot$ iff the outermost quantifier block of $\Phi$ is existential.

\begin{restatable}{lemma}{EssentiallyPA}
  \label{lemma:essentially-PA}
  Consider a prenex formula $\Pi.\Phi$, with $\Phi$ from $\cT$, in which all variables from the quantifier prefix $\Pi$ appear only linearly. When running the Master procedure on 
$\Pi.\Phi$,~\funcSem is never invoked.
  Moreover, if no variable in $\Phi$ occurs in a power term \textup{(}i.e., it is a formula from~$\PA$\textup{)}, then the quantifier-free formula returned by the procedure is in~$\PA$.
\end{restatable}

%% file: complexity.tex
\section{Decision procedures and their complexity}
\label{section:complexity}\label{sec:complexity}

In this section, we provide our top-level decision procedures, which make use of
the algorithms presented in Section~\ref{sec:procedure}.
We then provide a  complexity analysis that  establishes \Cref{theo:pow-3exp,theo:exexp-nexp}. 
To simplify the exposition, the growth of the formulae returned by the procedure is described with the help of ``parameter tables'' having the following shape:
\begin{center}
  {\renewcommand\arraystretch{1.2}
  \setlength{\tabcolsep}{3pt}
  \begin{tabular}{|g|c|c|c|c|c|c|}
      \hline
      \rowcolor{light-gray}
      & $\param_1(\cdot)$ 
      & $\param_2(\cdot)$
      & \mathcdots
      & $\param_n(\cdot)$ \\
      \hline
      \hline
      $\phi$
      & $a_1$ 
      & $a_2$
      & \mathcdots
      & $a_n$
      \\
      \hline
      \hline
      $\psi_1$
      & $f_{1,1}(a_1,\dots,a_n)$
      & $f_{1,2}(a_1,\dots,a_n)$
      & \mathcdots
      & $f_{1,n}(a_1,\dots,a_n)$
      \\ 
      \hline
      \mathcdots & \mathcdots & \mathcdots & \mathcdots & \mathcdots\\ 
      \hline
      $\psi_m$
      & $f_{m,1}(a_1,\dots,a_n)$
      & $f_{m,2}(a_1,\dots,a_n)$
      & \mathcdots
      & $f_{m,n}(a_1,\dots,a_n)$\\ 
  \hline
  \end{tabular}}
\end{center}

\noindent
In this table, $\phi$,$\psi_1$,\dots,$\psi_m$ are formulae,  $\param_1(\cdot)$,\dots,$\param_n(\cdot)$ are parameter functions 
from formulae to $\Nat$, the \emph{bounds on the input} $a_1,\dots,a_n \in \Nat$ are strictly positive, and all $f_{j,k}$ are functions from $\Nat^n$ to $\Nat$. The table states that 
\begin{center}
  if $\param_i(\phi) \leq a_i$ for all $i \in [1,n]$, 
  then $\param_k(\psi_j) \leq f_{j,k}(a_1,\dots,a_n)$ for all $j \in [1,m]$ and $k \in [1,n]$.
\end{center}
We stress the fact that $a_1,\dots,a_n$ are all at least $1$.
We sometimes assume greater lower bounds on these values (see, e.g., $h \geq 2$ and $a \geq 2$ in the table 
of~\Cref{lemma:presburger-one-round}) in order to simplify the definition of the functions $f_{j,k}$. 
Note that this does not change the semantics of the table. 
We sometimes write the ditto mark \ditto~inside a cell of the table. In that case, the ditto mark represents the value of the cell directly above it (e.g., the rightmost \ditto~appearing in the table of~\Cref{lemma:presburger-one-round} is short for $b+2 \cdot v + 1$).
Unfilled cells correspond to quantities that are not relevant for our bounds.

\paragraph*{\Cref{theo:exexp-nexp}: \nexptime upper bound for existential $\presexp$}
\label{sect:complexity-nexp}

Before arguing for a \nexptime decision procedure for $\exists\presexp$, we analyze the growth of formulae resulting from calls to~\funcPA, \funcSem and \funcLinearise. 

Our analysis of~\funcPA simply merges the analysis of Weispfenning's quantifier elimination for Presburger arithmetic from~\cite{Weispfenning90} (implemented in lines~\ref{pres:line1} to~\ref{pres:line-sub} of \funcPA) with an analysis of~\simplify. Here are the resulting bounds:

\begin{restatable}{lemma}{PresburgerOneRound}
  \label{lemma:presburger-one-round}
  On input $(x,\vec x, \phi(\vec x, \vec z))$ where $x$ only occurs linearly in $\phi$, $\funcPA$ returns a set $\{(\vec x, \psi_1),\dots,(\vec x,\psi_k)\}$
  whose formulae satisfy the parameter table below ${(i \in [1,k])}$:
  \begin{center}
      \renewcommand\arraystretch{1.2}
      \setlength{\tabcolsep}{3pt}
      \begin{tabular}{|g|c|c|c|c|c|c|c|}
          \hline
          \rowcolor{light-gray}
          & $\card\homterms$ 
          & $\card{\linterms}$
          & $\maxvars$ 
          & $\norminf{\homterms(\cdot)}$ 
          & $\norminf{\linterms(\cdot)}$ 
          & $\fmod$
          & $\boolnum$\\
          \hline
          \hline
          $\phi$
          & $h$
          & $q$
          & $v$ 
          & $a$ 
          & $c$ 
          & $m$
          & $b$
          \\
          \hline
          \hline
          $\psi_i$
          & $h$
          & $q$
          & $\min(2v, \card{(\vec x \cup \vec z)})$ 
          & $2 a^2$
          & $a^2 \cdot m + 2 a \cdot c$
          & $a \cdot m$
          & $b + 2v + 1$
          \\ 
          \hline
          $\bigvee_{j=1}^k\psi_j$
          & $h^2 + h$
          & \cellcolor{light-gray}
          & \ditto
          & \ditto
          & \ditto
          & $a^h \cdot m$
          & $k \cdot (\ditto\, + 1)$\\ 
      \hline
      \end{tabular}
  \end{center}

  \noindent
  and $k \leq 4 \cdot q \cdot (a \cdot m)^{2v + 1}$. The running time is in ${(\len{\phi} \cdot a \cdot m)}^{\poly{v}}$.
\end{restatable}

A simple analysis of~\funcSem yields the following bounds. 

\begin{restatable}{lemma}{LemSemOne}
  \label{lemma:sem-one}
  Let $\Theta = \{(\vec x, \theta_1),\dots,(\vec x, \theta_k)\}$ be the output of $\funcSem(\vec x,\phi(\vec x, \vec z))$, where $\vec x = (x_1,\dots,x_n)$ with $n \geq 1$. Then, the following parameter table holds, where $i \in [1,k]$: 
  \begin{center}
    \renewcommand\arraystretch{1.2}
    \setlength{\tabcolsep}{3pt}
    \begin{tabular}{|g|c|c|c|c|c|c|}
        \hline
        \rowcolor{light-gray}
        & $\card\homterms$ 
        & $\maxvars$ 
        & $\norminf{\linterms(\cdot)}$ 
        & $\fmod$
        & $\boolnum$\\
        \hline
        \hline
        $\phi$
        & $h$
        & $v \geq 2$ 
        & $c \geq 2$ 
        & $m$
        & $b$
        \\
        \hline
        \hline
        $\theta_i$
        & $h \cdot (v+10) + n$
        & $v$ 
        & $2^{12} v^2 c^3$
        & $m$
        & $b + h \cdot (v+15) + n$
        \\ 
        \hline
        $\bigvee_{j=1}^k\theta_j$
        & 
        $h \cdot 2^7  v^3 \log(c) + n^2$
        & \ditto
        & \ditto
        & \ditto
        & $k \cdot (\ditto + 1)$\\ 
    \hline
    \end{tabular}
  \end{center}

  \noindent
  where $k \leq (v+1)^{10h} \cdot \log(c)^h \cdot n$. Moreover, 
  \begin{enumerate}
    \renewcommand{\theenumi}{(\roman{enumi})}
    \renewcommand{\labelenumi}{\theenumi}
    \item\label{semcover-add-item1} at most $h$ universal quantifiers are added to the global variable $\Pi'$;
    \item\label{semcover-add-item2} the running time is in $(\len{\phi} \cdot n)^{\poly{h}}$; and
    \item\label{semcover-add-item3} for every $i \in [1,k]$ there are at most $h$ terms~$t \in \homterms(\theta_i)$
    that contain some variable from $\vec x$ and satisfy $(t + c' < 0) \in \linterms(\theta_i)$ with $t+c' < 0$ not in $\PC$, for some~${c' \in \Zed}$.%
  \end{enumerate}
\end{restatable}

\vspace{8pt}

Note that an estimate of~$\norminf{\homterms{(\theta_i)}}$ is missing from \Cref{lemma:sem-one}. For~\funcSem, this parameter grows similarly to~$\norminf{\linterms{(\theta_i)}}$, which by definition always bounds $\norminf{\homterms{(\theta_i)}}$.
We also note that~\Cref{lemma:sem-one} gives an upper bound on the number of global variables added to $\Pi'$. This bound is later required to analyze~\prespower, but is not needed in the context of deciding sentences from $\exists\presexp$. Indeed, from~\Cref{l:termination-and-correctness}, 
in this latter case $\Pi'$ is empty.

In computing the upper bounds on $\norminf{\homterms{(\theta_i)}}$ and $\norminf{\homterms{(\bigvee_{j=1}^k\theta_j)}}$
keep in mind that in lines~\ref{sem:case-1} to~\ref{sem:case-8} of~{\funcSem}
we perform ``tailored substitutions'': we only replace~$\pow{x}$ in linear terms $\alpha \in A$ with either a constant, a unique (given $\alpha$) expression $\lambda(\sigma)$, 
or a multiple of a power $\pow{y}$, where $y$ appears in $\alpha$. Our analysis tracks the impact of iterating these types of 
replacements on the number of homogeneous terms.

We continue by analyzing~\funcLinearise. Here the bounds are quite simple, but one observation is in order: line~\ref{lin:mod} might require iterating through all the residue classes of $q$ in order to find suitable~$q'$ and~$r'$. Since $q$ is encoded in binary, this yields an exponential running time for \funcLinearise (see $m$ below), as shown in the following lemma.

\begin{restatable}{lemma}{LinBounds}\label{lemma:lin-bounds}
  Consider a set $S = \{(\vec x, \theta_1),\dots,(\vec x, \theta_k)\}$ where $\vec x = (x_1,\dots,x_n)$, and let $\ell$ be the maximum number of variables appearing in some~$\theta_i$. On input $S$, \funcLinearise returns a set $\{(\vec x, \theta_1'),\dots,(\vec x, \theta_k')\}$
  with bounds as in the following table, for all $j \in [1,k]$:
  \begin{center}
    \renewcommand\arraystretch{1.2}
    \setlength{\tabcolsep}{3pt}
    \begin{tabular}{|g|c|c|c|c|c|c|}
        \rowcolor{light-gray}
        \hline
        & $\card\homterms$ 
        & $\maxvars$ 
        & $\norminf{\homterms(\cdot)}$ 
        & $\norminf{\linterms(\cdot)}$ 
        & $\fmod$
        & $\boolnum$\\
        \hline
        \hline
        $\theta_j$
        & $h$
        & $v$ 
        & $a$ 
        & $c \geq 2$ 
        & $m \geq 2$
        & $b$
        \\
        \hline
        \hline
        $\theta_j'$
        & $h + (6 \cdot \ell+2) \cdot n$
        & $v$ 
        & $a$
        & $\max(c,m)$
        & $m^2$
        & $24 \cdot b$
        \\ 
        \hline
    \end{tabular}
\end{center}

\noindent
The running time is in $\poly{k,\max_{i=1}^k\len{\theta_i}, m, n}$.
\end{restatable}

We now complete the description of the non-deterministic algorithm deciding~$\exists \presexp$ in~$\nexptime$.
As a preliminary step, the algorithm runs~\simplify on the matrix of the input existential sentence~$\Phi$, guessing a residue class for each variable and power, and obtaining an existential sentence where all the divisibilities are simple. The algorithm puts that sentence in prenex form. 
Afterwards, the algorithm follows~\Cref{algo:top-level} with $\cT = \QF$. The algorithm
returns a quantifier-free sentence, which we can then evaluate in polynomial time to determine the final output.

The description above is deterministic, and would be doubly-exponential. We now modify it to obtain a  non-deterministic
exponential-time algorithm.
We replace $\text{pop}(Q)$ in line~\ref{top:inner-loop}, 
as well as other forms of iteration inside~\funcPA and~\funcSem, with non-deterministic guesses.
More precisely, when \funcSem is called, it guesses a variable $x \in \vec x$ in line~\ref{sem:outer-loop}, iterates (deterministically) over every $(\eta,\sigma) \in H$ in line~\ref{sem:inner-loop}, and guesses only one of the cases in lines~\ref{sem:case-1} to~\ref{sem:case-8}. As a result, in the non-deterministic version of~\funcSem, the variable $\Gamma$ in line~\ref{sem:union-gamma} contains a single formula~$\gamma$. Since $\Phi$ is an existential sentence, the various terms $\sigma$ considered by~\funcSem are always $0$. Hence, $\lambda(\sigma)$ is $0$ and lines~\ref{sem:delambda-start} to~\ref{semenov:lambda-sub}
have no effect on the subroutine, which simply returns a singleton set containing the pair $(\vec x, \gamma)$.
For~\funcPA, non-deterministic guesses replace the iterations done in line~\ref{pres:line-sub}, as well as the ones performed in line~\ref{simp:loop} of~\simplify (as done in the aforementioned preliminary step of the algorithm). 
As usual, the overall algorithm returns true if one such (non-deterministically derived) formula~$\Psi$ is valid.

We now analyze the complexity.
The non-deterministic versions of~\funcPA and \funcSem described above always return singleton sets containing a pair of the form~$(\vec x, \theta)$. Then, by the correctness of \funcPA and \funcSem, we conclude that on an input sentence $\Phi$ containing $n$ quantified variables, 
the non-deterministic version of~\Cref{algo:top-level} never calls each of the subroutines~\funcPA, \funcSem and \funcLinearise more than $n$ times. By looking at the bounds on~$\psi_i$, $\theta_i$ and $\theta_j'$ 
from~\Cref{lemma:presburger-one-round,lemma:sem-one,lemma:lin-bounds} we conclude that these $3n$ subroutine calls (non-deterministically) return a formula that never requires more than exponential space to be represented. 
Since $\Phi$ is a sentence and $\cT = \QF$, the non-deterministic algorithm will eventually obtain a formula~$\Psi$ with no variables, only constants, which can be evaluated in exponential time.

\paragraph*{\Cref{theo:pow-3exp}: \threeexptime upper bound for \prespower}
\label{sec:threexpcomp}

We now move to~\prespower. Let $\Phi$ be obtained by translating a sentence of~\prespower into a prenex sentence of~\presexp without divisibility constraints (i.e., replace each~$\pp(x)$ with~$\exists y .\, x = \pow{y}$, each $q \mid t$ with $\exists z .\, t = q \cdot z$, and bring the resulting sentence in prenex form). Note that this translation is in polynomial time, and that each variable in $\Phi$ appears either always linearly or always in a power.
The algorithm to decide $\Phi$ is described below:
\begin{center} 
\begin{algorithmic}[1]
  \State $\Psi_1$ $\gets$ run~\Cref{algo:top-level} with $\cT = \QFPC$, on $\Phi$
  \Comment as $\Phi$ is a sentence, $\Psi_1 \,{\in}\hspace{3pt}\PowComp$.
  \State $\Pi.\Psi_2(\vec x)$ $\gets$ prenex form of $\Psi_1$ \Comment $\Psi_2$ q.f.; $\vec x$ are the variables appearing in $\Pi$
  \State $\{(\vec x,\Psi_3)\}$ $\gets$ $\funcLinearise(\{(\vec x,\Psi_2)\})$ with $\cT = \QF$
  \Comment $\Psi_3(\vec x)$ belongs to~\Oct
  \State\label{texp:line4}$\Omega$ $\gets$ run~\Cref{algo:top-level} with $\cT = \QF$, on $\Pi.\Psi_3$ 
  \Comment $\Omega$ does not contain variables
  \State evaluate truth of $\Omega$ 
\end{algorithmic}
\end{center}
Above we highlight the fact that, after the first invocation of~\Cref{algo:top-level}, 
we obtain a formula from~$\PowCmp$ which is then manipulated by~\funcLinearise 
into a formula from integer octagon arithmetic (\Oct).
Then, in order to estimate the running time of this algorithm, we need to study the running time of~\Cref{algo:top-level} on inputs that either come from~$\prespower$ or are from~\Oct. Let us discuss the latter case first. 

Since~\Oct is a fragment of Presburger arithmetic, 
line~\ref{texp:line4} above fundamentally runs Weispfenning's quantifier elimination procedure for Presburger arithmetic (see~\Cref{lemma:essentially-PA}), plus calls to~\simplify. It turns out that, on 
formulae from~\Oct, this procedure only runs in exponential time, as summarized in the following proposition.

\begin{restatable}{proposition}{PropOctaAlt}
  \label{prop:octa-alt}
  Let $\cT = \QF$.
  Consider a formula $\Phi(\vec z)$ from integer octagon arithmetic \textup{(\Oct)} in prenex form and having $\alt(\phi) = \ell \geq 1$ quantifier blocks, each with at most $n\geq 1$ variables. On input $\Phi$,~\Cref{algo:top-level}  returns a formula $\Psi$ with bounds:
  \begin{center}
    \renewcommand\arraystretch{1.2}
    \setlength{\tabcolsep}{3pt}
    \begin{tabular}{|g|c|c|c|c|c|c|}
        \hline
        \rowcolor{light-gray}
        & $\card\homterms$ 
        & $\maxvars$ 
        & $\norminf{\homterms(\cdot)}$ 
        & $\norminf{\linterms(\cdot)}$ 
        & $\fmod$
        & $\boolnum$\\
        \hline
        \hline
        $\Phi$
        & $h \geq 2$
        & $2$ 
        & $1$ 
        & $c \geq 2$ 
        & $m \geq 2$
        & $b$
        \\
        \hline
        \hline
        $\Psi$
        & $4 \cdot \card{\vec z}^2$
        & $2$ 
        & $1$
        & $2^{\ell \cdot n}(c + m)$
        & $m$
        & $(m + c + \card{\vec z})^{5 \cdot n^2 (\ell+1)^2} \cdot b$
        \\ 
        \hline
    \end{tabular}
\end{center}

\noindent
The running time of the procedure is in ${(\len{\Phi} \cdot m)}^{\poly{\ell,n}}$.
\end{restatable}

The proof of this proposition is by induction on $\alt(\phi)$, and essentially follows the standard arguments 
to bound the running time of the quantifier elimination procedure for Presburger arithmetic. 
The key ingredient that leads to the bounds above is that, for $\Oct$, the natural numbers $a$ in line~\ref{pres:line-sub} of~\funcPA 
are always $1$. This has two effects. 
Firstly, it shows that $\Oct$ admits quantifier elimination, i.e., while running the procedure no atomic formulae outside $\Oct$ can arise. This is best witnessed by looking at line~\ref{pres:line-sub} in~\funcPA. 
There, 
the divisibility constraints $a \mid t+k$ are trivially satisfied,
and we are replacing $x$ with a term of the form $\pm y + c$ for some $c \in \Zed$. From these substitutions, only constraints from~$\Oct$ or constraints of the form $\pm 2 \cdot y < b$ can arise, and the latter are 
normalized to $y \leq \floor{\frac{b-1}{2}}$
           or $y \geq \ceil {\frac{1-b}{2}}$ as explained in~\Cref{section:preliminaries}. 
The second effect is on the growth of the constants.  The variable $r$ in line~\ref{pres:line-sub} only depends on $\fmod(\Phi)$, which now does not grow during the procedure, and on $\norminf{\linterms(\Phi)}$, which grows only exponentially in the number of variables in $\Phi$. 

We now move to the running time of~\Cref{algo:top-level} on inputs that come from~$\prespower$. The properties of this procedure are summarized 
in the next proposition.

\begin{restatable}{proposition}{PASEMALL}\label{prop:PA-Sem-All}
  Let $\cT = \QFPC$.
  Let $\Phi(\vec y)$ be a formula from~$\presexp$ in prenex normal form, with no divisibility constraints, and in which each quantified variable appears either only linearly or only in powers.
  Suppose $\Phi$ has $\alt(\Phi) = \vB$ quantifier blocks, each with at most $\vL \geq 1$ variables occurring linearly and each block having at most
$\vE \geq 1$ variables occurring 
  in powers.  
  On input~$\Phi$,~\Cref{algo:top-level} returns a formula~$\Psi$ with bounds as in the following table:
  \begin{center}
    \renewcommand\arraystretch{1.4}
    \setlength{\tabcolsep}{4pt}
    \begin{tabular}{|g|c|c|c|c|c|c|}
        \hline
        \rowcolor{light-gray}
        & $\card\homterms$ 
        & $\maxvars$ 
        & $\norminf{\linterms(\cdot)}$ 
        & $\fmod$
        & $\boolnum$\\
        \hline
        \hline
        $\Phi$
        & $h \geq 2$
        & $v \geq 2$ 
        & $c \geq 4$ 
        & \cellcolor{light-gray}
        & $b$
        \\
        \hline
        \hline
        $\Psi$
        &
          $\vH \coloneqq (E \cdot h \cdot \log c) \uparrow {(2 \cdot v)^{2^4 \cdot L \cdot B^2}}$
        & $2^{B \cdot L} v$
        & $2^{\vH}$
        & $2^{\vH}$
        & $b \cdot 2^{\vH}$\\
        \hline
    \end{tabular}
  \end{center}

  \noindent
  and the number of quantifiers added to $\Pi'$ is at most $\vH$. The running time of the procedure is in
    $\len{\Phi} \uparrow (E \cdot h \cdot \log c) \uparrow v \uparrow \poly{L,B}$.
\end{restatable}
In the above proposition, $a \uparrow b \coloneqq a^b$ is the exponentiation function and, following Knuth's up-arrow notation, it is right-associative.
In view of our bounds for one iteration of~\funcSem given in~\Cref{lemma:sem-one}, the bound on $\card{\homterms(\Psi)}$ should 
seem somewhat surprising. We know from the correctness of \funcSem that, after a call to~\funcSem, 
one of the variables appearing in powers will only occur in constraints from~\PowCmp. 
Since all these variables need to have this property before moving to the next quantifier block, \funcSem must be chained at least~$E$ times within a block, $E$ being the number of variables occurring in powers in the block.
Then, from the bound $\card{\homterms(\theta_j)} \leq h \cdot (v+10)+n$ in~\Cref{lemma:sem-one}, one might expect $\card{\homterms(\Psi)}$ to be roughly
$h \cdot v^E$, thus exponential in~$E$ even for a single block of quantifiers.
\Cref{prop:PA-Sem-All}, however, proves otherwise: $\card{\homterms(\Psi)}$ is only polynomial in $E$.
For this, we need to sharpen the correctness statement for \funcSem.

We already know that
in each output formula some variable~$x$ is made unproblematic, and
no new problematic occurrences of variables are created.
We prove a stronger statement: namely, that every \emph{rewriting} of
a homogeneous term (by a substitution in lines~\ref{sem:case-1}--\ref{sem:case-8})
makes the number of problematic variables in that term decrease.
This is possible thanks to the tailoring of these substitutions to individual
inequalities, 
which allows us to track
the evolution of each term independently of the rest of the formula.
As a result:
\begin{itemize}
\item
a formula that is placed into $D$ at the end of  processing a single quantifier block
can be a result of $E$ chained calls to~\funcSem, but
\item
the number of calls to~\funcSem that rewrite an individual term during
its evolution is bounded by
the heft.
\end{itemize}
Therefore, we can iterate
the above-mentioned bound $\card{\homterms(\theta_j)} \leq h \cdot (v+10)+n$
just $2^L \cdot v$ times instead of $E$ times.
Thus $\card{\homterms(\Psi)}$ in~\Cref{prop:PA-Sem-All} is found to be exponential in $v$
and only polynomial in~$E$.
We remark that if $\card{\homterms(\Psi)}$ were instead found to be exponential 
in~$E$, then the algorithm would  have no hope of running in elementary time. 
This is because, after a block of quantifiers is considered, 
$E$ increases by the number of variables introduced in~\funcSem, which from \Cref{lemma:sem-one} is roughly the number of homogeneous terms.

To prove~\Cref{theo:pow-3exp} it suffices to chain the bounds and running times of~\Cref{prop:PA-Sem-All}, \Cref{lemma:lin-bounds} and~\Cref{prop:octa-alt}, according to the algorithm given at the beginning of the section.

\subparagraph*{Avoiding quadruply exponential numbers.} 
It may not be immediately evident from the bounds in the various tables 
why we do not perform quantifier elimination eagerly and instead
run \Cref{algo:top-level} without fully eliminating quantifiers first (in mode~$\QFPC$),
then call $\funcLinearise$, and 
only afterwards eliminate the remaining quantifiers by
running \Cref{algo:top-level} again (now in mode~$\QF$).
In fact, this sequence is fundamental for obtaining a $\threeexptime$ procedure.
Consider the formula 
$\Psi \coloneqq q \mid \pow{x} - r \land y \geq \pow{x}$.
For specific values of $q$ and $r$, the smallest $\abs{x}$ satisfying $\Psi$
might be $q-1$.
If $\Psi$ is a subformula obtained during quantifier elimination, then,
according to~\Cref{prop:PA-Sem-All}, $q$ might have a triply exponential
magnitude relative to the input size.
This means that the smallest $y$ satisfying $\Psi$ might have a quadruply
exponential magnitude.
Eliminating $x$ and $y$ in this case would lead to a quadruply exponential
blow-up in the number of disjuncts to be considered during quantifier
elimination. 
Our strategy avoids this problem by delaying (if necessary) the elimination of
$x$ and $y$ until we obtain a formula in~$\powcomp$.
Calling $\funcLinearise$ reduces the reasoning to the exponents, which are
triply exponential at worst.%

\subparagraph*{An observation on~$\Oct$.} The bounds for integer octagon arithmetic presented in~\Cref{prop:octa-alt} reveal not only that this logic admits an exponential-time quantifier elimination procedure, but also that the satisfaction problem for this theory can be solved in~{\pspace}.
Indeed, observe that the bound 
on $\norminf{\linterms(\Psi)}$ given in~\Cref{prop:octa-alt} implies that all constants and coefficients appearing in the output formula $\Psi$ have polynomial bit length. Then, one can apply the standard quantifier relativization 
algorithm from~$\PA$ to obtain a \pspace procedure for $\Oct$. 
Briefly, the quantifier relativization procedure for~$\PA$
first replaces every quantifier $\exists x . \phi$ in the input formula with a \emph{bounded quantifier} $\exists x \in [-f(\phi),f(\phi)] .\, \phi$, where $f \colon \PA \to \Nat$, and then iterates through all (finitely many) values the quantified variable can take, searching for a solution to the formula. The bound on~$\norminf{\linterms(\Psi)}$ obtained for~$\Oct$ implies that $f(\Psi)$ has bit-length that is at most polynomial in $\len{\Psi}$.
See~\cite{ReddyLoveland78} for more information on quantifier relativization.

\subparagraph{On the non-elementary bound for~$\presexp$.}
We can use our method to decide the full theory of $\presexp$. We proceed as in the existential case, first simplifying divisibility constraints
and putting the sentence in the normal form required to run the Master procedure.  We run the Master procedure with the quantifier-free fragment, to get an equivalent quantifier-free sentence. 
We evaluate this sentence and return its truth value.

Of course  \presexp is a \tower-complete logic, so the procedure above must run in non-elementary time.
We provide some insights on why this is the case.
One of the ingredients that guarantee that our procedure for $\prespower$ runs in \threeexptime is that we are able to postpone 
calls to~\funcLinearise to after~\Cref{algo:top-level}.
In $\presexp$ this is not possible: since each variable can appear both linearly and in powers,~\funcLinearise must be invoked after each call to~\funcSem, in order to ``linearize'' a variable, and then eliminate it with~\funcPA. 
However,~\funcSem adds, in the worst case, a number of additional variables that is roughly the number~$h$ of homogeneous terms in the formula 
(see~\Cref{lemma:sem-one}). When the next quantifier block is considered, these variables must all be linearized and eliminated with~\funcPA. 
As indicated in  the table of~\Cref{lemma:presburger-one-round} (leftmost column of the last row), in eliminating one variable the number of homogeneous
terms can square. Thus in eliminating $h$ variables,  the number of homogeneous terms of the resulting formula becomes exponential in $h$. This ``$h^h$'' dependency makes the algorithm  run in non-elementary time (in fact~$\tower$).

%% file: conc.tex
\section{Conclusion} \label{sec:con} We have proven new
elementary upper bounds for~$\prespower$, and for the
existential fragment of $\presexp$. We believe this is a
step towards understanding which decidable arithmetic
theories have elementary bounds, and moreover that our
method extends to provide elementary bounds for any prefix
class of $\presexp$, but we leave this for future work. 
Our results open several research directions, which we now summarize.

\subparagraph*{Quantifier elimination for open formulae.}
Our technique provides the same bounds for eliminating quantifiers
from open formulae. In the case of $\presexp$ formulae with free variables $\vec x$, our procedures
will produce a formula with one block of definitional quantifiers over
variables $z_1 \ldots z_n$ --- but where the definition of $z_{i+1}$ may involve
earlier $z_j$ in addition to $\vec x$.
Such formulae can be made quantifier-free by moving to an appropriate signature,
e.g., including the function~$\lambda$.

\subparagraph*{Tighter bounds for~$\prespower$.}
It is well-known that, using the bounds on the formulae
returned by quantifier elimination procedures for \PA,
one can derive a $\twoaexppol$
quantifier relativization procedure for \PA~\cite{Weispfenning90}.
Here $\twoaexppol$ is the
class of all problems that can be decided with an
alternating Turing machine running in doubly exponential
time and performing a polynomial number of alternations. In
fact, $\PA$ is complete for this class
under polynomial-time reductions~\cite{Berman80}.
Our $\threeexptime$
procedure for~$\prespower$ shows that, in terms of
deterministic time complexity, this theory is not harder to
decide than $\PA$. However, at this stage obtaining a
$\twoaexppol$ quantifier relativization algorithm from the
bounds of our procedure seems not easy.

\subparagraph*{Automata-based decision procedures.} As \prespower is a fragment of B\"uchi arithmetic,
it also admits a representation 
by finite automata.
It appears plausible that the automata-based procedure
for $\PA$~\cite{Klaedtke08,DurandH10} could be
adapted to $\prespower$.
However, having the procedure
run in $\threeexptime$ might be very challenging. 
This is due to the fact that, as observed above,
$\prespower$ formulae may require numbers of quadruply
exponential magnitude; instead of triply exponential as in the case of~$\PA$.

\subparagraph*{Geometric decision procedures.} A class of regular expressions corresponding to $\prespower$
was already defined by Semenov~\cite[Theorem~5]{semenovprespower}.
These expressions can
be seen as an extension of semilinear sets~\cite{Par66,GinsburgS66}, so it is
conceivable that there is an elementary decision
procedure for $\prespower$ which is based on geometry and manipulates these objects directly.
However,
similarly to the automata-based approach, making such a procedure run in $\threeexptime$,
as the recent one for \PA~\cite{0001HM22} does, appears challenging.

\subparagraph*{Tighter bounds for~$\exists\presexp$.}
An obvious question is whether
our upper bound for the existential fragment can be improved. For comparison,
the existential fragment of {\buchi} arithmetic is known to be in
{$\np$}~\cite{christophnparith}. While the same may be true for $\exists\presexp$, 
it would be very surprising if such a result were to be proved 
with a technique similar to the one in our paper.
In our \nexptime algorithm for
$\exists\presexp$, the main source of blow-up is
the use of Weispfenning's quantifier elimination procedure to
eliminate linearly occurring variables. Quantifier elimination 
is known to be often non-optimal when it comes to deciding 
existential fragments of logics, and this is the case for 
$\exists\PA$, the existential fragment of Presburger arithmetic.
A possible avenue to improve the \nexptime upper bound 
would be to look at geometric procedures, which in 
the context of $\exists\PA$ perform much better.

Improving the $\np$ lower bound is also challenging.
There are several extensions of $\exists\PA$ that 
currently fall between $\np$ and $\nexptime$. These include $\exists\PA$ with pre-quadratic constraints~\cite{GIVAN2002105,RayaHK23} and $\exists\PA$ with divisibility constraints~\cite{LechnerOW15}.
One idea is to exploit the ability of $\exists\presexp$ to  express a pairing function, that is,
an injection from $\Nat \times \Nat$ to $\Nat$, with, e.g.,
the formula $z = 2^{\abs{2x}} + 2^{\abs{2y+1}}$ \cite[p.~55]{comptonhenson}. 
Pairing functions are known to lead to non-elementary lower bounds in the presence of quantifier alternation, and
an interesting direction would be to study their effect on existential theories.

%% file: correctness.tex
\section{Correctness of the core procedure} \label{sec:correctness}

In this section we demonstrate that our core procedure is
correct.

\subsection{Definitions and notation}

By $\freevars(\phi)$ we denote the set of free variables of a formula $\phi$.

The \emph{semantics} $\sem{\Phi}$ of a formula $\Phi$
from $\presexp$ is a subset of all variable assignments
$\nu \in [\V \to \Zed]$, and it is recursively defined
as follows: 
\begin{align*}
 \sem{\top}
   &\coloneqq [\V \to \Zed],
 &\sem{\bot} &\coloneqq \emptyset,\\
 \sem{t_1 < t_2} 
   &\coloneqq \{ \nu \in [\V \to \Zed] : \nu(t_1) < \nu(t_2) \},
 &\sem{\lnot \Phi} 
   &\coloneqq [\V \to \Zed] \setminus \sem{\Phi},\\
 \sem{q \mid t_1} 
   &\coloneqq \{ \nu \in [\V \to \Zed] : q \mid \nu(t_1) \},
 &\sem{\Phi \land \Psi}
   & \coloneqq \sem{\Phi} \cap \sem{\Psi},\\ 
 \sem{\exists x\, \Phi} 
   &\coloneqq \{ \nu \in [\V \to \Zed] : \nu\substitute{n}{x} \in \sem{\Phi} \text{ for some } n \in \Zed \},
 & \sem{\forall x\, \Phi} &\coloneqq \sem{\lnot \exists x \lnot \Phi},
\end{align*}
where $\nu\substitute{n}{x}$ stands for the variable
assignment obtained from $\nu$ by updating $\nu(x)$
to~$n$; and 
given a term $t = \sum_{i \in I} a_i
\cdot \pow{{x}_i} + \sum_{j \in J} b_j \cdot {x}_j + c$, 
$\nu(t)$ stands for the integer $\sum_{i \in I} a_i \cdot
2^{\abs{\nu({x}_i)}} + \sum_{j \in J} b_j \cdot
\nu({x}_j) + c$.
We say that $\Phi$ is satisfiable whenever
$\sem{\Phi} \neq \emptyset$.

\subsection{Correctness of \simplify, \funcPA, and \funcSem}
\label{sec:correctness:simplify,PA,Sem}

We begin with proofs that subroutines correctly implement their specifications. Note that
termination of these subroutines is clear, since they have no unbounded loops.

\begin{lemma}
\label{l:correct-simplify}
The function \simplify correctly implements its specification:
\begin{itemize}
\item For $\Fragment = \QF$: given a formula $\phi \in \QF$, it outputs
a cover for $\phi$ that consists of formulae from $\QF$, where all divisibility
constraints are simple.
\item For $\Fragment = \Sem$: given a formula $\phi$ in which
every variable appears either always linearly or always as a power, and
moreover every bound variable appears only in atomic formulae from $\PT$,
it outputs a cover for $\phi$ that consists of formulae from $\Sem$.
\end{itemize}
\end{lemma}

\begin{proof}
A direct inspection verifies that \simplify always produces a cover for $\phi$:
indeed, the \textbf{for} loop in its body enumerates all possible assignments
of residues modulo~$d$ to variables and powers appearing in non-simple divisibility
constraints. For each assignment, a formula is output where these remainders
are asserted and every non-simple divisibility constraint is evaluated
(to $\true$ or $\false$).
Notice that this transformation is correct
by our definition of $\Sem$: indeed, by our assumption,
every variable that is bound by a quantifier inside~$\phi$ must
only appear in atomic formulae from \PowComp.

Further, as \simplify never introduces quantifiers, it is immediate that
in the case $\Fragment = \QF$ the output satisfies the specification.
In the case $\Fragment = \Sem$, it remains to observe that
the constraints imposed on the input formula~$\phi$ must extend
to every formula~$\psi$ in the output of the function.
This completes the proof.
\end{proof}

We now turn our attention to the function \funcPA. 
Its input contains a formula of the form $\phi(x, \vec y)$, 
so in particular $x \not\in \vec y$. 
Since each output pair is of the form $(\vec x, \psi(\vec y))$, 
it follows that each formula $\psi$ contains no
occurrences of the variable~$x$, that is, this variable has
been eliminated. 

The correctness of \funcPA is a standard fact going back to
Presburger~\cite{Pre29}, and in particular a very similar
statement appears in Weispfenning's
paper~\cite[Lemma~2.6]{Weispfenning90}. 
There are, however, a few differences:

\begin{itemize}
  \item 
  The choice of parameter~$r$ in
  line~\ref{pres:line-sub} of \funcPA is tighter than that
  of Weispfenning's. 
  More precisely, to define $r$, Weispfenning's procedure relies on computing a natural number~$n$ defined as the absolute value of the product of all coefficients of the variable $x$ that appear in terms
  (for each collection of terms that are only distinct due to their constant, the coefficient is only taken once).
  Our definition of $r$ does not require this number~$n$. 
  \item Weispfenning's procedure is written for linear
  arithmetic only and  does not handle powers. But
  our assumption that the variable~$x$ appears only linearly
  is sufficient for its correctness in the extension to
  {\presexp}. This common property of quantifier elimination procedures for \PA is well-known and relied upon in,
  e.g., Semenov's paper~\cite{semenovprespower}.
  \item The formula $\phi$ in the input to \funcPA is only assumed to belong to the fragment~\Fragment and, as such, may not be quantifier-free in the case $\Fragment = \Sem$. 
\end{itemize} 

\begin{lemma}
\label{l:correct-presburger}
The function \funcPA correctly implements its specification.
More precisely, suppose that it is
given a variable $x$, a vector of variables $\vec x$ with $x \in \vec x$,
and
a formula $\phi(x, \vec y)$ of $\cT$ where $\vec x \setminus \{x\} \subseteq \vec y$ and $x$ appears only linearly in atomic formulae.
Then it outputs
a set of pairs of the form $(\vec x, \psi(\vec y))$ where $\psi \in \cT$ and the set of all $\psi$ is a cover for $\exists x . \phi$.
\end{lemma}

\begin{proof}

The main step in our \funcPA is substitution made to the
input formula~$\phi$ in line~\ref{pres:line-sub}, the
variable~$x$ is replaced with~$\frac{t+k}{a}$ where $a, k
\in \Zed$, $a > 0$, and $t$ is a term in which $x$ does not appear.
Let us first check that all variables of the following two kinds are free variables in $\phi$:
\begin{itemize}
  \item all variables appearing in the term~$t$, and
  \item all variables in all atomic formulae in which substitutions
  occur.
\end{itemize}
In the case $\Fragment = \QF$, the check is trivial.
Consider $\Fragment = \Sem$. 
Recall that $\phi \in \Fragment$,
and that inequalities of the form $t_1 < t_2$ are converted into $t_1 - t_2 < 0$
under our normalization convention, so that instead of $t_1, t_2 \in \linterms(\phi)$ we only have $t_1 - t_2 \in \linterms(\phi)$.
By the definition of \Sem, all variables bound inside~$\phi$ may only
appear in powers. We will show that
\begin{equation}
  \label{proof12:iff}
  \exists x. \phi \fequiv \bigvee_{\gamma \in \Gamma} \gamma.
\end{equation}

\bigskip

\noindent 
$(\Leftarrow)$: Consider $\gamma \in \Gamma$. There are 
$(a,t) \in T$ and $k \in [-r,r]$ with $r = a \cdot \fmod(\phi)$ such that $\gamma = \phi\substitute{\frac{t+k}{a}}{x} \land (a \divides t+k)$. Take a valuation $\nu \in \sem{\gamma}$, assigning in particular an integer in $\Zed$ to each free variable of $\phi$. Note that $\gamma$ implies $a \divides t+k$, and therefore the integer $\nu(t+k)$ is a multiple of $a$.
Let $\mu \coloneqq \nu\substitute{\frac{\nu(t+k)}{a}}{x}$.
We show that $\mu \in \sem{\phi}$; which directly implies $\nu \in \sem{\exists x. \phi}$.
Since $\gamma$ asserts $\phi\substitute{\frac{t+k}{a}}{x}$,
it suffices to check that, for every atomic formula $\alpha$, we have 
\[ 
  \mu \in \sem{\alpha} 
  \text{ if and only if }
  \nu \in \sem{\alpha{\textstyle\substitute{\frac{t+k}{a}}{x}}}.
\]
For atomic formulae~$\alpha$ not involving $x$ this is trivial. 
For all other formulae, 
this follows from the definitions of $\mu$, $\nu$ and $\sem{\cdot}$.
For $\alpha = (b \cdot x + t' < 0)$ we have: 
\begin{align*}
  \mu \in \sem{b \cdot x + t' < 0}  
  &\ \text{ if and only if } \
  b \cdot \frac{\nu(t+k)}{a} + \nu(t') < 0\\
  &\ \text{ if and only if } \
  b \cdot \nu(t+k) + a \cdot \nu(t') < 0\\ 
  &\ \text{ if and only if } \ 
  \nu \in \sem{b \cdot (t+k) + a \cdot t' < 0}\\
  &\ \text{ if and only if } \ 
  \nu \in \sem{(b \cdot x + t' < 0)\substitute{\textstyle\frac{t+k}{a}}{x}}.
\end{align*}
The case of $\alpha = (q \divides c \cdot x + r)$ is analogous: 
\begin{align*}
  \mu \in \sem{q \divides c \cdot x + r}  
  &\ \text{ if and only if } \
  q \mid c \cdot \frac{\nu(t+k)}{a} + \nu(r)\\
  &\ \text{ if and only if } \
  a \cdot q \mid c \cdot \nu(t+k) + a \cdot \nu(r)\\
  &\ \text{ if and only if } \ 
  \nu \in \sem{a \cdot q \divides c \cdot (t+k) + a \cdot r}\\
  &\ \text{ if and only if } \ 
  \nu \in \sem{(q \divides c \cdot x + r)\substitute{\textstyle\frac{t+k}{a}}{x}}.
\end{align*}

\bigskip

\noindent
$(\Rightarrow)$:
Below, let $g \coloneqq \Pi\{ a : (a,t) \in T \text{ for some } t \}$, that is, $g$ is the product of all distinct coefficients of $x$ appearing in some linear inequality of $\phi$.
Observe that $T \neq \emptyset$.
We start by performing three manipulations on the formula~$\phi$, in this order:
\begin{quote}
  \begin{enumerate}
    \renewcommand{\labelenumi}{\theenumi.}
    \renewcommand{\theenumi}{M\arabic{enumi}}
    \item\label{corr-pa-M1} Update every inequality $b \cdot x + s \sim 0$, where $b > 0$, $s$ is a term and ${\sim} \in \{{<},{>}\}$, and every divisibility constraint $m \divides c \cdot x + r$, 
    where $r$ is a term, as follows: 
    \begin{align*}
      b \cdot x + s \sim 0 \quad &\mapsto \quad g \cdot x + \frac{g}{b} \cdot s \sim 0,\\
      m \divides c \cdot x + r \quad &\mapsto \quad g \cdot m \divides c g \cdot x + g \cdot r.
    \end{align*}
    \item\label{corr-pa-M2} Let $x'$ be a fresh variable. Replace each occurrence of $g \cdot x$ with $x'$, that is:
    \begin{align*}
      g \cdot x + \frac{g}{b} \cdot s \sim 0 \quad &\mapsto \quad x' + \frac{g}{b} \cdot s \sim 0,\\
      g \cdot m \divides c g \cdot x + g \cdot r \quad &\mapsto \quad 
      g \cdot m \divides c \cdot x' + g \cdot r.
    \end{align*}
    \item\label{corr-pa-M3} Conjoin the resulting formula with the divisibility constraint $g \divides x'$.
  \end{enumerate} 
\end{quote}

Let $\phi' \land (g \divides x')$ be the formula obtained after performing these three steps. The following claim is straightforward.

\begin{claim}
\label{corr-pa-claim-1}
$\exists x. \phi \fequiv \exists x'(\phi' \land (g \divides x'))$. All coefficients of $x'$ in the inequalities of $\phi'$ 
are $\pm 1$.
\end{claim}

We define:
\begin{align*}
  \ell &\coloneqq g \cdot \fmod(\phi),\\
  T' &\coloneqq \{-t'(\vec y) : x' + t' \in \linterms(\phi') \} \cup \{t'(\vec y) : -x' + t' \in \linterms(\phi') \} \cup \{0\},\\
  \Gamma' &\coloneqq \{ 
    (\phi' \land (g \divides x'))\substitute{t'+k'}{x'}
    : 
    t' \in T', 
    k' \in [-\ell,\ell]
  \}.
\end{align*}
Thus, every inequality in $\phi'$ has the form either $x' < \tau$ or $x' > \tau$ where $\tau \in T'$.
(Then $\tau = \frac{g}{b} \cdot (-s)$. Since $b > 0$, in the case $x' < \tau$ we have $\frac{g}{b} \cdot (-s) \in T'$,
 so $t = -s$. In the case $x' > \tau$, the inequality is $-x' + \frac{g}{b} \cdot (-s) < 0$, and so $\frac{g}{b} \cdot (-s) \in T'$.)

\begin{claim}
  $\exists x'(\phi' \land (g \divides x'))$ 
  implies $\bigvee_{\gamma' \in \Gamma'} \gamma'$.
\end{claim}

\begin{claimproof}

We apply the standard argument due to Presburger. 
Let $\nu$ be a valuation satisfying the formula $\phi' \land (g \divides x')$. Consider the set of inequalities in $\phi'$ in which $x'$ occurs. By~Claim~\ref{corr-pa-claim-1}, they are of the form $x' < t'$ and $x > t'$, where $t'$ ranges over $T'$. 
Let $T' = \{t_1',\dots,t_n'\}$ such that $\nu(t_1') \leq \dots \leq \nu(t_n')$. There are two cases.

If $\abs{\nu(x') - \nu(t_i')} \leq \ell$ for some $i \in [1,n]$, then $\nu$ satisfies $(\phi' \land (g \divides x'))\substitute{t'+k'}{x'}$ for $t' = t_i'$ and $k' = \nu(x') - \nu(t_i')$.

Otherwise $\abs{\nu(x') - \nu(t_i')} > \ell$ for every $i \in [1,n]$. Consider an index $j$ such that $\nu(t_j')$ has value closest to $\nu(x')$. Depending on whether $\nu(x') < \nu(t_j')$ or $\nu(x') > \nu(t_j')$, we can increase or decrease the value of $\nu(x')$ 
by an integer multiple of $\ell$ so that the new value would fall into the previous case. More formally, let us assign to $x'$ the value $\nu(x') + d \cdot \ell$ where $d \in \Zed$ is the largest or smallest such that  
$\nu(x') + d \cdot \ell < \nu(t_j')$ or $\nu(x') + d \cdot \ell > \nu(t_j')$, respectively. 
Observe that shifting $x'$ in this way does not change the truth values of inequalities and divisibility constraints in $\phi' \land (g \divides x')$, because we already know that 
every divisibility constraint involving $x'$ in this formula has divisor of the form $g \cdot m$, where $m \divides \fmod(\phi)$. 
Therefore, $\nu\substitute{\nu(x')+d \cdot \ell}{x'}$ satisfies $(\phi' \land (g \divides x'))$.

By our choice of $d$,  the number
$\nu(x') + d \cdot \ell$ is equal to $\nu(t_j') + k'$ for some $k' \in [-\ell,\ell]$.
Hence, $\nu$ satisfies $(\phi' \land (g \divides x'))\substitute{t_j'+k'}{x'}$.
\end{claimproof}

In the following claim, the set $\Gamma$ is from line~\ref{pres:line-sub} of~\funcPA, and we recall that its description 
involves a quantity $r$ defined in the same line.

\begin{claim}
  \label{corr-pa-claim-3}
  $\bigvee_{\gamma' \in \Gamma'} \gamma'$ implies $\bigvee_{\gamma \in \Gamma} \gamma$.
\end{claim}

\begin{claimproof}
  Consider a formula $\gamma' \in \Gamma'$. 
  We show that there is a formula $\gamma \in \Gamma$ implied by $\gamma'$ (in fact, the two formulae are equivalent). 
  Assume that $\gamma' = (\phi' \land (g \divides x'))\substitute{t'+k'}{x'}$ 
  with $t' \in T'$ and $k' \in [-\ell,\ell]$.
  Notice that $(g \divides x')\substitute{t'+k'}{x'}$ 
  is the formula $g \divides t'+k'$. 

  \proofsubparagraph{Replacing $k' \in [-\ell,\ell]$ with $k \in [-r,r]$.}
  Thanks to the manipulation steps above, $t'$ is of the form $\frac{g}{a} \cdot t$ with $a > 0$ and $t$ a term.
  More precisely, these $a$ and $t$ are exactly the divisor $b$ and term $-s$ that appear in the constraint $x' + \frac{g}{b} \cdot s \sim 0$ after step~\ref{corr-pa-M2} (naturally, $b$ divides $g$, so the number $b$ itself does not have to feature in the constraint).
  Therefore, $g \divides t'+k' \fequiv g \divides \frac{g}{a} \cdot t + k'$; 
  this constraint can only be satisfied if $k'$ is a multiple of $\frac{g}{a}$. 
  Thus, for our chosen $t' \in T'$, in the definition of $\Gamma'$ 
  it suffices to restrict the range of $k'$ to the set $[-\ell,\ell] \cap \frac{g}{a} \Zed$. Here, note that $a$ depends on $t'$.
  
  It remains to observe that every $k' \in [-\ell,\ell] \cap \frac{g}{a} \Zed$
  can be factorized as $k' = \frac{g}{a} \cdot k$, where $k$ 
  ranges over the set $[-\ell/\frac{g}{a},\ell/\frac{g}{a}] = [-r,r]$.
  
  \proofsubparagraph{Replacing $t' \in T'$ with $(a,t) \in T$.}
  Take some $k' \in [-\ell,\ell] \cap \frac{g}{a} \Zed$, say $k' = \frac{g}{a} \cdot k$. Consider each inequality and each divisibility constraint 
  in the formula $\gamma'$. Again due to the manipulations above, 
  each inequality either is copied verbatim from $\phi$, or has the form 
  \[
    \left(\frac{g}{a} \cdot t + \frac{g}{a} \cdot k\right) + \frac{g}{b} \cdot s \sim 0,
  \]
  which simplifies to 
  \[
    b \cdot \left(t + k\right) + a \cdot s \sim 0.
  \]
  Similarly, each divisibility constraint either is copied verbatim from $\phi$, or has the form 
  \[ 
    g \cdot m \divides c \cdot \left(\frac{g}{a} \cdot t + \frac{g}{a} \cdot k\right) + g \cdot r
  \]
  and simplifies to 
  \[ 
    a \cdot m \divides c \cdot \left(t + k\right) + a \cdot r.
  \]
  Observe that the simplified formula coincides with $\phi\substitute{\frac{t+k}{a}}{x} \land (a \divides t+k)$.
  It remains to show that $(a,t) \in T$, where $T$ is defined in line~\ref{pres:line1} of~\funcPA. 

  Recall that $t' = \frac{g}{a} \cdot t$ with $a > 0$ and $t$ a term, 
  and moreover, by the definition of $T'$, the formula $\phi'$ contains an inequality 
  of the form $x' - t' \sim 0$.
  This inequality features $x'$ and thus must have arisen from the manipulation steps above. 
  In the description of the steps, we wrote such inequalities as
  $x'+\frac{g}{b} \cdot s \sim 0$. 
  Therefore, we have $a = b$ and $s = -t$ because $t' = \frac{g}{a} \cdot t$. Undoing step~\ref{corr-pa-M2}, the inequality 
  was $g \cdot x + \frac{g}{a} \cdot (-t) \sim 0$ 
  before this step. Now undoing step~\ref{corr-pa-M1}, 
  the inequality was originally $a \cdot x - t \sim 0$ 
  before the manipulations. In each of the two cases ${\sim} \in \{{<},{>}\}$, we see that $(a,t) \in T$ by line~\ref{pres:line1} of~\funcPA. 
  This completes the proof of~\Cref{corr-pa-claim-3}.
\end{claimproof}

The three claims above complete the proof of the left to right direction 
of the double implication in~\eqref{proof12:iff}. Therefore, we conclude that 
$\Gamma$ is a cover for $\exists x. \phi$.
  
As a final part of this proof, we show that,
thanks to the correctness of the function \simplify,
every formula appearing in the output of \funcPA belongs to the fragment~$\Fragment$. We consider two cases.
\begin{description}
  \item[$\Fragment = \Sem$:] 
  In each
  formula $\phi\substitute{\frac{t+k}{a}}{x} \land (a \divides t+k)$,
  every variable appears either only linearly or only as a power,
  just because this is true for~$\phi$. Also, every bound variable
  can only feature in formulae from \PowComp. This means that the requirements
  imposed by \simplify on its input are satisfied. Therefore, by the correctness of \simplify
  (Lemma~\ref{l:correct-simplify}), the cover produced in the output
  will consist of formulae from $\Fragment$ only.
  \item[$\Fragment = \QF$:] 
  In this case \simplify eliminates all non-simple divisibility constraints
  introduced by term substitutions in line~\ref{pres:line-sub};
  this is correct by Lemma~\ref{l:correct-simplify}.
\end{description}
This completes the proof of~\Cref{l:correct-presburger}.
\end{proof}

We are now ready to provide the proof of Lemma \ref{l:only-simple-div} from the body, which we now recall:

\OnlySimpleDiv*

\begin{proof}
The original input of our Master procedure is assumed
to satisfy this requirement.
In all subsequent steps of the algorithm, non-simple divisibility constraints
may only be introduced by \funcPA in line~\ref{pres:line-sub};
however, these are then promptly removed by \simplify before
the formulae are added back to the set $Q$;
the correctness of \simplify is Lemma~\ref{l:correct-simplify}.
\end{proof}

We turn to the proof of correctness of procedure \funcSem.
We will need several auxiliary statements which will help us argue
that the case split
in lines~\ref{sem:case-1}--\ref{sem:end-case} is valid.

\begin{claim}
\label{c:powx-to-powx-div-x}
Let $b \ge 1$ and $x$ be two real numbers.
If $2^x \ge 64 b^2$, then
$2^x / x \ge 8 b$.
\end{claim}

\begin{proof}
Taking the logarithm on both sides of the target inequality,
we obtain $f(x) \ge \log b + 3$, where $f(u) = u - \log u$.
Notice that $f'(u) = 1 - 1 / u \ln 2$; so, assuming $u > 0$,
the value of $f'(u)$ is nonnegative iff $u \ln 2 \ge 1$;
here $\ln$ denotes the natural logarithm.
This means that the function~$f$ is nondecreasing on $u \in [\log_2 e, +\infty)$.
Recall that $x$ is chosen such that $x \ge 2 \log b + 6$
with $b \ge 1$; this means in particular that $x \ge 6$.
Clearly $\log_2 e \le 6$; so, to ensure that $f(x) \ge \log b + 3$
for any appropriately chosen $x$, it will suffice to check
that $f(2 \log b + 6) \ge \log b + 3$.
But this inequality can be rewritten as $f(2 v) \ge v$, where
$v = \log b + 3$.
Expanding the definition of~$f$, we obtain
$2 v - \log( 2 v ) \ge v$, or equivalently $v - \log v \ge 1$.
But if $b \ge 1$, then $v \ge 3$ and we already saw that
$v - \log v$ is a nondecreasing function
of $v \in [\log_2 e, +\infty)$. So we just need to check
that $f(3) \ge 1$; indeed, $f(3) = 3 - \log 3 \ge 3 - \log 4 = 1$.
This completes the proof.
\end{proof}

\begin{claim}
\label{c:var-vs-power}
Let $x$ and $b$ be positive integers.
If $2^x \ge 64 b^2$, then
$\lambda(b x) \le 2^{x-3}$.
\end{claim}

\begin{proof}
Observe that the desired inequality follows
from the inequality $b x \le 2^x / 8$, which is
just another way of writing $2^x / x \ge 8 b$.
But then this is just a restatement of Claim~\ref{c:powx-to-powx-div-x}.
\end{proof}

\begin{claim}
\label{c:sem-split}
Let $\eta(\vec x)$ be
a homogeneous term in which some $x \in \vec x$
occurs in a power (as well as possibly linearly).
Let $c \in \Zed$ and suppose that
$\vec x^0$ is a vector of values from~$\Zed$ for variables~$\vec x$;
in particular $x^0$ is the value for~$x$.
Suppose $\pow{x^0} > 2^7 \cdot \lambda(\normone\eta + \abs{c})^2$ and $\pow{x^0} > (2^7 \cdot \lambda(\normone\eta + \abs{c})^2)  \cdot \pow{u^0}$
for all $u \in \vec x \setminus \{x\}$, where $u^0$ is the value for variable $u$.
Let $a$ be the coefficient of $\pow x$ in $\eta$.
Then:
\begin{equation*}
\frac{1}{2} \cdot
\lambda\left(a \cdot \pow{x^0}\right)
\le
\lambda(\eta(\vec x^0) + c)
\le
\lambda\left(a \cdot \pow{x^0}\right)
.
\end{equation*}
\end{claim}
Note that $2^7 \cdot \lambda(\normone\eta + \abs{c})^2$ is the quantity $2^g$ in line~\ref{sem:g} of~\funcSem.

\begin{proof}
Let the term $\eta$ be given by the formula
$\eta(\vec x) = a \cdot \pow{x} + b \cdot x + \eta'(x_1, \ldots, x_k)$
where $a \ne 0$ and
      $\eta' = \sum_{i=1}^{k} (a_i \cdot \pow{x_i} + b_i \cdot x_i)$.

We first show that each of $\lambda(b x^0)$ and $\lambda(\eta'(\vec x^0) + c)$
is at most $\pow{x^0} / 8$.
Denote $N = \normone\eta + \abs{c}$, $N \ge 1$.
We know that
$\pow{x^0} > 2^7 \cdot (\lambda(N))^2$, so in particular
$\pow{x^0} \ge 256 \cdot (N/2)^2 = 64 N^2 \ge 64 b^2$.
Assume $b > 0$, otherwise $\lambda(b x^0) = 0$.
By Claim~\ref{c:var-vs-power}, $\lambda(b x^0) \le 2^{\abs{x^0} - 3} = \pow{x^0} / 8$.
Proceeding to $\eta' + c$, notice that
if $k = 0$, then
\begin{equation*}
\lambda(\eta'(\vec x^0) + c) =
\lambda(c) \le
\lambda(\normone\eta + \abs{c}) <
\frac{\pow{x^0}}{2^7 \cdot \lambda(\normone\eta + \abs{c})} \le
\frac{\pow{x^0}}{2^7},
\end{equation*}
and otherwise
\begin{align*}
|\eta'(\vec x^0) + c|
&=  \left|\,\sum_{j=1}^{k}\left( a_j \cdot \pow{x_j^0} + b_j \cdot x_j^0 \right) + c\, \right| &&\\
&\le        \sum_{j=1}^{k}\left(|a_j|\cdot \pow{x_j^0} +|b_j|\cdot|x_j^0|\right) +|c|\cdot 1 &&\\
&\le \left( \sum_{j=1}^{k}\left(|a_j|                  +|b_j|            \right) +|c| \right) \cdot \max_{1 \le j \le k} \pow{x_j^0} &&\\
&\le N \cdot \max\limits_{1 \le j \le k} \pow{x_j^0} &&\text{(putting back $a \cdot \pow{x} + b \cdot x$)}\\
&< (2 \lambda(N)) \cdot\max\limits_{1 \le j \le k} \pow{x_j^0} &&\text{(by definition of $\lambda$)}\\
&< (2 \lambda(N)) \cdot \frac{\pow{x^0}}{2^7 \cdot \lambda(\normone\eta + \abs{c})^2} &&\text{(by assumption on $\pow{x^0}$)}\\
&=
\pow{x^0} \cdot \frac{2 \lambda(N)}{2^7 (\lambda(N))^2} &&\text{(by definition of~$N$)}\\
&=
\pow{x^0} \cdot \frac{1}{64} \cdot \frac{1}{\lambda(N)} &&\\
&\le
\pow{x^0} \cdot \frac{1}{64}, &&\text{(since $N \ge 1$)}
\end{align*}
which gives the desired bound on $\lambda(\eta'(\vec x^0) + c)$.

Observe that the two bounds above imply
$\lambda(b x^0 + \eta'(\vec x^0) + c) \le \pow{x^0} \cdot \frac{1}{4}$.
In this new inequality,
the term appearing on the left under $\lambda$  only
differs from $\eta(\vec x^0) + c$ by (the absence of)
$a \cdot \pow{x^0}$.
But then the integer $b x^0 + \eta'(\vec x^0) + c$ has
fewer digits in binary expansion than
the integer
$\pow{x^0}$; and moreover
the difference between the numbers of digits is at least~$2$.
(One can think of this as ``significantly smaller''.)
Since $a \cdot \pow{x^0}$ is a multiple of $\pow{x^0}$, we conclude that
the sum
$\eta(\vec x^0) + c$ is ``close to'' $a \cdot \pow{x^0}$. That is, the sum:
\begin{itemize}
\item
cannot have more binary digits than $a \cdot \pow{x^0}$, and at the same time
\item
either has the same number of  binary digits
as $a \cdot \pow{x^0}$, or exactly one fewer. %
\qedhere
\end{itemize}
\end{proof}

\begin{lemma}
\label{l:initial-semcover}
In line~\ref{sem:I} of \funcSem,
no inequality of $I$ can contain bound variables.
\end{lemma}

\begin{proof}
The absence of bound variables in the case $\Fragment = \QF$ is trivial.
Furthermore, in the case $\Fragment = \Sem$,
we notice that the set $I$ is defined based on the formula~$\phi$ input to \funcSem
(without any rewritings applied to it).
Since $\phi \in \Sem$, it follows that
bound variables can only occur (within $\phi$)
in constraints from \PowComp.
But all inequalities from $I$ are outside \PowComp by our choice of $I$.
Therefore, no inequality from $I$ can contain any bound variables.
\end{proof}

For brevity, we refer to term substitutions in lines~\ref{sem:case-1}--\ref{sem:case-2}
and \ref{sem:case-5}--\ref{sem:case-6}, replacing $\pow x$ with other terms,
as \emph{Semenov substitutions}.

\begin{lemma}
\label{l:properties-semenov-sub}
For every Semenov substitution $\substitute{t}{\pow x}$\textup{:}
\begin{enumerate}[(a)]
\item\label{sem-sub:target-free}
the variable $x$ is free;
\item\label{sem-sub:replacement-free}
all variables in the term~$t$ are free;
\item\label{sem-sub:no-interference}
for every atomic formula $\alpha$ to which this substitution is applied,
$\alpha$ is outside $\PT$ and contains no bound variables.
\end{enumerate}
\end{lemma}

\begin{proof}
We prove part~\eqref{sem-sub:target-free} first.
The variable~$x$ is chosen among the variables of the vector~$\vec x$,
which are all free in~$\phi$.
(Recall that we never reuse variables.)
It remains to notice that no quantifiers are introduced
by the \textbf{for} loop in lines~\ref{sem:outer-loop}--\ref{sem:end-case}.

Let us now prove part~\eqref{sem-sub:no-interference} next.
This part is a consequence of the fact that
only inequalities from $A \subseteq I$ are rewritten in lines~\ref{sem:case-1}--\ref{sem:end-case}.
These inequalities are outside \PowComp by the choice of $I$.
As for occurrences of bound variables,
it remains to apply Lemma~\ref{l:initial-semcover}.

Finally, for part~\eqref{sem-sub:replacement-free}, observe that
all replacements for $\pow x$ can only contain variables from terms $\eta$ and $\sigma$,
possibly as part of $\lambda(\sigma)$ in lines~\ref{sem:case-5}--\ref{sem:case-6}.
All these variables occur in inequalities from $I$,
so once again it suffices to apply Lemma~\ref{l:initial-semcover}.
This completes the proof.
\end{proof}

\begin{lemma}
\label{l:semenov-cover-in-fragment}
Suppose function \funcSem
is given a vector $\vec x$ of variables and a formula $\phi(\vec x, \vec z)$ of $\cT$,
containing $\pow{x}$ for each $x \in \vec x$.
Then,
for every pair $(\vec x, \psi(\vec x, \vec z, \vec w))$ returned by \funcSem,
$\psi \in \cT$.
\end{lemma}

\begin{proof}
If $\Fragment = \QF$, the conclusion is immediate.
Consider the case $\Fragment = \Sem$.
We first check that every occurrence of a quantifier,
say $\exists x$,
in the formula $\psi$ must satisfy the requirement from
the definition of $\QFPC$, namely that
the variable $x$ appears in $\psi$ only in atomic formulae from $\PT$.
Recall that the same requirement holds for the input formula~$\phi$.
First note that divisibility (modulo) constraints are entirely
unaffected by \funcSem.
Second,
by Lemma~\ref{l:properties-semenov-sub}, part~\eqref{sem-sub:no-interference},
inequalities from \PowComp are left unchanged by Semenov substitutions,
as are all inequalities that contain bound variables.
In addition, by parts~\eqref{sem-sub:target-free}
and~\eqref{sem-sub:replacement-free} of the same lemma,
these substitutions can neither rewrite nor introduce bound variable occurrences.
Thus, throughout {\funcSem} and in the output formula~$\psi$, the variable~$x$ must
still only feature in atomic formulae from \PowComp.

We also need to
verify that all divisibility constraints in $\psi$ are simple
and
each variable appears either always linearly or always as a power.
The first condition holds, because it holds for $\phi$ and, as mentioned
in the previous paragraph, divisibility constraints are unaffected by \funcSem.
Let us prove the second condition.
Steps that may potentially take the formula outside \Sem
are substitutions and constructions of formulae that involve~$\pow{v}$ in line~\ref{sem:case-2} and
introduction of inequalities $\pow x \ge \pow y$ in line~\ref{sem:union-gamma}.
However, all variables that get a new mention in either of these cases
are contained in the vector $\vec x$. By the assumption on the input of \funcSem,
every such variable must appear at least once as a power in $\phi$, so any further
constraints that contain a power of such a variable cannot bring the formula outside \Sem.
Therefore, $\psi \in \Sem$ as required.
\end{proof}

The following claim
is pivotal in our proof of the correctness of \funcSem.

\begin{claim}
\label{c:semenov-split-phi-into-gamma}
In \funcSem, $\phi(\vec x, \vec z) \fequiv \bigvee_{\gamma \in \Gamma} \gamma(\vec x, \vec z)$,
where formulae $\gamma \in \Gamma$ are interpreted in the structure
extended by function~$\lambda$.
\end{claim}

\begin{proof}
The overall idea is simple:
the outer \textbf{for} loop at line~\ref{sem:outer-loop} picks
a variable $x \in \vec x$; all formulae produced in the corresponding
iteration are ``guarded'' in line~\ref{sem:union-gamma} by constraints
asserting that $x$~is the largest variable among~$\vec x$.
This is a case split.
To prove the Claim, we need to show that
updates to the set $\Gamma_x$ in lines~\ref{sem:case-1}--\ref{sem:end-case}
are case splits too; more precisely that the cases are exhaustive.

Recall that a key element of \funcSem is repeated substitutions made to the input formula~$\phi$
in lines~\ref{sem:case-1}--\ref{sem:end-case}.
In particular, line~\ref{sem:case-1} replaces $\pow{x}$ with a constant,
line~\ref{sem:case-2} with a term $2^j \cdot \pow{v}$, where $v$ is another free variable,
and the following six lines with $\top$, $\bot$, or terms depending on $\lambda(\sigma)$
with $\sigma$ a homogeneous term in variables $\vec z$.

To argue correctness of such replacements, first
observe that, by Lemma~\ref{l:properties-semenov-sub}, parts~\eqref{sem-sub:target-free}
and~\eqref{sem-sub:replacement-free}, all variables that participate in these replacements are free
in~$\phi$.

Consider the effect of our formula manipulation on other
occurrences of~$x$.
In lines~\ref{sem:A}--\ref{sem:end-case} \funcSem does not modify 
any inequalities from \PowCmp,
any inequalities in which the variable~$x$ occurs only linearly,
or any divisibility (modulo) constraints.
It is easy to see that replacements can only cause issues
if a variable bound inside~$\phi$ is directly affected by them, %
i.e., if this variable occurs in an atomic formula which is being rewritten or
if it gets introduced into an atomic formula as an effect of a rewriting.
For the inner \textbf{for} loop in lines~\ref{sem:inner-loop}--\ref{sem:end-case},
this is impossible by Lemma~\ref{l:properties-semenov-sub}.

Each case in lines~\ref{sem:case-1}--\ref{sem:end-case} is introduced in a separate line.
The first case (line~\ref{sem:case-1}) specifies that
the value of the power~$\pow{x}$ is small, enumerating
all possibilities up to a threshold, $2^g$, chosen earlier.
All subsequent cases, accordingly, include the assertion that the value
of $\pow x$ is higher than the threshold.
The second case (line~\ref{sem:case-2}) specifies that
the value of another power, $\pow v$, is close to the value
of $\pow x$. Again, all possible choices for $v \in V$ are enumerated
(see line~\ref{sem:end-case}) as are all possible multiplicative distances
for powers of two up to the same threshold.
All the following cases include the assertion~$\beta$, stating that
$\pow x$ is far away from $0$ and from other participating powers.

We now turn to the two cases in the middle, lines~\ref{sem:case-5}--\ref{sem:case-6}.
These specify a concrete value for $\pow x$, using $\lambda(\sigma)$ as
the reference point. (While $\lambda$ is not inside the main signature,
these occurrences of the symbol stay in the formulae until the final
loop of \funcSem at line~\ref{sem:for-sigmaprime}, which eliminates
them all.) As in the previous case analysis, the substitutions
in lines~\ref{sem:case-5}--\ref{sem:case-6} bind the value of $\pow x$
in an unambiguous way.

The most interesting cases appear in lines~\ref{sem:case-3}--\ref{sem:case-4}
and~\ref{sem:case-7}--\ref{sem:case-8}. The intention of the split
is clear from the guards
$\lambda(a) \cdot \pow{x} < \lambda(\sigma)$
and
$\lambda(a) \cdot \pow{x} > 2 \cdot \lambda(\sigma)$
introduced in these lines. What may be less obvious
is why the atomic formulae $\alpha$ can be replaced with Boolean constants.
This, however, is the subject of our auxiliary Claim~\ref{c:sem-split} proved above.
Indeed, consider lines~\ref{sem:case-3}--\ref{sem:case-4}.
Take the constraint $\eta + \sigma + c < 0$, from the set $A$.
The inequalities
$\pow{x^0} > 2^7 \cdot \lambda(\normone\eta + \abs{c})^2$ and $\pow{x^0} > (2^7 \cdot \lambda(\normone\eta + \abs{c})^2)  \cdot \pow{u^0}$
are satisfied thanks to the guards $\beta$
appearing in lines~\ref{sem:case-3}--\ref{sem:case-4}.
Observe that $\lambda(a \cdot \pow{x}) = \lambda(a) \cdot \pow{x}$,
where $a$ is the coefficient of $\pow x$ in $\eta$.
Thus, by Claim~\ref{c:sem-split},
\begin{equation*}
\lambda(\eta + c) \in \left\{ \lambda(a \cdot \pow{x}),\ \tfrac{1}{2} \,
                              \lambda(a \cdot \pow{x}) \right\}.
\end{equation*}
But $\lambda(a) \cdot \pow{x} < \lambda(\sigma)$ from the guard
in lines~\ref{sem:case-3}--\ref{sem:case-4}.
In words, the value of $a \cdot \pow{x}$ has strictly
fewer binary digits than the value of $\sigma$, and so the value of $\eta + c$
also has strictly fewer binary digits than the value of $\sigma$.
But then the sign of $\eta + \sigma + c$ coincides with the sign of $\sigma$.

Lines~\ref{sem:case-7}--\ref{sem:case-8} are considered similarly:
the guard is $\lambda(a) \cdot \pow{x} > 2 \cdot \lambda(\sigma)$ and so
the sign of $\eta + \sigma + c$ coincides with the sign of $a \cdot \pow{x}$,
which is the same as that of $\eta + c$.
This way the truth value of all inequalities in~$A$ is determined.
\end{proof}

For the following lemma,
we recall that the global variable $\Pi'$ stores
a sequence of (universal) quantifiers. Function \funcSem appends new quantifiers to $\Pi'$
when it introduces fresh variables $w \in \vec w$.

\begin{lemma}
\label{l:correct-semenov-cover}
Suppose function \funcSem
is given a vector $\vec x$ of variables and a formula $\phi(\vec x, \vec z)$ of $\cT$,
containing $\pow{x}$ for each $x \in \vec x$.
Then it returns a set of pairs $(\vec x, \psi(\vec x, \vec z, \vec w))$ such that
$\exists \vec x. \phi(\vec x, \vec z) \fequiv \forall \vec w.\bigvee \exists \vec x. \psi(\vec x, \vec z, \vec w)$,
where the disjunction is over all formulae $\psi$ in the output.
\end{lemma}

\begin{proof}
Intuitively, formulae placed
into the set $\Theta$ after the main \textbf{for} loop are responsible
for binding the values of fresh variables $w_\sigma$ to the corresponding
terms $\sigma \in \Sigma$.
Indeed, the first substitution and the guards introduced in line~\ref{semenov:lambda-sub}
achieve exactly that; the choice over possible subsets $\Sigma'$ of $\Sigma$
corresponds to the option that the value of $\sigma$ is $0$
(in which case it cannot be sandwiched between two powers of~$2$).

We now make this intuition into a rigorous argument.
As in the proof of Claim~\ref{c:semenov-split-phi-into-gamma},
we first observe that, by Lemma~\ref{l:properties-semenov-sub},
all variables that participate in the substitutions made to the input formula~$\phi$
are free in~$\phi$.
Lines~\ref{sem:outer-loop}--\ref{sem:union-gamma} have already been considered.
In the subsequent lines~\ref{sem:delambda-start}--\ref{sem:return},
substitutions only affect terms that contain the function $\lambda$, and newly-introduced formulae
mention variables among~$\vec z$ and $\vec w$ only. This means that correctness
of the procedure is not affected by possible quantification inside~$\phi$.

We need to demonstrate that
set of all $\exists \vec x.\psi$, where $(\vec x, \psi)$ is in the output,
covers $\exists \vec x.\phi$ under $\forall \vec w$,
i.e., $\exists \vec x. \phi(\vec x, \vec z) \fequiv \forall \vec w.\bigvee \exists \vec x. \psi(\vec x, \vec z, \vec w)$.
We can rely on Claim~\ref{c:semenov-split-phi-into-gamma} when proving this equivalence.


\proofsubparagraph{Left to right.}


Let us fix $\vec c$, a tuple of integer values assigned to $\vec z$,
and let $\vec a$ be a tuple of integer values such that $\phi(\vec a, \vec c)$ is true.
We would like to show that the formula $\forall \vec w. \bigvee \psi(\vec a, \vec c, \vec w)$ is true.
By Claim~\ref{c:semenov-split-phi-into-gamma},
we can assume that $\gamma(\vec a, \vec c)$ is true for some $\gamma \in \Gamma$
at line~\ref{sem:union-gamma}.

Let $\vec b = (b_1, \ldots, b_m)$ be a tuple of integer values assigned to $\vec w = (w_1, \ldots, w_m)$;
here $m = |\Sigma|$ where the set $\Sigma$ is defined at line~\ref{sem:delambda-start}.
Denoting $\Sigma = \{\sigma_1, \ldots, \sigma_m\}$,
we write $w_i$ instead of $w_{\sigma_i}$ for $1 \le i \le m$.
Recall that each $\sigma_i$ is a homogeneous term with variables $\vec z$.
Thus, since the values $\vec c$ for $\vec z$ are already fixed,
each $\sigma_i(\vec c)$ is a concrete integer.
We consider several scenarios.

In the first scenario,
some $\pow{b_i}$ is different from $\lambda(\sigma_i(\vec c))$
and moreover $\sigma_i(\vec c) \ne 0$.
Consider the chained inequality $\pow{b_i} \leq \abs{\sigma_i(\vec c)} < 2 \cdot \pow{b_i}$,
which is asserting that $\pow{b_i}$ is equal to $\lambda(\sigma_i(\vec c))$.
In the present scenario, it fails:
either $\pow{b_i} > \abs{\sigma_i(\vec c)}$ or $\abs{\sigma_i(\vec c)} \ge 2 \cdot \pow{b_i}$ is true;
therefore,
the corresponding formula in line~\ref{semenov:guard-for-lambda} is true.
Note that $\sigma_i(\vec c) \ne 0$ by assumption.

In the second scenario,
$\pow{b_i} = \lambda(\sigma_i(\vec c))$
whenever $\sigma_i(\vec c) \ne 0$ for the same index~$i$.
In other words, there might be some $\pow{b_i}$ different from the corresponding $\lambda(\sigma_i(\vec c))$,
but $\sigma_i(\vec c)  = 0$ for all such indices~$i$.
In this scenario,
let $\Sigma'$ be the set of all $\sigma \in \Sigma$ such that
$\sigma(\vec c) \ne 0$;
in particular, 
the formula $\hspace{-3pt}\bigwedge\limits_{\sigma \in \Sigma \setminus \Sigma'}\hspace{-9pt} \sigma = 0$
in line~\ref{semenov:lambda-sub}
is true.
Also, in this scenario, 
for all~$i$ such that $\sigma_i(\vec c) \ne 0$, we have $\pow{b_i} = \lambda(\sigma_i(\vec c))$,
and thus the formula
$\hspace{-2pt}\bigwedge\limits_{\sigma \in \Sigma'}\hspace{-2pt} \pow{w_\sigma} \leq \abs{\sigma} < 2 \cdot \pow{w_\sigma}$
at the beginning of line~\ref{semenov:lambda-sub}
is also true.
Since the true equalities
\begin{align*}
&\pow{b_i} = \lambda(\sigma_i(\vec c)) && \text{for $\sigma_i \in \Sigma'$, and} \\
&\sigma_i(\vec c) = 0 && \text{for $\sigma_i \in \Sigma \setminus \Sigma'$,}
\end{align*}
are between concrete numbers, and since,
as observed above, the formula $\gamma(\vec a, \vec c)$ is true,
this formula will remain true after
each occurrence 
of $\lambda(\sigma_i)$, $\sigma_i \in \Sigma'$, is replaced with $\pow{b_i}$, and
each occurrence
of $\sigma_i$, $\sigma_i \in \Sigma \setminus \Sigma'$, with~$0$.
Recalling that $b_i$ is the value assigned to $w_i$ for every index~$i$,
we conclude
that the remaining conjunct in line~\ref{semenov:lambda-sub},
$
         \gamma\,
         \substitute{\pow{w_\sigma}}{\lambda(\sigma) : \sigma \in \Sigma'}
         \substitute{0}{\lambda(\sigma) : \sigma \in \Sigma \setminus \Sigma'}
$,
is true.
Therefore, the entire formula added to $\Theta$ in this line
is true, completing the proof of the left-to-right direction.


\proofsubparagraph{Right to left.}


As previously, let us fix $\vec c$, a tuple of integer values assigned to $\vec z$.
We would like to show that the formula
$\phi(\vec a, \vec c)$ is true for some choice of $\vec a$.
By Claim~\ref{c:semenov-split-phi-into-gamma}, it suffices to find
some $\gamma \in \Gamma$ for which $\gamma(\vec a, \vec c)$ is true,
but now even choice of $\vec a$ is not obvious.
This is because different values for $\vec x$ might be needed
depending on the assignment to $\vec w$.

Recall once again that each $\sigma_i$ is a homogeneous term with variables $\vec z$.
Since the values $\vec c$ for $\vec z$ are fixed,
each $\sigma_i(\vec c)$ is a concrete integer.
Denote by $\Sigma'$ the set of all $\sigma \in \Sigma$ such that
$\sigma(\vec c) \ne 0$.
For each $\sigma \in \Sigma'$,
choose $b_\sigma \in \Nat$ such that $\pow{b_\sigma} = \lambda(\sigma(\vec c))$.
Also, for each $\sigma \in \Sigma \setminus \Sigma'$, let $b_\sigma = 0$.
Let $\vec b$ be the tuple of integers $b_\sigma$ for all $\sigma \in \Sigma$.
Since the formula
$\forall \vec w. \bigvee \exists \vec x. \psi(\vec x, \vec c, \vec w)$ is true,
so is the disjunction $\bigvee \exists \vec x. \psi(\vec x, \vec c, \vec b)$.
Here we have assumed that
the ordering of variables within $\vec w$ matches
the ordering of integers~$b_\sigma$ within $\vec b$.

By the argument above, there is at least one
true disjunct in $\bigvee \exists \vec x. \psi(\vec x, \vec c, \vec b)$.
Let us write simply $\exists \vec x. \psi(\vec x, \vec c, \vec b)$ to denote it.
However, consider formulae
$\{ (\sigma \neq 0 \land \lnot (\pow{w_\sigma} \leq \abs{\sigma} < 2 \cdot \pow{w_\sigma}))  : \sigma \in \Sigma \}$
from line~\ref{semenov:guard-for-lambda} of \funcSem.
By our choice of $\Sigma'$ and $\vec b$,
the chained inequality $\pow{b_\sigma} \leq \abs{\sigma(\vec c)} < 2 \cdot \pow{b_\sigma}$ is true
for every $\sigma$ such that $\sigma(\vec c) \ne 0$.
Thus, none of the formulae from line~\ref{semenov:guard-for-lambda} are true.

It follows that the disjunct $\exists \vec x. \psi(\vec x, \vec c, \vec b)$
corresponds to a
formula from line~\ref{semenov:lambda-sub} of \funcSem.
Let $\Sigma'' \subseteq \Sigma$ be arbitrary and consider the formula parameterised
by this choice of subset in line~\ref{sem:for-sigmaprime}; call it
$\theta_{\Sigma''}(\vec x, \vec z, \vec w)$.
We show that $\Sigma'' = \Sigma'$ is necessary for this formula to be true.
Indeed, for every $\Sigma'' \subseteq \Sigma$ different from $\Sigma'$,
either $\Sigma''$ contains some $\sigma$ that is not in $\Sigma'$,
    or $\Sigma' $ contains some $\sigma$ that is not in $\Sigma''$.
In the former case, $\theta_{\Sigma''}$ contains a conjunct
$\pow{w_\sigma} \leq \abs{\sigma} < 2 \cdot \pow{w_\sigma}$ for which
$\sigma(\vec c) = 0$; therefore $\exists \vec x. \theta_{\Sigma''}(\vec x, \vec c, \vec b)$ is false.
In the latter case, $\theta_{\Sigma''}$ contains a conjunct
$\sigma = 0$ such that $\sigma(\vec c) \ne 0$; again this means
therefore $\exists \vec x. \theta_{\Sigma''}(\vec x, \vec c, \vec b)$ is false.
We have thus proved that the disjunction
\begin{equation*}
\bigvee_{\gamma \in \Gamma}
\gamma\,
\substitute{\pow{w_\sigma}}{\lambda(\sigma) : \sigma \in \Sigma'}
\substitute{0}{\lambda(\sigma) : \sigma \in \Sigma \setminus \Sigma'}
\end{equation*}
is true when
vectors of variables $\vec x$, $\vec z$, $\vec w$ are replaced by
vectors of integers $\vec a$, $\vec c$, $\vec b$, respectively.
Here $\vec b$ and $\vec c$ were already fixed, and a suitable choice $\vec a$
for $\vec x$ exists because the disjunct
$\exists \vec x. \psi(\vec x, \vec c, \vec b)$ is true.
We remark that, by our choice of $\vec b$,
\begin{equation*}
\gamma\,
\substitute{\pow{w_\sigma}}{\lambda(\sigma) : \sigma \in \Sigma'}
\substitute{0}{\lambda(\sigma) : \sigma \in \Sigma \setminus \Sigma'}
\substitute{\vec a}{\vec x}
\substitute{\vec c}{\vec z}
\substitute{\vec b}{\vec w}
=
\gamma\,
\substitute{\vec a}{\vec x}
\substitute{\vec c}{\vec z},
\end{equation*}
because, under the composition
$
\substitute{\vec a}{\vec x}
\substitute{\vec c}{\vec z}
\substitute{\vec b}{\vec w}
$,
the first two substitutions on the left-hand side
of the equation are not changing any integers in the formula.
Therefore, 
some formula $\gamma\,\substitute{\vec a}{\vec x}\substitute{\vec c}{\vec z}$,
$\gamma \in \Gamma$,
must be true.

By Claim~\ref{c:semenov-split-phi-into-gamma},
this ends the proof of the right-to-left direction.
\end{proof}

\begin{lemma}
\label{l:correct-semenov-nice-var}
Suppose function \funcSem
is given a vector $\vec x$ of variables and a formula $\phi(\vec x, \vec z)$ of $\cT$,
containing $\pow{x}$ for each $x \in \vec x$.
Then for every pair $(\vec x, \psi(\vec x, \vec z))$ in its output,
there is a variable $x \in \vec x$ such that
$\pow x$ only occurs in constraints from \PowComp in $\psi$.
\end{lemma}

\begin{proof}
Every formula $\psi$ appearing in the output either arises from line~\ref{semenov:guard-for-lambda},
in which case it contains no variable from $\vec x$ (and so every $x \in \vec x$ fits the bill),
or can be traced back
to one of the sets $\Gamma_x$, $x \in \vec x$.
Let us show that, in the latter case, this variable~$x$ satisfies the requirement.
Indeed, first consider the set of all atomic formulae in~$\phi$ in which
$x$ appears.
Beyond constraints from \PowComp,
these can be:
\begin{enumerate}
\renewcommand{\labelenumi}{\theenumi)}
\item inequalities outside \PowComp in which $x$ appears as a power;
\item inequalities outside \PowComp in which $x$ does not appear as a power
but appears linearly;
\item divisibility (modulo) constraints outside \PowComp.
\end{enumerate}
All occurrences of the constraints of the first type undergo substitutions
in lines~\ref{sem:case-1}--\ref{sem:end-case}. In the resulting atomic
formulae, $x$ may still feature but all these appearances will be linear.
In particular, the requirement on $\pow x$ is irrelevant for them.
Constraints of the second type are unaffected by the procedure, and
the requirement on $\pow x$ is not applicable to them either.
Finally, no new divisibility constraints are introduced or changed
in \funcSem, and thanks to Lemma~\ref{l:only-simple-div}
all such constraints are simple; therefore, all occurrences
of $\pow x$ in them will be in \PowComp.

We should now consider occurrences of~$x$ introduced by the
procedure itself. Here, it is easy to see that any such occurrences
will be in atomic formulae from \PowComp. In particular,
comparisons between $\lambda(a) \cdot \pow x$ and $\lambda(\sigma)$
(or $2 \cdot \lambda(\sigma)$) introduced by repeated updates to $\Gamma_x$
will be brought to that fragment after the lambda term is replaced
with $0$ or with $\pow{w_\sigma}$ in line~\ref{semenov:lambda-sub}.
This completes the proof.
\end{proof}

\begin{lemma}
\label{l:correct-semenov}
The function \funcSem correctly implements its specification.
\end{lemma}

\begin{proof}
This is a consequence of Lemmas~\ref{l:semenov-cover-in-fragment},
\ref{l:correct-semenov-cover},
and~\ref{l:correct-semenov-nice-var}, as well as the observation that
variables $\vec w$ are fresh.
\end{proof}

Based on the preceding lemmas,
we give a simple argument that the inner while loop is correct, assuming
that the subroutines meet their specifications.


\subsection{Correctness of function \funcLinearise}

In this subsection we will show that function {\funcLinearise} meets its specification.
For the substitutions involving modulus, this will involve some easy number-theoretic
facts.

The following claims can be deduced using properties of rings;
we give proofs that only rely on two properties of Euler's totient function~$\totient$.
We recall that $\totient(m)$, $m \ge 1$, is the number of integers between $1$ and $m$
that are coprime to~$m$, that is, whose greatest common divisor with~$m$ is~$1$.

\begin{claim}
\label{c:sufficient-period}
Let $b, r, m$ be positive integers.
Suppose there exists a positive integer $u$ for which
$r \cdot (b^u - 1) \equiv 0 \mod m$.
Then $u \eqdef \totient(m)$ is also a solution to this congruence, that is,
$r \cdot (b^{\totient(m)} - 1) \equiv 0 \mod m$.
\end{claim}

\begin{proof}
Denote $d = \gcd(r, m)$.
If $d > 1$, then we note that
the two congruences
$r \cdot (b^u - 1) \equiv 0 \mod m$ and
$\frac r d \cdot (b^u - 1) \equiv 0 \mod \frac m d$
have equal sets of solutions.

So let us assume first that $d = 1$.
Rewrite the congruence as a divisibility constraint:
$m \divides r \cdot (b^u - 1)$;
since $\gcd(r, m) = 1$, this is in fact equivalent to
$m \divides (b^u - 1)$.
We have assumed that a solution~$u > 0$ exists; this means
that $b$ belongs to the multiplicative group of integers modulo~$m$;
i.e., the set of all nonnegative integers between $0$ and $m - 1$ that
are coprime to $m$.
This group has order $\totient(m)$.
Now, if $m \divides b^u - 1$, then $u$ is a multiple
of the order of $b$ in this group, which means that $\totient(m) \divides u$.
Clearly, $\totient(m)$ is in this case also a solution to the original congruence.

Let us now consider the scenario $d > 1$.
By the argument above, $\totient(\frac m d)$ is a solution to the congruence.
By a standard property of $\totient$, if $x \divides y$ then $\totient(x) \divides \totient(y)$;
see, e.g.,~\cite[Exercise~5.21]{NT}.
Therefore, $\totient(m)$ must also be a solution to the congruence.
\end{proof}

\begin{claim}
\label{c:progression}
Let $b, r, m$ be positive integers.
The set of all nonnegative integer solutions \textup{(}in $u$\textup{)} to the congruence
$r \cdot (b^u - 1) \equiv 0 \mod m$ is either $\{0\}$
or an arithmetic progression $\{0, D, 2D, \ldots \}$, for some $D > 0$.
\end{claim}

\begin{proof}
Similarly to the previous claim, we can assume without loss of generality
that $\gcd(r, m) = 1$ and, moreover, that $r = 1$.
It suffices to show that the difference of two solutions is also a solution.
If $b^u \equiv b^v \equiv 1 \mod m$ and $u > v$, then
$b^u - b^v = b^v \cdot (b^{u-v} - 1) \equiv 1 \cdot (b^{u-v} - 1) \equiv 0 \mod m$,
which completes the proof.
\end{proof}

\begin{lemma}
\label{l:exp-congruence}
Let $b, r, m$ be positive integers.
Let $S$ be the set of all nonnegative integer solutions \textup{(}in $x$\textup{)} to
the congruence $b^x \equiv r \mod m$.
Then, if $S \ne \emptyset$,
\begin{gather*}
\text{either} \qquad S = \{ s \}
\qquad \text{or} \qquad
S = \{ s + i t:\ i \ge 0 \},
\\
\text{where}\qquad
s = \min S,
\qquad
t = \min \{ u > 0 :\ r \cdot (b^u - 1) \equiv 0 \mod m \}.
\end{gather*}
Moreover, in the second case $t \divides \totient(m)$.
\end{lemma}

\begin{proof}
First suppose that the number $t$ from the statement of the lemma is well-defined,
i.e., there exists some $u > 0$ for which $r \cdot (b^u - 1) \equiv 0 \mod m$.
We show that in this case
all elements of the set $\{s + i t : i \ge 0\}$ are solutions to the congruence.
Indeed, notice that for $t$ we have
$r \cdot b^t \equiv r \mod m$.
Consider $s + i t$ for some $i \ge 0$.
If $i = 0$, this is clearly a solution to the congruence by our choice of $s$.
Otherwise 
$b^{s + i t} = b^s \cdot b^t \cdot (b^t)^{i-1} \equiv r \cdot b^t \cdot (b^t)^{i-1}
 \equiv r \cdot (b^t)^{i-1}$;
continuing in the same way, we arrive at $b^{s + i t} \equiv r \mod m$,
which means that $s + i t$ is indeed a solution to the congruence.

To complete the proof,
suppose the congruence in question has
some two solutions in nonnegative integers: say,
$b^x \equiv b^y \equiv r \mod m$ with $x \ge y \ge 0$.
Then
$b^x - b^y = b^y \cdot (b^{x-y} - 1) \equiv r \cdot (b^{x-y} - 1) \equiv 0 \mod m$.
By Claim~\ref{c:progression},
the number $x-y$ belongs to an arithmetic progression with difference~$t$.
Therefore, $x - y$ is a multiple of $t$.
Finally,
$t$ is a divisor of $\totient(m)$ by Claim~\ref{c:sufficient-period}.
\end{proof}

For Lemma~\ref{l:correct-linearise}, recall that
by $\freevars(\phi)$ we denote the set of free variables of a formula $\phi$.

\begin{lemma}
\label{l:correct-linearise}
The function \funcLinearise correctly implements its specification:
assuming $\cT = \QF$,
for each pair $(\vec x, \theta)$, where $\vec x \subseteq \freevars(\theta)$,
it outputs
a pair $(\vec x, \theta')$ where $\theta \fequiv \theta'$ and, for all $x \in \vec x$,
    if $\pow x$ only occurs in constraints from \PC in $\theta$, then
    $x$ only occurs linearly in $\theta'$.
Moreover,
every formula $\theta'$ in the output belongs to $\Fragment$
as long as $\theta \in \Fragment$ in the input.
\end{lemma}

\begin{proof}
For atomic formulae that are comparisons, the correctness of the substitutions
is straightforward.
For divisibility constraints, the correctness is a consequence of Lemma~\ref{l:exp-congruence}
with $b = 2$:
\begin{itemize}
\item The third case corresponds to  $S$ in the lemma being empty.
\item The second case
corresponds to the possibility that  $S$ in the lemma is a singleton: $r'$ is defined to be the minimum value of $|x|$ that is a solution.
\item
The initial case corresponds to  $S$ in the lemma being a non-trivial arithmetic progression. Here $r'$ will be the initial value of
the progression, while $q'$ will be the difference. The first conjunct states that $|x|$ differs from the initial value
by a multiple of the difference.
\end{itemize}
Notice that, in the search for~$s$ and~$t$ for divisibilty constraints,
all computations can be done modulo~$q$; for example, the powers
$\pow{s}$ with $s \ge q$ will necessarily repeat some of the previous powers.
Because of this, it is possible to stop the search at~$q$.

Let us demonstrate that the output of \funcLinearise adheres to
the specification.
Let $x \in \vec x$ be such that $\pow{x}$ occurs in constraints from \PC only, in $\theta$.
In this case, all of these constraints will be replaced in the \textbf{for} loop,
and thus all occurrences of~$x$ in $\theta'$ will be linear.
All other occurrences of $x$ will be left unchanged.
Finally, $\theta' \in \Fragment$, because
no quantifiers are introduced.
\end{proof}

\subsection{Termination and correctness of the Master procedure}
Throughout this subsection we refer to Algorithm \ref{algo:top-level}.

\begin{lemma}
\label{l:dnf-like}
In between iterations of the inner \textbf{while} loop (line~\ref{top:inner-loop})
the equivalence
\begin{equation*}
\exists \vec u . \Psi \ \fequiv\ %
\Pi'.
\left[
\bigvee_{(\vec x, \phi) \in Q} \exists \vec x. \phi
\lor
\bigvee_{\phi \in D} \phi
\right]
\end{equation*}
holds.
\end{lemma}

\begin{proof}
Due to line~\ref{top:init-q-and-d},
the equivalence holds trivially at the beginning
of each iteration of the outer \textbf{while} loop.
When the inner \textbf{while} loop is reached,
we need to show that the removal of each pair $(\vec x, \phi)$ from $Q$
in the loop header
is compensated by the new pairs added to $Q$ and $D$ by the procedure.

\begin{itemize}

\item
The case of $x \in \vec x$ not appearing in $\vec x$ is trivial:
$\exists  \vec x                 . \phi \fequiv
 \exists (\vec x \setminus \{x\}). \phi$.

\item
The case $(\exists x. \phi) \in \cT$ is also easy because
$\exists  \vec x                 . \phi \fequiv
 \exists (\vec x \setminus \{x\}). (\exists x. \phi)$.

\item
The cases of calls to Presburger and Semenov subroutines
are considered separately in Lemmas~\ref{l:correct-presburger}
and~\ref{l:correct-semenov},
as well as Lemma~\ref{l:correct-linearise} for \funcLinearise,
which demonstrate the correctness
of these subroutines.
\qedhere

\end{itemize}
\end{proof}

\begin{lemma}
\label{l:d-and-q-are-good}
All formulae in $D \cup \{ \phi : \text{$(\vec x, \phi) \in Q$ for some $\vec x$}\}$
belong to the fragment~$\cT$.
\end{lemma}

\begin{proof}
Let us track formulae that are added to the set $D$, as well as  formulae $\phi$ for which
pairs $(\vec x, \phi)$ are added to the set $Q$.

At the very beginning, $D$ is empty and $Q$ contains the pair $(\vec u, \Psi)$ in which
$\Psi \in \Fragment$ by line~\ref{top:decompose-prefix}.
Note that at least one decomposition
$\Phi = \Pi . \exists  \vec u . \Psi$ where $\Psi$ is in $\cT$
always exists, because
a possible choice for $\Psi$ is the quantifier-free part of
the input formula~$\Phi$.
Indeed,
the quantifier-free part of $\Phi$ is in \Fragment
by the specification of Algorithm~\ref{algo:top-level}.
We remark that if the innermost quantifier block is universal, then the required
format can be ensured by rewriting this block as $\lnot \exists \vec u. \lnot$.

Within the inner \textbf{while} loop,
lines~\ref{top:move-to-d}, \ref{top:remove-variable}, and \ref{top:fall-into-fragment}
require no special consideration.
Calls to \funcPA in line~\ref{top:call-presburger} ensure that all outputs only contain formulae
from \Fragment, by the specification of \funcPA. The spec is guaranteed to be observed, by
Lemma~\ref{l:correct-presburger}.
As for calls to \funcSem in line~\ref{top:call-semenov}, they follow the specification
of \funcSem, which ensures (Lemma~\ref{l:correct-semenov}) that all output formulae are in \Fragment.
Notice that in the case $\Fragment = \Sem$ there is no linearization involved:
\funcLinearise is a no-op.
In the case $\Fragment = \QF$, %
we then apply
the final part of Lemma~\ref{l:correct-linearise}.

We can conclude at this point that, by the time the formula $\Phi$ is updated in line~\ref{top:update-phi},
all formulae in $D \cup \{ \phi : \text{$(\vec x, \phi) \in Q$ for some $\vec x$}\}$
belong to the fragment~$\cT$. If the algorithm terminates after this, we are done.
Otherwise we observe that a disjunction of formulae from $D$ belongs to~$\cT$ as well,
and so at the beginning of the next iteration of the outer \textbf{while} loop
at least one decomposition
$\Phi = \Pi . \exists  \vec u . \Psi$, where $\Psi$ in $\cT$,
again exists at line~\ref{top:decompose-prefix}. So, once again
$D$ becomes empty and $Q$ contains the pair $(\vec u, \Psi)$ in which
$\Psi \in \Fragment$. By induction on the number of iterations, we conclude that
the statement is true throughout the run of Algorithm~\ref{algo:top-level}.
\end{proof}

\begin{lemma}
\label{l:pi-prime}
The string of quantifiers $\Pi'$ only ever gets populated by
universal quantifiers.
If $\vec y$ is empty and $\Pi$ is empty \textup{(}or contains $\lnot$ only\textup{)}
in line~\ref{top:decompose-prefix}, then
$\Pi'$ stays empty as well in lines~\ref{top:inner-loop}--\ref{top:update-phi}.
\end{lemma}

\begin{proof}
New quantified variables, $w_\sigma$, are only introduced by the \funcSem subroutine.
They are used
to remove occurrences of $\lambda$-terms, of the form $\lambda(t)$.
Inspecting the update to $\Gamma_x$ in that subroutine, we observe
that there are two types of these terms:
firstly, those with $t$ an explicitly given integer and,
secondly, those with $t = \sigma(\vec z)$ where $\vec z$ is a vector
of free variables (those not appearing in $\vec x$).
The former can be evaluated directly and require no introduction
of extra variables or quantifiers; so only the latter count.

For the second assertion of the lemma,
suppose that the outer \textbf{while} loop reaches an iteration
in which the quantifier prefix $\Pi$ contains no quantifiers.
Also suppose that the original formula $\Phi$ has no free variables (i.e., $\vec y$ is empty).
Then, in all iterations of the inner \textbf{while} loop,
the pair $(\vec x, \phi)$ popped from the set $Q$ has the property that
$\freevars(\phi) \subseteq \vec x$. This property is inductive, because:
\begin{itemize}
\item variables can only be removed from $\vec x$ if they don't appear in $\phi$ (line~\ref{top:remove-variable}),
      or if the quantifier that binds them gets moved to the front of $\phi$ (line~\ref{top:fall-into-fragment});
\item calls to the \funcPA subroutine introduce no new variables and keep $\vec x$
      unchanged;
\item calls to the \funcSem subroutine have no free variables (more precisely, $\vec z$ is empty)
      and thus all occurrences of $\lambda$ in the output can be evaluated directly ---
      that is, no fresh variables or quantifiers get introduced, by \funcSem or \funcLinearise.
\qedhere
\end{itemize}
\end{proof}

We see that, if the input formula $\Phi$ has free variables,
$\Phi$ can be converted into either an existentially quantified or a universally quantified formula
(modulo $\Fragment$).

We are now ready to prove Lemma \ref{l:termination-and-correctness} from the body, which we now restate:

\TerminationAndCorrectness*

\begin{proof}
Let us first show that each iteration of the outer \textbf{while} loop in Algorithm~\ref{algo:top-level} terminates.
The idea of the proof is that, in formulae $\phi$ where $(\vec x, \phi) \in Q$, the number
of variables in $\vec x$ decreases. A single pair $(\vec x, \phi)$ popped from~$Q$
may create a significant number of other pairs with the same vector of variables~$\vec x$,
say $(\vec x, \psi_1), \ldots, (\vec x, \psi_k)$,
but in this case all the new formulae $\psi_i$ will in some sense be simpler
and their further processing will decrement the number of variables.
Eventually all the formulae will be brought to $D$, when the vectors
of variables $\vec x$ are exhausted.

Let us show how this idea works in detail.
We need to consider the pairs $(\vec x, \phi)$ added to $Q$ during the iterations
of the inner \textbf{while} loop in Algorithm~\ref{algo:top-level}.
Lines~\ref{top:remove-variable} and~\ref{top:fall-into-fragment} are unproblematic,
and in fact we show below that no pair $(\vec x, \phi)$ can ``evolve''
for long without triggering either of these two lines, thus decrementing
the number of variables in~$\vec x$.
\begin{itemize}
\item
In line~\ref{top:call-presburger}, the function \funcPA is called; by Lemma~\ref{l:correct-presburger}
this function correctly implements its specification. In particular, no formula $\psi(\vec y)$
output by the algorithm can contain the variable~$x$. Therefore, all pairs
$(\vec x, \psi(\vec y))$ added by \funcPA back to~$Q$ have the following property:
when they are next popped from~$Q$, they will trigger line~\ref{top:remove-variable}.
\item
In line~\ref{top:call-semenov},
the function \funcSem is called; by Lemma~\ref{l:correct-semenov}
this function correctly implements its specification. In particular,
for every pair $(\vec x, \psi(\vec x, \vec z, \vec w))$ produced by \funcSem,
there is a variable $x \in \vec x$ such that
$\pow x$ only occurs in constraints from \PowComp in $\psi$.
Notice that %
pairs $(\vec x, \psi(\vec x, \vec z, \vec w))$ output by \funcSem will be fed to the
\funcLinearise function.
We consider two scenarios:
\begin{itemize}
\item
$\Fragment = \QF$. In this case, we rely on Lemma~\ref{l:correct-linearise}
asserting that \funcLinearise correctly implements its specification.
Since, for $x \in \vec x$,
$\pow x$ only occurs in constraints from \PC in $\theta$, we know
$x$ only occurs linearly in $\theta'$, which is the output of \funcLinearise
that corresponds to $\theta$.
All such pairs $(\vec x, \theta')$ are added back to $Q$.
And they all satisfy the following property: when they are next popped from~$Q$,
they will trigger line~\ref{top:call-presburger}, and later, by the argument above,
line~\ref{top:remove-variable}.
In fact, strictly speaking, some other variable might be eliminated from~$\vec x$
altogether, in which case line~\ref{top:remove-variable} is triggered immediately.
\item
$\Fragment = \Sem$. In this case, function \funcLinearise passes its input directly
to the output, and
pairs $(\vec x, \psi)$ are added straight back to $Q$.
However, recall that, again by Lemma~\ref{l:correct-semenov},
$\psi \in \Sem$. Since
we already know that $\pow x$ only occurs in constraints from \PowComp in $\psi$,
this means that $(\exists x. \psi) \in \Sem$. Therefore, when the pair $(\vec x, \psi)$
is next popped from~$Q$, it will trigger line~\ref{top:fall-into-fragment}.
\end{itemize}
\end{itemize}
As we have just seen, pairs from the worklist~$Q$, when they are popped
and processed, are always replaced with new pairs that either have
strictly fewer variables in the vector $\vec x$ or have the property
that all their descendants will have strictly fewer variables
within at most two steps (pops).
As we start with a finite $Q$, this process will necessarily terminate.
And when it does, the current iteration of the outer \textbf{while} loop will come to an end.

To show termination of the entire procedure,
we use the invariants established by Lemmas~\ref{l:d-and-q-are-good} and~\ref{l:pi-prime}.
Note that, by Lemma~\ref{l:pi-prime},
the alternation depth \emph{inside} inside the prefix $\Pi. \Pi'$
cannot increase within any single iteration of the outer \textbf{while} loop.
Indeed, only universal quantifiers can get added to $\Pi'$.
If $\Pi$ ended with an existential quantifier, this quantifier could have
been moved out of $\Pi$ and into the block $\exists \vec u$ in the first place.
Thus, since $\Pi$ was chosen as the shortest prefix, it may only end with a universal
quantifier, or it may contain no quantifiers at all.
In the former case, appending new universal quantifiers ($\Pi'$) does not
increase the alternation depth.
In the latter case,
the outer \textbf{while} loop terminates thanks to line~\ref{top:return}.

To sum up, on each iteration of the outer \textbf{while} loop of our Master procedure
one block gets consumed at line~\ref{top:decompose-prefix}.
In total, the outer \textbf{while} loop
will run at most $\alt(\Phi)$ iterations, one per each quantifier
block in the prenex normal form.
We note that there are two possible scenarios
for the last iteration of this loop:
\begin{itemize}
\item
Suppose $\freevars(\Phi) = \emptyset$;
that is, $\vec y$ is empty.
Again by Lemma~\ref{l:pi-prime}, $\Pi'$ remains
empty and so there are no quantifiers at the end of the outer \textbf{while} loop iteration.
This means that the procedure will return the (possibly negated, if $\Pi = \lnot$)
disjunction of all $\phi \in D$,
which is in $\cT$ by Lemma~\ref{l:d-and-q-are-good}.
\item
Now suppose that $\Phi$ has at least one free variable.
By the argument above, the concatenation $\Pi . \Pi'$ will be alternation-free
and will be returned followed by the disjunction $\bigvee_{\phi \in D} \phi$.
As this disjunction is, again, in $\cT$,
the output of the procedure is indeed alternation-free modulo $\cT$
in this case too.
\end{itemize}
This justifies the first bullet point in the statement of the lemma.
The second bullet point holds by Lemma~\ref{l:pi-prime}.

It remains to check the equivalence of $\Phi$ and $\Phi'$.
The invariant from Lemma~\ref{l:dnf-like} implies that
\begin{equation*}
\Phi(\vec y) \ \fequiv\ %
\Pi. \Pi'.
\left[
\bigvee_{(\vec x, \phi) \in Q} \exists \vec x. \phi
\lor
\bigvee_{\phi \in D} \phi
\right].
\end{equation*}
Since the exit from the inner \textbf{while} loop at line~\ref{top:update-phi} is on $Q = \emptyset$,
this equation simplifies to
$\Phi(\vec y) \fequiv \Pi. \Pi'. \bigvee\limits_{\phi \in D} \phi$.
This ensures that line~\ref{top:update-phi} only ever rewrites $\Phi$
into an equivalent formula.
For $\Pi$ without quantifiers, the equation matches the return value of the procedure.
\end{proof}

\subsection{Supplementary facts about the procedures}

We prove two lemmas on our procedures that are later required for the complexity analysis.

\begin{lemma}
\label{l:sem-no-accidental-linear}
Let $\Fragment = \Sem$.
Suppose that, for some iteration of the outer \textbf{while} loop in \TopAlgo,
every variable from the vector $\vec u$ appears only in powers within the formula $\Psi$.
Then the subroutine \funcPA is never invoked during this iteration.
\end{lemma}

\begin{proof}
Under the assumptions of the lemma, the \funcLinearise subroutine is a no-op.
Therefore, it suffices to show that the formulae output by repeated calls
to \funcSem cannot be linear in any of the variables $u \in \vec u$
(unless $u$ disappears completely from such a formula).
Consider the updates to $\Gamma_x$ in lines~\ref{sem:case-1}--\ref{sem:end-case} first:
\begin{itemize}
\item
Constraints introduced as part of $\beta$ clearly  have the variables of $\vec x$ only within powers.
The same holds for all ``guards'' appearing in the updates before the results of substitutions,~$\gamma[\ldots]$.
\item
Constraints that are introduced in~line~\ref{sem:case-2}
as a result of substitutions in inequalities $\alpha$, $\alpha \in A$,
may see a cancellation of power terms. 
Let $a \ne 0$ and $b \ne 0$ be the original coefficients in $\alpha$ at the power
terms $\pow{x}$ and $\pow{v}$, respectively.
If $a \cdot 2^j - b = 0$, then the power term $\pow{v}$ disappears from
the inequality. However, the term $\pow{v}$ is still present in the newly introduced
constraint $\pow{x} = 2^j \cdot \pow{v}$ in this case, so the variable $v$ retains
at least one power occurrence.
(It may also be the case that \emph{all} occurrences of $v$ disappear in a particular $\psi$,
for some pair $(\vec x, \psi)$ from the output of the procedure.)
\item
Constraints that are introduced in lines~\ref{sem:case-5} and~\ref{sem:case-6}
will, after the removal of $\lambda$-terms in line~\ref{semenov:lambda-sub},
have the term $\pow{x}$ replaced either by a constant multiple of a fresh $\pow{w_\sigma}$,
or by $0$. Neither option can lead to other variables becoming linear.
\end{itemize}
In a similar fashion, formulae introduced within $\Gamma$ in line~\ref{sem:union-gamma}
and to $\Theta$ in lines~\ref{semenov:guard-for-lambda}--\ref{semenov:lambda-sub},
cannot lead to any variable becoming linear.
This completes the proof.
\end{proof}

\EssentiallyPA*

\begin{proof}
  First, if a bound variable appears within a power in $\Phi$ and it is selected in line~\ref{top:decompose-prefix} as part of $\vec u$, then it is 
  handled by line~\ref{top:fall-into-fragment}. Hence, we can assume that all variables from $\vec u$ are from $\Pi$. In this case, 
  note that lines~\ref{top:move-to-d} to~\ref{top:call-presburger} do not introduce  variables in powers. in particular,~$\funcPA$ implements a rather standard quantifier-elimination procedure for~$\PA$, \emph{\`a la} Weispfenning~\cite{Weispfenning90}.
  The lemma then follows from the fact that~\funcSem requires a variable to occur as a power in the formula~$\phi$ of line~\ref{top:inner-loop} in order to be invoked.
\end{proof}

%% file: appendix-complexity.tex
\section{Omitted proofs from Section~\ref{section:complexity}}
\label{appendix:complexity}

In this appendix, given a formula $\Phi$ we define $\norminf{\Phi} \coloneqq \max(\norminf{\linterms{(\Phi)}},\fmod(\Phi))$.

\subsection{Proof of Theorem~\ref{theo:exexp-nexp}}

Recall the following lemma from the body of the paper, Lemma~\ref{lemma:presburger-one-round},
which handles a single call to~$\funcPA$:

\PresburgerOneRound*

\begin{proof} 
  We discuss the growth of each parameter in the table. 

  \proofsubparagraph{Bounds on {\rm{$\maxvars$}} and {\rm{$\norminf{\homterms{(\cdot)}}$}}.}
  The analyses for $\maxvars$ and $\norminf{\homterms{(\cdot)}}$ is simple, 
  and essentially follows from 
  the substitutions $\phi\substitute{\frac{t_1+k}{a_1}}{x}$ performed in
  line~\ref{pres:line-sub} of~\funcPA. 
  In particular, these substitutions cause 
  the parameter~$\maxvars$ to
  double (but, of course, the $\maxvars$ is always bounded by the number of variables in the formula), and $\norminf{\homterms{(\cdot)}}$ to increase to $2 \cdot a^2$.
  Indeed, note that given an inequality $a_2 \cdot x + t_2 < 0$ such that $x$ does not occur in $t_2$, 
  we have
  $(a_2 \cdot x + t_2)\substitute{\frac{t_1+k}{a_1}}{x} \, = \, a_2 \cdot (t_1+k) + a_1 \cdot t_2 < 0$; and in both $a_2 \cdot (t_1+k)$ and $a_1 \cdot t_2$ the absolute value of the coefficients of the variables is bounded by $a^2$. Lastly, note that~\simplify does not have any effect on inequalities as it only manipulates the divisibility constraints.

  \proofsubparagraph{Bound on {\rm{$\norminf{\linterms(\cdot)}$}}.}
  This bound is found with an analysis similar to the one for $\norminf{\homterms{(\cdot)}}$. Again, the key ingredient is given by the substitutions $\phi\substitute{\frac{t_1+k}{a_1}}{x}$ performed in
  line~\ref{pres:line-sub} of~\funcPA. There, by definition, $\abs{k} \leq a \cdot m$ and the constant in $t_1$ is bounded, in absolute value, by $c$. Then, the constant in the inequality $a_2 \cdot (t_1+k) + a_1 \cdot t_2 < 0$ defined as above is bounded by $a(c + a \cdot m) + a \cdot c$.

  \proofsubparagraph{Bound on {\rm{$\card{\homterms}$}}.}
  Because of line~\ref{pres:line-sub} of~\funcPA, each $\psi_i$ ($i \in [1,k])$ has at most the number of homogeneous terms as in $\phi$. 
  This is due to the fact that all inequalities in $\psi_i$ are obtained form a single substitution $\phi\substitute{\frac{t_1+k}{a_1}}{x}$
  (\simplify does not have any effect on inequalities). 

  Let us now discuss $\card{\homterms}$ across all $\psi_i$. 
  First, observe that given $(a_1,t_1),(a_2,t_2) \in T$ such that $t_1$ and $t_2$ only differ by their constant (i.e., they have the same homogeneous term), 
  and given any $k_1,k_2 \in \Zed$, 
  the formulae $\phi\substitute{\frac{t_1+k_1}{a_1}}{x}$ and $\phi\substitute{\frac{t_2+k_2}{a_2}}{x}$ have the same set of homogeneous terms. 
  Up to constant factors in $t$, there are at most $h+1$ elements $(a,t) \in T$ (where the ``$+1$'' comes from the inclusion in $T$ of the pair $(1,0)$ required to make $T$ non-empty).
  From the bound $\card{\homterms(\psi_i)}$ obtained above, 
  we then conclude that $\card{\homterms(\bigvee_{j=1}^k \psi_j)} \leq h \cdot (h+1)$.

  \proofsubparagraph{Bound on {\rm{$\card{\linterms}$}}.} 
  The bound on $\card{\linterms(\psi_i)}$ follows the same reasoning as $\card{\homterms(\psi_i)}$.

  \proofsubparagraph{Bound on {\rm{$\fmod$}}.}
  First, note that the function~\simplify does not change $\fmod$.
  Then, the bounds on this parameter follow again line~\ref{pres:line-sub} of~\funcPA. Indeed, each $\psi_i$ ($i \in [1,k])$ 
  features a divisibility constraint of the form $a_1 \divides t_1 + k$, where $a_1$ is the coefficient of $x$ in some homogeneous term. Then, 
  $\fmod(\psi_i) \leq a \cdot m$, whereas $\fmod(\bigvee_{i=1}^k \psi_i) \leq a^h \cdot m$, as across different disjuncts a different homogeneous term might be considered. 

  \proofsubparagraph{Bounds on {\rm{$\boolnum$}} and {\rm{$k$}}.}
  For the parameter $\boolnum$, consider $\psi_i$ with
  $i \in [1,k]$. Line~\ref{pres:line-sub} of \funcPA
  adds one constraint $a \divides t + k$ to $\phi$,
  increasing $\boolnum$ from $b$ to $b+1$. Afterwards,
  \simplify adds one constraint $d \divides t' - r(t')$ for
  every $t'$ that is a variable or an exponential
  $\pow{y}$ appearing in a non-simple divisibility
  constraint (see~line~\ref{simp:add-to-gamma}). Since 
  all divisibility constraints in $\phi$ are assumed to be simple 
  (see Lemma~\ref{l:only-simple-div} and the text below this lemma), these variables and exponentials all come from the term~$t+k$
  considered
  in line~\ref{pres:line-sub} of the function~\funcPA. 
  This term has at most $v$ variables, and therefore 
  the number of
  divisibility constraints added in
  line~\ref{simp:add-to-gamma} of \simplify is bounded
  by $2 \cdot v$, leading to $\boolnum(\psi_i) \leq b' \coloneqq
  b+2\cdot v + 1$, as shown in the parameter table.
  Then, $\boolnum(\bigvee_{j =1}^k \psi_j) \leq
  \sum_{j=1}^k (\boolnum(\psi_j)+1) \leq k \cdot (b' +
  1)$ follows immediately.
  Lastly, to bound $k$ it suffices to count the number of
  iterations in \funcPA and \simplify. 
  Note that, by definition, $\card{\linterms(\phi)} \leq \card{\homterms(\phi)} \cdot (2 \cdot \norminf{\linterms(\phi)}+1)$.
  We get:
  \begin{align*}
    k &\leq \underbrace{\card{T} \cdot (2 \cdot \norminf{\homterms(\phi)} \cdot \fmod(\phi)+1)}_{\text{cases in line~\ref{pres:line-sub} of \funcPA}} \cdot \underbrace{\max_{1 \le j \le k}(\norminf{\homterms(\phi)} \cdot \fmod(\phi))^{2\,\maxvars(\phi)}}_{\text{line~\ref{simp:loop} of \simplify}}\\
    & \leq \card{\linterms(\phi)} \cdot (2 \cdot a \cdot m + 1)\cdot (m \cdot a)^{2v} \leq 4 \cdot q \cdot (a \cdot m)^{2v + 1}.
  \end{align*}
  The running time is simple to establish. It is essentially bounded by the sizes of the output formulae (when one disregards the ``automatic'' term normalizations discussed in~\Cref{section:preliminaries}, which might decrease the size of the formula). That is to say, the upper bounds in the parameter table, together with $\len{\phi}$ to parse the initial formula, are enough to estimate the running time of the procedure.
\end{proof}

The following lemma, Lemma \ref{lemma:sem-one} in the body,
 handles a single call to~\funcSem.

\LemSemOne*

\begin{proof}
  We discuss the growth of each parameter in the table. 

  \proofsubparagraph{Bound on {\rm{$\fmod$}}.} The function~\funcSem does not introduce or
  change the set of divisibility constraints in $\phi$,
  and so $\fmod(\theta_i) = \fmod(\phi)$. 
  
  \proofsubparagraph{Bound on {\rm{$\maxvars$}}.}
  The bound
  $\max(2,v)$ on $\maxvars$ is rather simple to
  establish: all constraints introduced by \funcSem are
  either from $\PC$ or they are computed by opportunely
  substituting in inequalities an exponentiated
  variable $\pow{x}$ with either a constant or another exponentiated
  variable $\pow{y}$, hence never increasing $\maxvars$
  --- note that the substitutions in
  lines~\ref{sem:case-5} and~\ref{sem:case-6} replace
  $\pow{x}$ with $\lambda(\sigma)$ and
  $\frac{\lambda(\sigma)}{\lambda(a)}$, respectively,
  but $\lambda(\sigma)$ is then replaced with either
  $0$ or $\pow{w_{\sigma}}$ in
  line~\ref{semenov:lambda-sub}.

  \proofsubparagraph{Bound on {\rm{$\norminf{\linterms(\cdot)}$}}.}
  We start by considering how
  $\norminf{\linterms(\cdot)}$  increases in
  lines~\ref{sem:outer-loop} to~\ref{sem:union-gamma}.
  First, note that the iterations of the outermost
  \textbf{for} loop in line~\ref{sem:outer-loop} act on
  different variables $\Gamma_x$ that are initialized
  as $\{\phi\}$. Hence, it suffices to bound
  $\norminf{\linterms(\cdot)}$ for elements in a single
  $\Gamma_x$. We also note that the innermost
  \textbf{for} loop does not modify the sets $I$
  and $H$ defined in lines~\ref{sem:I} and~\ref{sem:H},
  and that $A$ defined in line~\ref{sem:A} only depends
  on $I$ and the current pair $(\eta,\sigma) \in H$
  considered by the for loop. This implies that,
  throughout the iterations of the innermost
  \textbf{for} loop, the variable $2^g$ defined in
  line~\ref{sem:g} is always bounded by $2^7((2 \cdot
  \maxvars({\phi})+1) \cdot \norminf{\linterms(\phi)})^2 \leq 2^{11} \cdot v^2 \cdot c^2$.
  Similarly, $a$ in line~\ref{sem:a} and $\card V$ for
  line~\ref{sem:V} are bounded by
  $\norminf{\linterms{(\phi)}}$ and $\maxvars({\phi})$.
  Since the substitutions performed in
  lines~\ref{sem:case-1} to~\ref{sem:case-8} are local
  to the term $\alpha \in A$, we conclude that
  $\norminf{\linterms(\cdot)}$ can be bounded by looking at
  a single iteration of the innermost \textbf{for}
  loop. An analysis of lines~\ref{sem:case-1}
  to~\ref{sem:case-8} reveals that
  $\norminf{\linterms(\cdot)}$ increases the most in
  line~\ref{sem:case-2}, because of the substitution
  $\alpha\substitute{2^j \cdot \pow{v}}{\pow{x}}$. We
  have 
  \begin{align*}
    \norminf{\alpha\substitute{2^j \cdot \pow{v}}{\pow{x}}} 
    & \leq \norminf{\alpha} \cdot 2^j + \norminf{\alpha}\\
      &&\hspace{-6.3cm}\text{(since } (a_1 \cdot \pow{x} + a_2 \cdot \pow{v} + \dots)\substitute{2^j \cdot \pow{v}}{\pow{x}} = (a_1 \cdot 2^j + a_2) \cdot \pow{v} + \dots\text{\,)}\\
    & \leq \norminf{\linterms(\phi)} \cdot 2^{11} \cdot v^2 \cdot c^2 + \norminf{\linterms(\phi)} 
      &\hspace{-0.5cm}\text{(since $2^j \leq 2^g$)}\\ 
    & \leq 2^{12} \cdot v^2 \cdot c^3
  \end{align*}
  We conclude that, in line~\ref{sem:union-gamma}, all
  formulae in $\Gamma$ have $\norminf{\linterms(\cdot)}$
  bounded by $2^{12} \cdot v^2 \cdot c^3$. Subsequent
  formulae introduced in lines~\ref{sem:delambda-start}
  to~\ref{sem:return} do not change this upper bound.
  Indeed, these lines only consider inequalities of the
  form $\sigma = 0$ or $\pow{w_\sigma} \leq
  \abs{\sigma} < 2 \cdot \pow{w_\sigma}$ (possibly
  negated), which have an infinity norm bounded by
  $\max(2,\norminf{\sigma}) \leq c$, together with
  formulae obtained by substituting $\lambda(\sigma)$
  with $\pow{w_\sigma}$ or $0$, which do not cause an
  increase to $\norminf{\linterms(\cdot)}$. 
  \proofsubparagraph{Bound on {\rm{$\card{\homterms(\theta_i)}$}}.}
  Once more,
  we start by analyzing lines~\ref{sem:outer-loop}
  to~\ref{sem:union-gamma}, and note that iterations of
  the outermost \textbf{for} loop act on different
  variables $\Gamma_x$, so it suffices to analyze the
  growth of $\card{\homterms}$ for a single 
  $x \in \vec x$. Pick $(\eta,\sigma) \in H$. Note that
  $\card{\homterms(\beta)} \leq \maxvars(\phi)+1$,
  where $\beta$ is defined in line~\ref{sem:beta}. An
  analysis of lines~\ref{sem:case-1}
  to~\ref{sem:case-8} reveals that $\card{\homterms}$
  increases there by at most $\card{\homterms(\beta)} +
  2 \leq \maxvars(\phi)+3$ (note that the \emph{substitutions} 
  with respect to elements $\alpha \in A$
  performed in these lines do not change the number of
  homogeneous terms, by definition of~$A$;
  line~\ref{sem:A}).
  Since the innermost \textbf{for}
  loop is performed at most $\card{\homterms(\phi)}$
  times, we conclude that each formula~$\gamma \in
  \Gamma$ (line~\ref{sem:union-gamma}) is such that 
  \begin{align*}
    \card{\homterms(\gamma)} 
    &\leq \underbrace{\card{\homterms(\phi)}}_{\text{initial number of hom.~terms}} + 
    \underbrace{(\maxvars(\phi)+3) \cdot \card{\homterms(\phi)}}_{\text{growth due to the \textbf{for} loops}} + 
    \underbrace{\card{\vec{x}}}_{\text{formula $\bigwedge\limits_{y \in \vec x} \pow{x} \geq \pow{y} \land {}$}}\\
    & \leq \ (v+4) \cdot h + \card{\vec x}.
  \end{align*}
  We now consider lines~\ref{sem:delambda-start}
  to~\ref{sem:return}. 
  Given $\sigma \in \Sigma$ 
  (see~line~\ref{sem:delambda-start})
  consider
  the $7$ homogeneous terms stemming from the inequalities
  \[ 
    \sigma > 0, \quad \sigma < 0, \quad \sigma \geq 0, \quad \pow{w_\sigma} \le \sigma, \quad \sigma < 2 \cdot \pow{w_\sigma}, \quad \pow{w_\sigma} \le -\sigma, \quad -\sigma < 2 \cdot \pow{w_\sigma}.
  \]
  Each formula in $\Theta$ introduced
  in line~\ref{semenov:guard-for-lambda} 
  only uses some of the inequalities above for a fixed $\sigma \in \Sigma$, and thus has at most $7$ homogeneous terms. 
  Similarly, every formula
  in~line~\ref{semenov:lambda-sub} has either $6$ or
  $2$ of these $7$ homogeneous terms, for every
  $\sigma \in \Sigma$ (depending on whether $\sigma =
  0$ or $\pow{w_\sigma} \le \sigma < 2 \cdot
  \pow{w_\sigma}$ is considered), plus
  $\card{\homterms(\gamma)}$ many homogeneous terms,
  where $\gamma \in \Gamma$. Therefore, 
  \[
    \homterms(\theta_i) \leq 6 \cdot \card{\Sigma} + \max_{\gamma \in \Gamma}(\card{\homterms(\gamma)}) \leq 6 \cdot h + (v + 4) \cdot h + \card{\vec x} \leq (v+10) \cdot h + \card{\vec x}.
  \]

  \proofsubparagraph{Bound on {\rm{$\card{\boolnum(\theta_i)}$}}.}
  This case is very similar to the one of~$\card{\homterms(\theta_i)}$. Indeed, in computing an upper
  bound for $\card{\homterms(\theta_i)}$ we were essentially
  counting the number of atomic formulae added to a
  single formula, and in particular one can show that
  for every $\gamma \in \Gamma$
  (line~\ref{sem:union-gamma}), 
  \begin{align*}
    \boolnum(\gamma)
    &\leq \hspace{-5pt} \underbrace{\boolnum(\phi)}_{\text{initial number of Bool. connectives}} + 
    \underbrace{(\maxvars(\phi)+5) \cdot \card{\homterms(\phi)}}_{\text{growth due to the for loops}} + 
    \underbrace{\card{\vec{x}}}_{\text{expression $\bigwedge_{y \in \vec x} \pow{x} \geq \pow{y} \land$}}\\
    & \leq b + (v+5) \cdot h + \card{\vec x}.
  \end{align*}
  Afterwards, note that each formula in $\Theta$ introduced
  in line~\ref{semenov:guard-for-lambda} requires $13$
  connectives, since $\sigma \neq 0$ is short for
  $\lnot(-1 < \sigma \land \sigma < 1)$, and
  $\pow{w_{\sigma}} \leq \abs{\sigma} < 2 \cdot
  \pow{w_\sigma}$ is short for 
  \[ 
    \lnot (\sigma \geq 0 \land \lnot (\pow{w} \leq \sigma \land \sigma < 2 \cdot \pow{w})) \land \lnot (\sigma < 0 \land \lnot (\pow{w} \leq -\sigma \land -\sigma < 2 \cdot \pow{w}))
  \]
  (we avoid here using derived Boolean connectives such
  as $\implies$ or $\lor$), where $\sigma \in \Sigma$
  (see line~\ref{sem:delambda-start}). Each formula
  in~line~\ref{semenov:lambda-sub} is a conjunction of
  one of the two formulae above for each $\sigma \in
  \Sigma$ together with a formula from $\gamma \in
  \Gamma$ (up to substitution of terms that do not
  change the number of Boolean connectives
  in~$\gamma$). We obtain,
  \[ 
    \boolnum(\theta_i) \leq 10 \cdot \card \Sigma + \max_{\gamma \in \Gamma}(\boolnum(\gamma)) \leq 10 \cdot h + b + (v+5) \cdot h + \card{\vec x} \leq b + (v+15) \cdot h + \card{\vec x}.
  \]

  \proofsubparagraph{Bound on {\rm{$\card{\homterms(\bigvee_{i=1}^k
  \theta_i)}$}}.} 
  We start by working out how many
  homogeneous terms are generated by the various
  substitutions in lines~\ref{sem:case-1}
  to~\ref{sem:case-8} and
  line~\ref{semenov:lambda-sub}.
  Note that each $\rho \in \homterms(\phi)$ is considered at most once per variable $x \in \vec x$ in the \textbf{for} loop of line~\ref{sem:outer-loop}. 
  Moreover, the substitutions in lines~\ref{sem:case-1}
  to~\ref{sem:case-8} are simultaneously done on all inequalities having the same homogeneous term. 
  Then, below we say that a homogeneous term~$\rho'$ is a \emph{substitute} of $\rho \in \homterms(\phi)$ if $\rho'$ is the (single) homogeneous term generated from these inequalities, after substitution. We count the number of substitutes in lines~\ref{sem:case-1} to~\ref{sem:case-8} and
  line~\ref{semenov:lambda-sub}, for a fixed homogeneous term $\rho \in \homterms(\phi)$ but across all variables in $\vec x$. Below, note that when we perform a substitution on $\alpha$ we only ever consider variables appearing in it, or $\lambda(\sigma)$ (lines~\ref{sem:case-5} and~\ref{sem:case-6}). 
  \begin{enumerate}
    \item The substitutions in line~\ref{sem:case-1} generate at most $v \cdot (g+1)$ substitutes of~$\rho$, since 
    the substitution $\alpha\substitute{2^j}{\pow{x}}$ can take $\maxvars(\alpha) \leq v$ many values for $x$ and $g+1$ many values for $j$ (again, across all choices of variables in $\vec x$). 
    \item The substitutions in line~\ref{sem:case-2} generate at most $v^2 \cdot (g+1)$ substitutes of~$\rho$,
    \item The substitutions of lines~\ref{sem:case-3} and~\ref{sem:case-7} replace all inequalities having $\rho$ as a homogeneous term as $\true$,
    \item The substitutions of lines~\ref{sem:case-4} and~\ref{sem:case-8} replace all inequalities having $\rho$ as a homogeneous term as $\false$, 
    \item\label{s1} The substitutions in line~\ref{sem:case-5} generate at most $v$ substitutes for~$\rho$, since $\lambda(\sigma)$ is uniquely defined for every $\rho \in \homterms(\phi)$, no matter the variable we are iterating over in line~\ref{sem:outer-loop}, 
    \item\label{s2} The substitutions in line~\ref{sem:case-6} generate at most $v$ substitutes for~$\rho$ (for the same reason as in the previous point).
    \item The substitutions in line~\ref{semenov:lambda-sub} take all the substitutes generated in points~\ref{s1} and~\ref{s2} above, and generate for each of them $2$ substitutes.
  \end{enumerate}
  Hence, for each homogeneous term in $\homterms(\phi)$ the procedure generates at most 
  \[
    {v \cdot (g+1) + v^2 \cdot (g+1) + 2 \cdot 2 \cdot v}
  \] 
  many substitutes. 
  Note that
  \[
    g \ \leq \ \log(2^{11} \cdot v^2 \cdot c^2) 
    \ \leq \ 11 + 2 \cdot \log(v \cdot c)
    \ \leq \ 13 \cdot \log(v \cdot c),
  \]
  and therefore
  \begin{align*}
    v \cdot (g+1) + v^2 \cdot (g+1) + 2 \cdot 2 \cdot v
    &\leq{}
      v \cdot (g \cdot (v+1) + v + 5)\\ 
    &\leq{}
      v \cdot (13 \log(v \cdot c) \cdot (1+v)+v + 5)\\ 
    &\leq{}
      v \cdot \left(13 v \log(c) \cdot (2+v)\right)\\ 
    &\leq{}
      26 \cdot v^3 \log(c).
  \end{align*}
  Therefore, the number of homogeneous terms stemming from substitutions is at most $h \cdot 2^5 \cdot v^3 \cdot \log(c)$. 
  To conclude the analysis on
  $\card{\homterms(\bigvee_{i=1}^k \theta_i)}$ it
  suffices to count the number of homogeneous terms
  coming from inequalities outside the substitutions.

  Let us fix a
  homogeneous term $\rho \in \homterms(\phi)$ of an inequality $\rho+c'<0$ (with
  $c' \in \Zed$) that at some point during the run
  of~\Cref{algo:top-level} occurs in the set~$I$ of
  line~\ref{sem:I}. Associated to $\rho$ there is a single
  homogeneous term $\sigma(\vec z)$ such that $\rho = \tau(\vec x) + \sigma(\vec z)$, for some $\tau$. We count the number of
  inequalities added in lines~\ref{sem:case-1}
  to~\ref{sem:case-8} and
  lines~\ref{semenov:guard-for-lambda}
  and~\ref{semenov:lambda-sub} for that specific $\rho$
  and its associated $\sigma$.
  Let $G \coloneqq 13 \cdot \log(v \cdot c)$, the previously derived upper bound on $g$.
  \begin{enumerate}
    \item Line~\ref{sem:beta} adds $v + v^2$ many homogeneous terms overall (we are counting across all possible variables occurring in the fixed $\rho \in \homterms(\phi)$; and recall that $\maxvars(\phi) \leq v$).
    \item Line~\ref{sem:case-1} adds $2 \cdot v$ many homogeneous terms ($\pow{x}$ and $-\pow{x}$, for all $x \in \vec x$ appearing in $\rho$), $v$ of which were already considered in the previous step.
    \item Line~\ref{sem:case-2} adds $2 \cdot v^2 \cdot (G+1)$ new homogeneous terms ($\pow{x} > 2^g$ already counted in line~\ref{sem:beta}).
    \item\label{semcover-lambda-1} Line~\ref{sem:case-3} adds $v$ new homogeneous terms of the form $\lambda(a) \cdot \pow{x} < \lambda(\sigma)$, since $a \cdot \pow{x}$ must appear in $\rho$. This line also adds $1$ more homogeneous term (that is,~$\sigma$, from~$\sigma < 0$).
    \item Line~\ref{sem:case-4} adds $1$ new homogeneous terms (i.e., $-\sigma$).
    \item\label{semcover-lambda-2} Line~\ref{sem:case-5} adds $v$ new homogeneous terms. Observe that the homogeneous terms of the inequality $\lambda(a) \cdot \pow{x} \leq \lambda(\sigma)$ (which is used for expressing the equality $\lambda(a) \cdot \pow{x} = \lambda(\sigma)$) already appear in line~\ref{sem:case-4}.
    \item\label{semcover-lambda-3} Line~\ref{sem:case-6} adds $2 \cdot v$ many homogeneous terms.
    \item Line~\ref{sem:case-7} and~\ref{sem:case-8} do not add new homogeneous terms ($a < 0$ and $a > 0$ simplify as $\top$ or $\bot$, and every $2 \cdot \lambda(\sigma) - \lambda(a) \cdot \pow{x} < 0$ is already considered above).
    \item Line~\ref{semenov:guard-for-lambda} adds $4$ new homogeneous terms.
    \item Line~\ref{semenov:lambda-sub} modifies every 
    homogeneous term involving $\lambda(\sigma)$ considered in lines~\ref{sem:case-3} to~\ref{sem:case-6}.
    By~\Cref{semcover-lambda-1,semcover-lambda-2,semcover-lambda-3}, there are
    at most $4 \cdot v$ terms 
    of the form ${\pm (\lambda(a) \cdot \pow{x} - \lambda(\sigma))}$ 
    and $\pm (\lambda(a) \cdot \pow{x} - 2 \cdot \lambda(\sigma))$, 
    and line~\ref{semenov:lambda-sub}
    replaces the $\lambda(\sigma)$ appearing in these terms with $0$ or $\pow{w_\sigma}$.
    When replacing $\lambda(\sigma)$ with $0$, the inequalities in which these terms appear are evaluated to $\top$ or $\bot$.
    Then, only the substitutions with respect to $\pow{w_\sigma}$ are of interest, but these do not change the total number of the (at most) $4 \cdot v$ terms we are considering.
  \end{enumerate}
  This means that for every $\rho$ the procedure adds 
  \begin{align}
    & (v+v^2) + v + 2 v^2\cdot (G+1) + (v +1) + 1 + v + 2 v + 4 \notag\\
    {}\leq{}& 2 \cdot v^2\cdot G + 3 \cdot v^2 + 6v + 6 \label{lemma7:counting-hom}\\
    {}\leq{}& 26 \cdot v^2 \log(v \cdot c) + 3 \cdot v^2 + 6 v + 6 \notag\\
    {}\leq{}& 2^6 v^3 \log(c) \notag
  \end{align}
  many homogeneous terms.
  Accounting for the different homogeneous terms brings this bound to $h \cdot 2^6  v^3 \log(c)$. It then suffices to add the number of substitutes previously computed, which we found to be bounded by $h \cdot 2^5  v^3 \log(c)$, as well as $\card{\vec x}^2$ for the homogeneous terms in line~\ref{sem:union-gamma} (which are the same across all iterations of \funcSem), to obtain the upper bound $\card{\homterms(\bigvee_{j=1}^k \theta_j)}  \leq h \cdot 2^7  v^3 \log(c) + \card{\vec x}^2$.
  
  \proofsubparagraph{Bound on {\rm{$k$}}.} 
  What is left
  is to establish the bound on $k$, 
  from which we directly also get a bound on ${\boolnum(\bigvee_{j=1}^k
  \theta_j)}$. To do that, we
  simply look at the number of iterations done. Let us
  first look at the update to $\Gamma_x$ in
  line~\ref{sem:case-1}. Suppose that, before the
  update, $\card{\Gamma_x} = \xi$. Then, after the
  update we have 
  \begin{align*}
    \card{\Gamma_x} 
    &\leq 8 \cdot \xi \cdot (g+1) \cdot (v - 1)\\
    &\leq 8 \cdot \xi \cdot (13 \cdot \log(v \cdot c)+1) \cdot (v - 1)\\
    &\leq \xi \cdot 2^7 \cdot v^2 \cdot \log(c).
  \end{align*}
  The variable $\Gamma_x$ is initially set to
  $\{\phi\}$ and it is updated at most $h$ times, hence
  when the algorithm reaches line~\ref{sem:union-gamma},
  we get:
  \[ 
    \card{\Gamma} \leq \sum_{x \in \vec x} \card{\Gamma_x} \leq \card{\vec x} \cdot (2^7 \cdot v^2 \cdot \log(c))^h.
  \]
  In line~\ref{semenov:guard-for-lambda}, $\Theta$ has
  cardinality at most $\card{\Sigma} \leq h$. The
  \textbf{for} loop in line~\ref{sem:for-sigmaprime}
  adds to this set $2^{\card{\Sigma}} \cdot
  \card{\Gamma}$ many formulae. Hence, $k$ is bounded
  by 
  \begin{align*}
    h + 2^{\card{\Sigma}} \cdot \card{\Gamma} 
    &\leq h + 2^h \cdot \card{\vec x} \cdot (2^7 \cdot v^2 \cdot \log(c))^h\\ 
    &\le  h + v^{10 h} \cdot (\log(c))^{h} \cdot \card{\vec x}\\
    &\leq (v+1)^{10h} \cdot (\log(c))^{h} \cdot \card{\vec x}.
  \end{align*}

\proofsubparagraph{Items~\ref{semcover-add-item1}--\ref{semcover-add-item3} after the table of~\Cref{lemma:sem-one}.} \Cref{semcover-add-item1}, that is the bound on $\Pi'$, is immediate from $\card{\Sigma} \leq h$ (line~\ref{sem:delambda-start}). 
For~\Cref{semcover-add-item2}, 
as in the case of~\Cref{lemma:presburger-one-round} and the procedure~\funcPA, inspection of the function~\funcSem 
shows that the runtime of the procedure is at most polynomial in the size of the output formulae
(again, as stressed in the proof of~\Cref{lemma:presburger-one-round}, one has here to disregard the ``automatic'' term normalizations 
discussed in~\Cref{section:preliminaries}, which might decrease the size of the formula). Therefore, the upper bounds in the parameter table, together with $\len{\phi}$ to parse the initial formula, are enough to estimate the running time of the procedure.
\Cref{semcover-add-item3} follows from our analysis of $\card{\homterms(\theta_i)}$. Indeed, note that 
in each $\theta_i$ all homogeneous terms involving variable from $\vec x$ that do not correspond to inequalities from $\powcomp$ come from the various substitutions done in lines~\ref{sem:case-1}
  to~\ref{sem:case-8} and
  line~\ref{semenov:lambda-sub}. None of these substitutions increase the number of homogeneous terms. Hence, the number of homogeneous terms is at most $\card{\homterms(\phi)} \leq h$.
\end{proof}

We now move to the complexity of the procedure~\funcLinearise.
In order to prove Lemma \ref{lemma:lin-bounds} (in the body),
we need the following two claims stating properties of
Euler's totient function.

\begin{claim}[cf.~{\cite[Exercise~2.3.5(b)]{FreudGyarmati}}]
  \label{c:lcm-of-totient}
  If $a_1, \ldots, a_m$ are positive integers, then
  \[
  \lcm(\totient(a_1), \ldots, \totient(a_m)) \le \lcm(a_1, \ldots, a_m).
  \]
  \end{claim}
  
  \begin{proof}
  We rely on the standard fact that $\totient(N) = N \cdot \prod\limits_{p \divides N} (1 - 1/p)$,
  where the product is over all primes~$p$ that divide $N$;
  see, e.g.,~\cite[Corollary~5.7]{NT}.
  Let $p_1, \ldots, p_n$ be all primes that occur in prime factorizations
  of $a_1, \ldots, a_m$; and let $\alpha_{i j}$ be nonnegative integers
  for which $a_i = p_1^{\alpha_{i 1}} \cdot \ldots \cdot p_n^{\alpha_{i n}}$,
  for $1 \le i \le m$.
  We then have
  \begin{equation*}
  \totient(a_i) = \prod_{j : \alpha_{i j} > 0} \left[ p_j ^ {\alpha_{i j} - 1} \cdot (p_j - 1) \right].
  \end{equation*}
  Therefore,
  \begin{align*}
  \lcm(\totient(a_1), \ldots, \totient(a_m)) \ \divides\ %
  &\prod_{j=1}^{n} \left[ p_j ^ {\max\nolimits_i \alpha_{i j} - 1} \cdot (p_j - 1) \right]\\ \le\,
  &\prod_{j=1}^{n} p_j ^ {\max_i \alpha_{i j}} \cdot \prod_{j=1}^{n} \frac{p_j - 1}{p_j},
  \end{align*}
  where the maxima are over $i$, $1 \le i \le m$.
  It remains to observe that the final product does not exceed
  $ \lcm(a_1, \ldots, a_m) \cdot 1 $.
  \end{proof}
  
  \begin{claim}
  \label{c:subset-totient}
  For any two finite sets
  $\{a_1, \ldots, a_m\}$ and $\{b_1, \ldots, b_n\}$ of positive integers,
  \begin{equation}
  \label{eq:subset-totient}
  \lcm\left(
      \totient(a_1), \ldots, \totient(a_m),
      b_1, \ldots, b_n
      \right)
  \le
  \lcm(a_1, \ldots, a_m, b_1, \ldots, b_n)^2.
  \end{equation}
  \end{claim}
  
  \begin{proof}
  This is a consequence of Claim~\ref{c:lcm-of-totient}.
  Denote
  $A = \lcm(\totient(a_1), \ldots, \totient(a_m))$ and
  $B = \lcm(b_1, \ldots, b_n)$.
  We already know that $A \le \lcm(a_1, \ldots, a_m)$,
  so the left-hand side of Equation~\eqref{eq:subset-totient}
  is equal to
  $\lcm(A, B) \le A \cdot B \le \lcm(a_1, \ldots, a_m) \cdot B$.
  Each of the two factors is upper-bounded by the least common
  multiple of all $m+n$ integers, which completes the proof.
  \end{proof}
  
We remind the reader that, in a parameter table, the bounds on the input are assumed to be positive integers. In particular, in the lemma below $b \geq 1$.

\LinBounds*

\begin{proof}
  We discuss the growth of each parameter in the table. 

  \proofsubparagraph{Bound on {\rm{$\boolnum(\theta_j')$}}.}
  We analyze each substitution done by the three cases of line~\ref{linearise:cases}. 
  For the first two cases, 
  the number of Boolean connectives corresponds to the number of connectives required to express formulae of the form $\abs{x} - \abs{y} + c < 0$. 
  This formula is equivalent to:
  \begin{align*}
    & \Big(x < 0 \implies ((y < 0 \implies -x + y + c < 0) \land (y \geq 0 \implies -x - y + c < 0)) \Big)\\
    & \land  \Big( x \geq 0 \implies ((y < 0 \implies x + y + c < 0) \land (y \geq 0 \implies x - y + c < 0)) \Big),
  \end{align*}
  which requires $24$ Boolean connectives, as each implication $\phi \implies \psi$ is a shortcut for~${\lnot(\phi \land \lnot \psi)}$, which requires three Boolean connectives, 
  and $x \geq 0$ is a shortcut for $\lnot(x < 0)$, which requires one Boolean connective.
  The third of the cases of line~\ref{linearise:cases} requires less than $24$ Boolean connectives. Indeed, 
  $(q' \divides \abs{x} - r') \land (\abs{x} \ge r')$ is equivalent to  
  \[ 
    \big(x < 0 \implies (q' \divides x +r' \land x+r' \leq 0)\big)
    \land
    \big( x \ge 0 \implies (q' \divides x -r' \land r'-x \leq 0)\big),
  \]
  which uses $12$ Boolean connectives. Lastly, the formula $\abs{x} = r'$ is equivalent to 
  \[ 
    \big(x < 0 \implies (x \geq r' \land  x \leq r')\big)
    \land
    \big( x \ge 0 \implies (-x \geq r' \land  -x \leq r')\big)
  \]
  which requires $14$ Boolean connectives. 
  Noting that the number of atomic formulas in $\theta_j$ is bounded $\boolnum(\theta_j)$, we then conclude that  
  $\boolnum(\theta_j') \leq 24 \cdot \boolnum(\theta_j)$.

  \proofsubparagraph{Bounds on {\rm{$\maxvars(\theta_j')$}}, {\rm{$\norminf{\homterms(\theta_j')}$}} and  {\rm{$\norminf{\linterms(\theta_j')}$}}.}
  The bounds on $\maxvars(\theta_j')$
  and $\norminf{\homterms(\theta_j')}$
  are trivial, as each inequality considered by~\funcLinearise 
  is replaced with inequalities having the same number of variables and a smaller (in fact, logarithmic) magnitude of constants, 
  and otherwise~\funcLinearise only adds inequalities of the form $\pm (x - r') < 0$ or $\pm(x+r') < 0$, arising from divisibility constraints.
  In this case, note that $r'$ is bounded by $\fmod(\theta_j)$, and thus we also conclude that $\norminf{\linterms(\theta_j')} \leq \max(\norminf{\linterms(\theta_j)},\fmod(\theta_j)) \leq \max(c,m)$.

  \proofsubparagraph{Bound on {\rm{$\card\homterms(\theta_j')$}}.}
  Given a variable $x \in \vec x$, at most two homogeneous terms are added in line~\ref{lin:in-1} (i.e., $x$ and $-x$) and at most $6 \cdot \ell$ terms are added in line~\ref{lin:in-2} (i.e., $\pm x \pm y$ and $\pm y$ for each of at most $\ell$ variables~$y$). Line~\ref{lin:mod} introduces yet again the homogeneous terms $x$ and $-x$. Hence, the total number of new homogeneous terms is bounded by $(6\cdot \ell + 2) \cdot n$.

\proofsubparagraph{Bound on {\rm{$\fmod(\theta_j')$}}.}
  The bound $\fmod(\theta_j') \leq \fmod(\theta_j)^2$ follows directly from~\Cref{l:exp-congruence} and~\Cref{c:subset-totient}. Indeed, suppose that $\theta_j$ contains divisibility constraints with divisors $q_1,\dots,q_\eta$. \funcLinearise modifies some of these divisors in line~\ref{lin:mod}, obtaining new divisors $p_1,\dots,p_{g}$.
  By~\Cref{l:exp-congruence}, each $p_i$ divides $\totient(q_{u})$ for some $u \in [1,\eta]$. Then, 
  $\fmod(\theta_j') \leq \lcm(p_1,\dots,p_g,q_1,\dots,q_\eta) \leq \lcm(\totient(q_1),\dots,\totient(q_\eta),q_1,\dots,q_\eta) \leq \lcm(q_1,\dots,q_\eta)^2 = \fmod(\theta_j)^2$, 
  where the last inequality is a consequence of~\Cref{c:subset-totient}.
  
  \smallskip
  Regarding the running time of~\funcLinearise, it is clear that it runs in polynomial time with respect to $\card{S}$ (outermost loop), $n$ (innermost loop), and $\len{\theta_j}$ (to scan the formula~$\theta_j$ for affected inequalities and divisibilities). The length, $\len{\theta_j}$, upper-bounds $\log(a)$ and $r$ (required to compute the replacement of the inequalities). However, note that the procedure is \underline{not} polynomial in $\log(m)$ and thus in general~\funcLinearise runs in exponential time. 
  This is because computing the numbers $q'$ and $r'$ might require up to $m$ iterations (and $m$ is represented in binary).
\end{proof}

\begin{remark}\label{remark:non-deterministic-linearize}
  The computation of $q'$ and $r'$ from line~\ref{lin:mod} of procedure~\funcLinearise
  can be performed in non-deterministic polynomial time
  with respect to the bit lengths of $q$ and $r$.

  The non-deterministic machine proceeds as follows.
  It first guesses (and checks) the factorization of $q$, 
  which it uses to compute $\totient(q)$.
  If $q \mid (2^{\totient(q)} - 1)$,
  then the number $q' > 0$ is found by guessing a divisor of $\totient(q)$
  together with all prime divisors of $q'$.
  The machine verifies that $q' = \min\{t > 0 : q \divides r \cdot (2^t - 1)\}$
  by checking $q \mid r \cdot (2^{q'} - 1)$ and $q \nmid r \cdot (2^{\frac{q'}{d}} - 1)$
  for every prime divisor $d$ of $q'$.
  The number $r' \geq 0$ is obtained by guessing a number smaller than $q'$
  and verifying that $q \divides 2^{r'} - r$.

  If $q \nmid (2^{\totient(q)} - 1)$, 
  then $q'$ does not exist. 
  In this case, Lemma~\ref{l:exp-congruence} guarantees that
  there is at most one $r' \geq 0$ satisfying $q \divides 2^{r'} - r$. 
  The machine then guesses $r' \in [0,q]$
  and verifies whether $q \divides 2^{r'} - r$.
  Failing this verification corresponds to the third case in line~\ref{lin:mod}.
\end{remark}

\begin{proof}[Proof of~\Cref{theo:exexp-nexp}]
  Without loss of generality, 
  we consider a prenex sentence $\exists \vec x \Phi$ 
  having variables from $\vec x = (x_1,\dots,x_n)$, 
  and $\Phi$ quantifier-free.
  We can also assume that all divisibilities in $\Phi$ are simple.
  This can achieved by initially running on $\Phi$ 
  a non-deterministic variant of $\simplify$ 
  that in line~\ref{simp:loop} guesses a map~$r$, instead of iterating through all maps. Observe that this increases the size of $\Phi$ 
  only polynomially.

  In view of the discussion provided in~\Cref{sect:complexity-nexp}, it suffices to show that, starting from a pair $(\vec x, \Phi)$ such that $\vec x = (x_1,\dots,x_n)$,
  iterating $N$ calls to the non-deterministic version of the functions~\funcPA, \funcSem and \funcLinearise leads to formulae of
  sizes at most exponential in $n$, $\len{\Phi}$ and $N$. 
  Overall, the procedure will call these functions $N = O(n)$ times in order to eliminate each variable from $\Phi$.
  \Cref{lemma:presburger-one-round,lemma:sem-one,lemma:lin-bounds} give us the bounds for a single call to these non-deterministic procedures (see the first output row of the tables in all these lemmas), so in order to complete the proof it suffices to compose these bounds $N$ times. 
  For simplicity, from the 
  tables in~\Cref{lemma:presburger-one-round,lemma:sem-one,lemma:lin-bounds}
  we take an upper bounds for each parameter. According to these tables, each procedure among~\funcPA, \funcSem and \funcLinearise yields, when (non-deterministically) run on $(\vec x,\Phi)$ a single time, a formula $\Psi$ having the following parameters:

  \begin{center}
      \renewcommand\arraystretch{1.2}
      \setlength{\tabcolsep}{3pt}
      \begin{tabular}{|g|c|c|c|c|c|}
          \hline
          \rowcolor{light-gray}
          & $\card\homterms$ 
          & $\norminf{.}$
          & $\boolnum$\\
          \hline
          \hline
          $\Phi$
          & $h$
          & $c \geq 2$
          & $b$
          \\
          \hline
          \hline
          $\Psi$
          & $12 \cdot n^2 h$ 
          & $2^{12} \cdot n^2 \cdot c^3$
          & $24 \cdot (b + n \cdot h)$
          \\ 
          \hline
      \end{tabular}
  \end{center}
  Above, we recall that $\norminf{\Phi} \coloneqq \max(\norminf{\linterms{(\Phi)}},\fmod(\Phi)) \geq \norminf{\homterms{(\Phi)}}$,
  and that $\maxvars(\Phi) \leq n$.
  It is rather clear that, by calling these three procedures $N$ times, one obtains a formula $\Psi_N$ with the following bounds:
  \begin{center}
    \renewcommand\arraystretch{1.2}
    \setlength{\tabcolsep}{3pt}
    \begin{tabular}{|g|c|c|c|c|c|c|}
        \hline
        \rowcolor{light-gray}
        & $\card\homterms$ 
        & $\norminf{.}$
        & $\boolnum$\\
        \hline
        \hline
        $\Phi$
        & $h$
        & $c \geq 2$
        & $b$
        \\
        \hline
        \hline
        $\Psi_N$
        & $(12 \cdot n^2)^N \cdot h$
        & $(2^{12} \cdot n^2)^{{3}^{N}-1} \cdot c^{{3}^{N}}$
        & $24^N \cdot (b + 24 \cdot n \cdot h)- 24 \cdot n \cdot h$
        \\ 
        \hline
    \end{tabular}
\end{center}
The formal proof follows with a straightforward induction on $N$. 
Then, notice that the resulting formula $\Psi_N$ features $2^N \cdot O(n \cdot \len{\Phi})$ Boolean connectives, and atomic formulae with coefficients and constants whose bit length is ${\ceil{\log(\norminf{\Psi_N}+1)} \leq 2^{O(N)} \cdot \len{\Phi}}$. Hence, the overall length of the sequence of formulae $\Psi_1,\dots,\Psi_N$ is in 
\[ 
  N \cdot (\underbrace{2^N \cdot O(n \cdot \len{\Phi})}_{\text{number of Boolean connectives}} \cdot 
  \underbrace{(n+1) \cdot 2^{O(N)} \cdot \len{\Phi}}_{\text{bit length of an inequality}}) \leq 2^{O(n)} \cdot \len{\Phi}^2.
\]
Then, the non-deterministic algorithm runs in time $2^{O(n)} \cdot \len{\Phi}^2$, concluding the proof. 
We note that, while unnecessary to establish the exponential time upper bound, the described non-deterministic procedure uses 
the non-determinisitc version of~\funcLinearise, which runs in non-deterministic polynomial time with respect to the bit length of the input formula, as discussed in~\Cref{remark:non-deterministic-linearize}.
\end{proof}

\subsection{Towards a proof of Theorem~\ref{theo:pow-3exp}: \funcPA and Octagons}

We start by deriving bounds on iterated calls to~\funcPA.

\begin{restatable}{proposition}{PropQEPAmultiV}
  \label{prop:qePA-multiV}
  Consider $\Psi(\vec x, \vec z)$ in~$\QFPC$ in which variables $\vec x = (x_1,\dots,x_n)$ always occur linearly, and $n \geq 1$. Let $\cT \in \{\QF,\QFPC\}$. Run~\Cref{algo:top-level} from line~\ref{top:init-q-and-d} with $Q = \{(\vec x, \Psi)\}$. 
  After the {\rm\textbf{while}} loop in lines~\ref{top:inner-loop}--\ref{top:call-semenov} terminates, $D = \{\psi_1,\dots,\psi_k\}$ satisfies the parameter table
  \begin{center}
    \renewcommand\arraystretch{1.2}
    \setlength{\tabcolsep}{3pt}
    \begin{tabular}{|g|c|c|c|c|c|c|c|}
        \hline
        \rowcolor{light-gray}
        & $\card\homterms$ 
        & $\card\linterms$
        & $\maxvars$ 
        & $\norminf{\homterms(\cdot)}$ 
        & $\norminf{\linterms(\cdot)}$ 
        & $\fmod$
        & $\boolnum$\\
        \hline
        \hline
        $\Psi$
        & $h \geq 2$
        & $q$
        & $v$ 
        & $a \geq 2$ 
        & $c \geq 2$ 
        & $m \geq 2$
        & $b$
        \\
        \hline
        \hline
        $\psi_i$
        & $h$
        & $q$
        & $V$ 
        & $a^{2^{n+1}-1}$
        & $a^{2^{2n+1}}(c + m)$
        & $a^{2^{n+2}-n}m$
        & $b+n \cdot (2 \cdot V+1)$
        \\ 
        \hline
        $\bigvee_{j=1}^k\psi_j$
        & $h^{n+1}$
        & \cellcolor{light-gray}
        & \ditto
        & \ditto
        & \ditto
        & $a^{h^{2n+2}} m$
        & $k (\ditto\, + 1)$\\ 
    \hline
    \end{tabular}
\end{center}
where $V \coloneqq \min(2^n v, n + \card{\vec z})$ and $k \leq q^n \cdot (a^{2^{n+2}} \cdot m)^{n(2V+1)}$. Moreover, the algorithm runs in time $q^{O(n)} m^{O(n \cdot V)} \cdot a^{2^{O(n)} v}$.
\end{restatable}

\begin{proof}
  Note that, since all variables from $\vec x$ occur only linearly, from~\Cref{lemma:essentially-PA} the set $D = \{\psi_1,\dots,\psi_k\}$ coincides with the set of formulae in $\Gamma =  \{(\vec y, \psi_1),\dots,(\vec y,\psi_k)\}$, where $\Gamma$ is the set computed by the following procedure:
  {\rm{
  \begin{algorithmic}
    \State $\Gamma$ $\gets$ $\{\Psi\}$
    \For{$i \in [1,n]$}
      \State $\Gamma \gets \bigcup\limits_{\gamma \in \Gamma}
      \{\gamma' : (\vec w, \gamma') \in \funcPA(x_i,\vec x \setminus \{x_1,\dots,x_{i-1}\}, \gamma)\}$
      \Statex
      \Comment{In fact, $\vec w$ equals $\vec x \setminus \{x_1,\dots,x_{i-1}\}$}
    \EndFor
    \State \textbf{return} $\Gamma$
  \end{algorithmic}
  }}
  \noindent
  We show the bounds from the table by referring to this procedure.
  Note that in it the order of variables is the same for all formulae,
  whereas in Algorithm~\ref{algo:top-level} it might differ.
  This, however, will not affect our bounds, as our analysis is invariant 
  under permutations of variables.
  Let~$\Gamma_i$ be the value of $\Gamma$ after the
  $i$-th iteration of the for loop. That is, $\Gamma_0
  = \{\Psi\}$, and for $i \in [1,n]$, $\Gamma_{i} =
  \bigcup_{\gamma \in \Gamma_{i-1}} \{\gamma' : (\vec w, \gamma') \in \funcPA(x_{i}, 
  \vec x \setminus \{x_1,\dots,x_{i-1}\}, \gamma) \}$. 

  To start, we define a family of sets $H_0,\dots,H_n$ that will be useful to keep track of the growth of the number of homogeneous terms and of the moduli of the formulae in $\Gamma_i$. The set $H_i$ contains pairs of the form $(S,\eta)$, where $S$ is a set of homogeneous terms and $\eta$ is a positive integer. It satisfies the following properties: 
  \begin{enumerate}
    \renewcommand{\theenumi}{(\roman{enumi})}
    \renewcommand{\labelenumi}{\theenumi}
    \item\label{H:prop1} $\card H_i \leq h^i$,
    \item\label{H:prop2} for every $(S,\eta) \in H_i$, $\card S \leq h$,
    \item\label{H:prop3} for every $\gamma \in \Gamma_i$, there is $(S,\eta) \in H_i$ such that $\homterms(\gamma)\subseteq S$ and $\fmod(\gamma)$ divides $\eta$, and
    \item\label{H:prop4} if $i \geq 1$, for every $(S_1,\eta_1) \in H_i$,   
      there is $(S_2,\eta_2) \in H_{i-1}$ such that $\eta_1 = \eta_2 \cdot \widetilde{a}$ where $\widetilde{a}$ is a coefficient of $x_i$ in a homogeneous term of $S_2$.
  \end{enumerate}
  The set $H_0$ is defined as $\{(\homterms(\Psi),\fmod(\Psi))\}$. Given $i \geq 1$, the set $H_i$ is iteratively constructed as follows (starting from the empty set): 
  \begin{algorithmic}
    \For{$(S,\eta) \in H_{i-1}$}
      \For{each homogeneous term $a \cdot x_i + t$ from $S$ with $a \neq 0$}
        \If{$a < 0$}
          $(a,t) \gets (-a, -t)$
        \EndIf
          \State $S' \gets$ 
          \begin{minipage}[t]{0.8\linewidth} the set obtained from $S$ by replacing with $a \cdot t' - b \cdot t$ every term $b \cdot x_i + t'$ having $b \neq 0$ (terms not featuring $x_i$ are left unchanges).
          \end{minipage}
        \smallskip
        \State add $(S',\eta \cdot a)$ to $H_i$
      \EndFor
    \EndFor
  \end{algorithmic}
  
  \begin{claim}\label{claim:H-properties}
    For every $i \in [0,n]$, $H_i$ satisfies Properties~\ref{H:prop1}--\ref{H:prop4}.
  \end{claim}

  \begin{claimproof}
    The proof is by induction on $i$. The base case of $i = 0$ is trivial. Let $i \geq 1$, and assume by induction hypothesis that $H_{i-1}$ satisfies Properties~\ref{H:prop1}--\ref{H:prop4}. We show that $H_i$ satisfies these properties as well.
    \begin{description}
      \item[Property~\ref{H:prop1}.] By induction hypothesis, $\card{H_{i-1}} \leq h^{i-1}$, and each $(S,\eta) \in H_{i-1}$ contains at most $h$ homogeneous terms. Hence, the number of pairs $(S',\eta \cdot a)$ added to $H_i$ from each $(S,\eta) \in H_{i-1}$ is at most $h$, and thus $\card{H_i} \leq h^i$.
      \item[Property~\ref{H:prop2}.] By induction hypothesis, for every $(S,\eta) \in H_{i-1}$, $\card S \leq h$. The replacements performed when constructing a set $S'$ from $S$ do not increase the number of homogeneous terms. 
      Hence, $\card{S'} \leq \card S \leq h$.
      \item[Property~\ref{H:prop3}.] 
        Consider $\gamma \in \Gamma_i$. By definition, $(\vec w, \gamma) \in \funcPA(x_i,\vec x \setminus \{x_1,\dots,x_{i-1}\}, \gamma')$ for some $\gamma' \in \Gamma_{i-1}$ and vector of variables $\vec w$ (which is in fact equal to $\vec x \setminus \{x_1,\dots,x_{i-1}\}$).
        By induction hypothesis, there is $(S,\eta) \in H_{i-1}$ 
        such that $\hom(\gamma') \subseteq S$ and $\fmod(\gamma')$ divides~$\eta$. 
        Let $\substitute{\frac{t-k}{a}}{x_i}$ be the substitution considered by $\funcPA$ 
        when constructing $\gamma$ from $\gamma'$, where $a$ is a positive integer.
        The set $S$ contains either the homogeneous term $a \cdot x_i -t$ or the homogeneous term $-a \cdot x_i + t$. 
        Let $(S',\eta \cdot a)$ be the pair such that $S'$ is constructed from $S$ 
        by replacing with $a \cdot t' - b \cdot t$ every term $b \cdot x_i + t'$ having $b \neq 0$. By definition, $(S',\eta \cdot a) \in H_i$. 
        The update performed on $S$ in order to produce $S'$ is analogous 
        to the update performed by $\substitute{\frac{t-k}{a}}{x_i}$ on $\gamma'$. 
        Let $\tilde{\gamma} \coloneqq \gamma'\substitute{\frac{t-k}{a}}{x_i} \land (a \divides t + k)$.
        We thus have $\hom(\tilde{\gamma}) \subseteq S'$ and $\mod(\tilde{\gamma})$ divides $a \cdot \eta$. 
        The formula $\gamma$ is obtained by applying~\simplify to $\tilde{\gamma}$. 
        The updates performed by~\simplify guarantee $\hom(\gamma') \subseteq \hom(\tilde{\gamma})$ and $\fmod(\gamma')$ divides $\fmod(\tilde{\gamma})$.
        Then, $\hom(\gamma') \subseteq S'$ and $\fmod(\gamma')$ divides $a \cdot \eta$.

      \item[Property~\ref{H:prop4}.] This follows directly from the last line of the code constructing $H_i$.
      \qedhere
    \end{description}
  \end{claimproof}

  We now show, by induction on $i \in
  \PNat$, that all the formulae
  $\psi_1',\dots,\psi_\ell'$ in $\Gamma_i$ satisfy the
  bounds in the table below ($r \in [1,\ell]$):
  \begin{center}
    \renewcommand\arraystretch{1.2}
    \setlength{\tabcolsep}{3pt}
    \begin{tabular}{|g|c|c|c|c|c|c|c|c|}
        \hline
        \rowcolor{light-gray}
        & $\card\homterms$ & 
        $\card\linterms$ & $\maxvars$ &
        $\norminf{\homterms(\cdot)}$ &
        $\norminf{\linterms(\cdot)}$ & $\fmod$ &
        $\boolnum$\\
        \hline
        \hline
        $\phi$ & $h \geq 2$ & 
        $q$ &
        $v$ & $a \geq 2$ & $c
        \geq 2$ & $m \geq 2$ & $b$ \\
        \hline
        \hline
        $\psi_r'$ & $h$ & 
        $q$ & $V_i$ &
        $a^{2^{i+1}-1}$ & $a^{2^{2i+1}} (c+m)$ & 
        $a^{2^{i+2}-i}m$ & 
        $b+i \cdot (2 \cdot V_i+1)$ \\ 
        \hline
        $\bigvee_{j=1}^\ell\psi_j'$ & $h^{i+1}$ &
        \cellcolor{light-gray}
        &
        \ditto &
        \ditto & \ditto & $a^{h^{2i+2}} \cdot m$ & $\ell \cdot
        (\ditto + 1)$\\ 
    \hline
    \end{tabular}
  \end{center}
  where $V_i \coloneqq \min(2^i v, n + \card{\vec z})$ and $\ell \leq q^i \cdot (a^{2^{i+2}} m)^{i(2 V_i+1)}$. 
  This implies that the formulae
  $\psi_1,\dots,\psi_k$ in $\Gamma_n$ satisfy the
  bounds in the statement of the~lemma.

  The base case of $i = 1$ is direct
  from~\Cref{lemma:presburger-one-round}, since
  $\Gamma_1 = \{\gamma : (\vec x, \gamma) \in \funcPA(x_1,\vec x, \Psi)\}$. 
  In the induction step, let us assume by induction
  hypothesis that, for $i \geq 1$, $\Gamma_{i}$ contains
  formulae $\psi_1',\dots,\psi_\ell'$ satisfying the
  above bounds. We show that the formulae
  $\psi_1,\dots,\psi_k$ in $\Gamma_{i+1}$ satisfy the
  bounds in the table below ($s \in [1,k])$: 
\begin{center}
  \renewcommand\arraystretch{1.2}
  \setlength{\tabcolsep}{3pt}
  \begin{tabular}{|g|c|c|c|c|c|c|c|}
      \hline
      \rowcolor{light-gray}
      & $\card\homterms$ & 
      $\card\linterms$
      & $\maxvars$ &
      $\norminf{\homterms(\cdot)}$ &
      $\norminf{\linterms(\cdot)}$ & $\fmod$ & $\boolnum$\\
      \hline
      \hline
      $\phi$ & $h \geq 2$ & 
      $q$ & 
      $v$ & $a \geq 2$ & $c \geq
      2$ & $m \geq 2$ & $b$ \\
      \hline
      \hline
      $\psi_s$ & $h$ 
      & $q$ & $V_{i+1}$ &
      $a^{2^{i+2}-1}$ & 
      $a^{2^{2i+3}}(c+m)$ & 
      $a^{2^{i+3}-(i+1)} m$ & 
      $b + (i+1) (2 V_{i+1} + 1)$ \\ 
      \hline
      $\bigvee_{j=1}^k\psi_j$ & $h^{i+2}$ & 
      \cellcolor{light-gray}
      & \ditto &
      \ditto & \ditto & $a^{h^{2i+4}} \cdot m$ & $k
      \cdot (\ditto + 1)$\\ 
  \hline
  \end{tabular}
\end{center}
and moreover
$k \leq q^{i+1} \cdot (a^{2^{i+3}}m)^{(i+1)(2 V_{i+1}+1)}$.

By definition of $\Gamma$, for every $s \in
[1,k]$ there is $r \in [1,\ell]$ such that ${(\vec x \setminus
\{x_1,\dots,x_{i-1}\},
\psi_s)} \in \funcPA(x_{i},\vec x \setminus
\{x_1,\dots,x_{i-1}\},\psi_r')$. In
the analysis below, let $\psi_s$ and $\psi_r'$ be
formulae in such a relation.
\allowdisplaybreaks
\begin{description}
  \item[Bound on $\card\homterms$.] 
  By~\Cref{claim:H-properties}, there is $(S,\eta) \in H_{i+1}$ such that $\homterms(\psi_s) \subseteq S$ and $\card S \leq h$. 
  Hence, $\card{\homterms(\psi_s)} \leq h$.
  Again by~\Cref{claim:H-properties},  
  $\card{\homterms(\bigvee_{j=1}^k \psi_j)} \leq
  \card{H_{i+1}} \cdot h \leq h^{i+2}$.
  \item[Bounds on $\maxvars$.] Follows from the induction
  hypothesis and~\Cref{lemma:presburger-one-round}.
  \item[Bound on $\norminf{\homterms(\cdot)}$.] We have 
    \begin{align*}
      \norminf{\homterms(\psi_s)} &\leq 2\max(2,\norminf{\homterms(\psi_r')})^2 
        &\text{by~\Cref{lemma:presburger-one-round}}\\
      &\leq 2(a^{2^{i+1}-1})^2
        &\text{by the induction hypothesis (recall $a \geq 2$)}\\ 
      &\leq 2 \cdot a^{2^{i+2}-2}\\ 
      &\leq a^{2^{i+2}-1}
    \end{align*}
  \item[Bound on $\fmod$.] We have
    \begin{align*}
      \fmod(\psi_s) &\leq\max(2,\norminf{\homterms(\psi_r')}) \cdot  \fmod(\psi_r')
        &\text{by~\Cref{lemma:presburger-one-round}}\\
      &\leq a^{2^{i+1}-1} \cdot (a^{2^{i+2}-i}m)
        &\text{by the induction hypothesis}\\
      &\leq (2^{2^{i+1}+2^{i+2}-(i+1)}) \cdot m\\
      &\leq a^{2^{i+3}-(i+1)} m.
    \end{align*}
    Moreover, 
    \begin{align*}
      &\phantom{\leq{}} \textstyle\fmod(\bigvee_{j=1}^k\psi_j)\\
      &\leq{} \lcm_{(S',m') \in H_{i+1}}(m')
        &\text{by the definition of $H_{i+1}$}\\
      &\leq{} \lcm_{(S',m') \in H_{i+1}}(m'' \cdot \widetilde{a})
        &\hspace{-20pt}\text{where $m' \divides m'' \cdot \widetilde{a}$ for some
      $(S'',m'') \in H_i$}\\&&\text{with $\widetilde{a}$ coefficient
      of $x_i$ in a term of $S''$}\\
      &\leq{} \textstyle\fmod(\bigvee_{j=1}^\ell \psi_j') \cdot \norminf{\homterms{(\bigvee_{j=1}^\ell \psi_j')}}^{\card{H_{i+1}}}\\ 
      &\leq{} (a^{h^{2i+2}} m) \cdot (a^{2^{i+1}-1})^{h^{i+1}}
        &\text{by ind.~hyp.~and $\card{H_{i+1}} \leq h^{i+1}$}\\
      &\leq{} m \cdot a^{h^{2i+2} + h^{i+1}2^{i+1}}
        &\text{since $a \geq 2$}\\
      &\leq{} m \cdot a^{h^{2i+3}}
          &\text{since $h \geq 2$.}
    \end{align*}

  \item[Bound on $\norminf{\linterms(\cdot)}$.]
    \begin{align*}
      \norminf{\linterms(\psi_s)}
      &\leq{}
      \norminf{\homterms(\psi_r')}
      (\norminf{\homterms(\psi_r')} \cdot \fmod(\psi_r') + 2 \cdot \norminf{\linterms(\psi_r')})
      &\text{by \Cref{lemma:presburger-one-round}}\\
      &\leq{}
      (a^{2^{i+1}-1}) \cdot ((a^{2^{i+1}-1}) \cdot (a^{2^{i+2}-i}m) + 2 \cdot a^{2^{2i+1}}(c+m))
      &\text{by ind.~hyp.}\\
      &\leq{}
      a^{2^{i+3}}m + a^{2^{2i+2}}(c+m)
      &\text{since $a \geq 2$}\\
      &\leq{}
      a^{2^{2i+3}}(c+m).
    \end{align*}
  \item[Bound on $\boolnum$.]
    \begin{align*}
      \boolnum(\psi_j) 
      &\leq \boolnum(\psi_r') + 2 \cdot \maxvars(\psi_r') + 1
        &\text{by \Cref{lemma:presburger-one-round}}\\
      &\leq (b + i \cdot (2 \cdot V_i + 1)) + 2 \cdot V_{i} + 1 
        &\text{induction hypothesis}\\
      &\leq b + (i+1)(2 \cdot V_{i+1}+1)
        &\text{since $V_i \leq V_{i+1}$.}
    \end{align*}
    The bound on $\boolnum(\bigvee_{j=1}^k \psi_j)$ follows.
  \item[Bound on $k$.]
    Note that $\card{\linterms(\psi_r')} \leq \card{\homterms(\psi_r')} \cdot (2 \cdot \norminf{\linterms(\psi_r')}+1)$. Therefore,
    \begin{align*}
      k 
      &\leq \sum_{r=1}^\ell 
      \left(
        4 \cdot \card{\linterms(\psi_r')}
        \cdot
        (\norminf{\homterms(\psi_r')} 
        \cdot 
        \fmod(\psi_r'))^{2\cdot\maxvars(\psi_r')+1}
      \right) & \text{by def.~and~\Cref{lemma:presburger-one-round}}\\
      &\leq \ell \cdot
        4 \cdot q
        \cdot 
        (
          a^{2^{i+1}-1}
          \cdot
          a^{2^{i+2}-i}m
        )^{2 V_i +1}
      & \text{by ind.~hyp.}\\
      &\leq 
      4 \cdot \left( 
        q^i \cdot (a^{2^{i+2}}m)^{i(2V_i+1)}
      \right) 
      \cdot
      \left(
        q
        \cdot 
        (
          a^{2^{i+3}-(i+1)}
          m
        )^{2 V_i +1}
      \right)
      &\hspace{-1cm}\text{by def.~of~$\ell$}\\
      &\leq 
        q^{i+1} \cdot (a^{2^{i+3}}m)^{(i+1)(2 \cdot V_{i+1}+1)}.
      &\text{recall $V_i \leq V_{i+1}$.}
    \end{align*}
\end{description}
As in the case of~\Cref{lemma:presburger-one-round}, 
the procedure runs in time polynomial in the bit length of the output (again, as stressed in the proof of \Cref{lemma:presburger-one-round}, this is disregarding the ``automatic'' term normalizations discussed in~\Cref{section:preliminaries}, which might decrease the size of the formula).
\end{proof}

Before giving the proof of~\Cref{prop:octa-alt}, we study the elimination of a single quantifier and a block of quantifiers from an integer octagon arithmetic formula.

\begin{lemma}
  \label{lem:integer-octa-one-step}
  Let $\Gamma = \{(\vec x, \psi_1),\dots,(\vec x,\psi_k)\}$ be the output of $\funcPA(x,\vec x, \phi(\vec x, \vec z))$, 
  where $\phi$ is a q.f.~formula from integer octagon arithmetic.
  Then, the following parameter table holds, where $i \in [1,k]$:
  \begin{center}
      \renewcommand\arraystretch{1.2}
      \setlength{\tabcolsep}{3pt}
      \begin{tabular}{|g|c|c|c|c|c|c|c|}
          \hline
          \rowcolor{light-gray}
          & $\card\homterms$ 
          & $\card\linterms$
          & $\maxvars$ 
          & $\norminf{\homterms(\cdot)}$ 
          & $\norminf{\linterms(\cdot)}$ 
          & $\fmod$
          & $\boolnum$\\
          \hline
          \hline
          $\phi$
          & $h \geq 2$
          & $q$
          & $2$ 
          & $1$ 
          & $c$ 
          & $m$
          & $b$
          \\
          \hline
          \hline
          $\psi_i$
          & $h$
          & $q$
          & $2$ 
          & $1$
          & $2 \cdot c + m$
          & $m$
          & $b+1$
          \\ 
          \hline
          $\bigvee_{j=1}^k\psi_j$
          & $\min(h^2,4 \cdot \card{(\vec z \cup \vec x \setminus \{x\})}^2)$
          & \cellcolor{light-gray}
          & \ditto
          & \ditto
          & \ditto
          & \ditto
          & $k \cdot (b + 2)$\\ 
      \hline
      \end{tabular}
  \end{center}
  where $k \leq q \cdot (2m+1)$.
  The runtime of the procedure is in~$\poly{\len{\phi},m,\card{\vec x}}$.
\end{lemma}

\begin{proof}
  First of all, note that if $\phi$ is a q.f.~formula
  from integer octagon arithmetic, then each $\psi_i$
  also belongs to integer octagon arithmetic. Indeed,
  in this case, the
  formulae $\phi\substitute{\frac{t+k}{a}}{x} \land (a
  \divides t + k)$ in line~\ref{pres:line-sub} are of the
  form $\phi\substitute{\frac{t+k}{1}}{x} \land 1 \divides
  t + k$ and thus simplify to $\phi\substitute{t+k}{x}
  \land \true$. Here, $t = \pm y + c$ or $t = c$, and therefore
  the substitution $\phi\substitute{t+k}{x}$
  only produces atomic formulae from
  integer octagon arithmetic, plus formulae of the form
  $\pm 2 \cdot y < d$, which are normalized into
  $y \leq \floor{\frac{d-1}{2}}$ or
  $y \geq \ceil{- \frac{d-1}{2}}$, respectively.
  Therefore, all
  divisibility constraints in $\phi\substitute{t+k}{x}$
  are already simple, and thus the function $\simplify$ simply
  returns its input in a singleton
  set, i.e., $\simplify(\phi\substitute{t+k}{x} \land
  \true) = \{\phi\substitute{t+k}{x} \land \true\}$.
  
  The properties described above imply
  the bounds on $\card{\homterms}$, $\card{\linterms}$, $\maxvars(\cdot)$,
  $\norminf{\homterms(\cdot)}$, $\fmod(\cdot)$ and $\boolnum(\cdot)$ in
  the table. 
  To conclude the proof it suffices
  to establish the bounds on $\norminf{\linterms(\cdot)}$
  and~$k$. 
  The bound on $\norminf{\linterms(\psi_i)}$ 
  follows from~\Cref{lemma:presburger-one-round} 
  by substituting $a$ with $1$.
  To bound $k$, it suffices to count the number of
  formulae in the set $\Gamma$ (since \simplify does not
  change the number of formulae, as discussed above).
  It is
  \begin{align*}
    k &\leq \card{T} \cdot (2 \cdot\fmod(\phi)+1)
      &\text{from line~\ref{pres:line-sub} of~\funcPA}\\ 
      & \leq q \cdot (2m+1).
  \end{align*} 
  The running time of the procedure follows as in~\Cref{lemma:presburger-one-round}.
\end{proof}

We analyze the bounds on the set of formulae obtained by iteratively calling the function~\funcPA on a formula of integer octagon arithmetic. 

\begin{restatable}{proposition}{PropIntegerOcta}
  \label{prop:integer-octa}
  Suppose that $\phi(\vec x,\vec z)$ in~\Cref{prop:qePA-multiV} is a q.f.~formula from integer octagon arithmetic. Then, the parameter table of~\Cref{prop:qePA-multiV} can be refined as follows: 
  \begin{center}
    \renewcommand\arraystretch{1.2}
    \setlength{\tabcolsep}{3pt}
    \begin{tabular}{|g|c|c|c|c|c|c|c|}
        \hline
        \rowcolor{light-gray}
        & $\card\homterms$ 
        & $\card\linterms$ 
        & $\maxvars$ 
        & $\norminf{\homterms(\cdot)}$ 
        & $\norminf{\linterms(\cdot)}$ 
        & $\fmod$
        & $\boolnum$\\
        \hline
        \hline
        $\phi$
        & $h \geq 2$
        & $q$ 
        & $2$ 
        & $1$ 
        & $c \geq 2$ 
        & $m \geq 2$
        & $b$
        \\
        \hline
        \hline
        $\psi_i$
        & $h$
        & $q$ 
        & $2$ 
        & $1$
        & $2^n c + (2^n-1) \cdot m$
        & $m$
        & $b+n$
        \\ 
        \hline
        $\bigvee_{j=1}^k\psi_j$
        & $\min(h^{n+1},4 \cdot \card{\vec z}^2)$
        & \cellcolor{light-gray}
        & \ditto
        & \ditto
        & \ditto
        & $m$
        & $k (\ditto\, + 1)$\\ 
    \hline
    \end{tabular}
\end{center}
Moreover, $k \leq q^n \cdot (2\cdot m+1)^n$.
The runtime of the procedure is in~$(\len{\phi} \cdot m)^{\poly{n}}$.
\end{restatable}

\begin{proof}
  The proof is a straightforward adaptation of the
  proof of~\Cref{prop:qePA-multiV}. In fact, note that
  the bounds on $\card{\homterms}$, $\card{\linterms}$, $\maxvars$,
  $\norminf{\homterms(\cdot)}$, $\fmod$ and $\boolnum$ follow
  directly from~\Cref{prop:qePA-multiV}
  and~\Cref{lem:integer-octa-one-step}. For the bound
  on $\norminf{\linterms(\cdot)}$ it suffices to show that
  this parameter is bounded from above by $L_i \coloneqq 2^i c +
  (2^i-1) m$ when $i$~variables are eliminated.
  For $i = 1$,~\Cref{lem:integer-octa-one-step} yields
  a bound of $2 \cdot c + m = L_1$, as required.
  Inductively, assume $\norminf{\linterms(\cdot)}$ to be
  bounded by $L_{i-1}$ after the elimination of $i-1$ variables.
  Then,
  by~\Cref{lem:integer-octa-one-step}, after the
  elimination of the $i$-th variable
  $\norminf{\linterms(\cdot)}$ is bounded by 
  \[
    2 \cdot L_{i-1} + m 
    \ \leq \ 2 \cdot (2^{i-1} \cdot c + (2^{i-1}-1) \cdot m) 
    \ \leq \ 2^i \cdot c + (2^{i}-2) \cdot m + m
    \ \leq \ L_i.
  \]
  Finally, by the definition of $k$ and
  from~\Cref{lem:integer-octa-one-step}, we get 
  \begin{align*} 
    k &\leq \Big(
      \max_{1 \le i \le k}(\card{\linterms(\psi_i)}) \cdot ( 2 \cdot \max_{1 \le i \le k}(\fmod(\psi_i))+1) 
    \Big)^n
    \leq q^n \cdot (2 m + 1)^n.
  \end{align*}
  The running time of the procedure follows as in~\Cref{lemma:presburger-one-round}.
\end{proof}

We now turn to proving Proposition \ref{prop:octa-alt} from the body, which we now recall:

\PropOctaAlt* 

\begin{proof}
  This proposition follows from a straightforward
  induction on the number of quantifier blocks, by relying on the bounds
  in~\Cref{prop:integer-octa}. 
  In particular, the bounds on $\card\homterms$, 
  $\maxvars$, $\norminf{\homterms(\cdot)}$ and $\fmod$ are trivial to establish. 
  We focus on bounding $\norminf{\linterms(\Psi)}$ 
  and $\boolnum$. 
  Below, let $\Psi_r$ be the formula obtained from $\Phi$ after removing $r$ quantifier blocks.

  To prove the bound on $\norminf{\linterms(\Psi)}$, 
  it suffices to show that $\norminf{\linterms(\Psi_r)} \leq 2^{r
  \cdot n}c + (2^{r \cdot n}-1) \cdot m$, which
  is a sharper bound than the one given in the table,
  but one that can be easily maintained in the induction
  on $r$. 
  In the base case $r = 1$, this bound is given by \Cref{prop:integer-octa}.
  In the induction step ($r \geq 2$), 
  we have: 
  \begin{align*}
    \norminf{\linterms(\Psi_{r})} 
    &\leq 
    2^n \cdot \norminf{\linterms(\Psi_{r-1})} + 
    (2^n-1) \cdot \fmod(\Psi_{r-1})
    &\text{by \Cref{prop:integer-octa}}\\
    &\leq 
    2^n \cdot (2^{(r-1) \cdot n}c + (2^{(r-1) \cdot n}-1) \cdot m) + 
    (2^n-1) \cdot m
    &\text{by induction hypothesis}\\ 
    &= 
    2^{r \cdot n}c + (2^{r \cdot n}-1) \cdot m.
  \end{align*}

  For the bound on $\boolnum$, 
  note first that
  $\card{\homterms(\Psi_r) \leq 4 \cdot (\card {\vec z} + (\ell - r) \cdot n)^2}$
  and
  $\card{\linterms(\Psi_r)} \leq \card{\homterms(\Psi_r)} \cdot (2 \cdot \norminf{\linterms(\Psi_r)}+1)$. We prove by induction that 
  \begin{align*}
    \boolnum(\Psi_r) \leq  K^{r} \cdot b + (K^{r+1}-2) \cdot (n + 1),
  \end{align*}
  where $K \coloneqq 4^n(\card{\vec z} + \ell \cdot n)^{2n} \cdot (2m+1)^n \cdot (2^{\ell \cdot n}c + (2^{\ell \cdot n}-1) \cdot m)^n$.
  Note that then $\boolnum(\Psi_\ell) \leq 2 \cdot K^{\ell+1}(n+1) \cdot b$, 
  and one can verify $2 \cdot K^{\ell+1}(n+1) \leq (m + c + \card{\vec z})^{5 \cdot n^2 (\ell+1)^2}$ to obtain the bound in the table.
  In the base case ($r = 1$), the bound on $\boolnum(\Psi_r)$
  follows from~\Cref{prop:integer-octa}:
  \begin{align*}
    \boolnum(\Psi_1) 
      & \leq (4 \cdot (\card{\vec z} + (\ell -1) \cdot n)^2 \cdot (2 \cdot c + 1))^n \cdot (2 \cdot m + 1)^n \cdot (b + n + 1)\\
      & \leq K \cdot (b + n + 1) \leq K \cdot b + (K^2 - 2) \cdot (n + 1).
  \end{align*}
  
  For the induction step ($r \geq 2$), we have:
  \begin{align*}
    \boolnum(\Psi_r)
    &\leq{} 
    (\card{\linterms(\Psi_{r-1})} \cdot (2 \fmod(\Psi_{r-1})+1))^n 
    \cdot (\boolnum(\Psi_{r-1})+n+1)
    &\hspace{-5cm}\text{by \Cref{prop:integer-octa}}\\ 
    &\leq{} 
    \big(\card{\homterms(\Psi_{r-1})} \cdot (2 \cdot \norminf{\linterms(\Psi_{r-1})}+1) \cdot (2 \cdot \fmod(\Psi_{r-1})+1)\big)^n\\
    & \qquad \cdot (\boolnum(\Psi_{r-1})+n+1)
    &\hspace{-5cm}\text{bound on $\card{\linterms(\Psi_r)}$}\\
    &\leq{} 
    \big((4 \cdot (n \cdot \ell + \card{\vec z})^2) \cdot (2 \cdot (2^{(r-1)n}c + (2^{(r-1)n}-1)m)+1) \cdot (2 m+1)\big)^n\\
    & \qquad \cdot (\boolnum(\Psi_{r-1})+n+1)
    &\hspace{-5cm}\text{by previously established bounds}\\
    &\leq{} 
    K
    \cdot (\boolnum(\Psi_{r-1})+n+1)
    &\hspace{-1cm}\text{by def.~of $K$}\\
    &\leq{} 
    K
    \cdot (K^{r-1} \cdot b + (K^r-2) \cdot (n + 1)+n+1)
    &\hspace{-1cm}\text{by ind.~hyp.}\\
    &\leq{} K^{r} \cdot b + (K^{r+1}-2) \cdot (n + 1)
    &\hspace{-1cm}\text{since $K \geq 2$.}
  \end{align*}
The running time in the statement follows from iterating $\ell$ times
the bound on the running time of the procedure from~\Cref{prop:integer-octa}, noting that if that procedure runs in time $T$, then its output formula has size at most $T$. 
One last detail: the bound we provided for $\boolnum$ 
depends polynomially on $c$, which, however, does not appear in the running time of the procedure. This is because this dependency is an artifact of our analysis that avoids tracking $\card{\linterms}$ explicitly, 
and instead bounds it with $\card{\homterms} \cdot (2 \cdot \norminf{\linterms}+1)$.
\end{proof}

\subsection{Towards a proof of Theorem~\ref{theo:pow-3exp}: \funcSem}

The following proposition considers iterated calls to \funcSem.

\begin{restatable}{proposition}{PropSemRounds}
  \label{prop:Sem-rounds}
  Consider $\Psi(\vec y, \vec z)$ in $\QFPC$ in which variables $\vec y = (y_1,\dots,y_n)$ always occur in powers, and $n \geq 1$. 
  Let~${\cT = \QFPC}$, and run~\Cref{algo:top-level} from line~\ref{top:init-q-and-d} with $Q = \{(\vec y, \Psi)\}$. 
  At the end of the {\rm\textbf{while}} loop of line~\ref{top:inner-loop}, $D = \{\theta_1,\dots,\theta_k\}$ satisfies the parameter table
\begin{center}
  \renewcommand\arraystretch{1.3}
  \setlength{\tabcolsep}{4pt}
  \begin{tabular}{|g|c|c|c|c|c|c|}
      \hline
      \rowcolor{light-gray}
      & $\card\homterms$ 
      & $\maxvars$ 
      & $\norminf{\linterms(\cdot)}$ 
      & $\fmod$
      & $\boolnum$\\
      \hline
      \hline
      $\Psi$
      & $h$
      & $v \geq 2$ 
      & $c \geq 4$ 
      & $m$
      & $b$
      \\
      \hline
      \hline
      $\bigvee_{j=1}^k\theta_j$
      & $h \cdot 2^{20 \cdot v^2} \cdot \log(c)^{v+2}  + n^2$
      & $v$ 
      & $c^{2^{2(v+2)}}$
      & $m$
      & $k \cdot b'$\\ 
      \hline
  \end{tabular}
\end{center}
where 
$\boolnum(\theta_j) < b' \coloneqq b + h \cdot 2^{26 \cdot v^2} \cdot \log(c)^{v+2} \cdot n^3$, \ \ 
$k \leq 2 \uparrow ({h \cdot \log(c)^{30 \cdot v^2} \cdot n^3})$, and the number of quantifiers added to~$\Pi'$ is at most $h \cdot 2^{6 \cdot (v+1)^2} \cdot \log(c)^{v+1}$.
The runtime of this procedure is in~$\len{\Psi} \uparrow ({\poly{h,n} \cdot \log(c)^{\poly{v}}})$.
\end{restatable}

Above, $\uparrow$ is the exponentiation operator $(a \uparrow b) = a^b$, as in Knuth's up-arrow notation, and it is right-associative.

\begin{proof}
  By~\Cref{l:sem-no-accidental-linear}, \funcPA is
  never called in this case, hence the bounds follow by
  analyzing iterated calls to \funcSem. Below, let
  $\Theta \coloneqq \bigvee_{j=1}^k \theta_j$.

  The bounds on $\maxvars(\Theta)$ and $\fmod(\Theta)$
  follow directly from~\Cref{lemma:sem-one}. So, let us
  start by bounding $\norminf{\linterms(\Theta)}$.
  First, let us  once more note a key point of
  \funcSem: all inequalities that are generated by the
  procedure outside substitutions are either from
  $\powcomp$, or they only involve variables that are
  not in $\vec y$ (i.e., the introduced variables
  $w_\sigma$ or variables from $\vec z$). Because of
  the definition of $I$ in line~\ref{sem:I}, these
  inequalities are disregarded in subsequent rounds
  of~\funcSem. Hence, as already stressed in the proof
  of~\Cref{lemma:sem-one} when deriving the bound on $\norminf{\linterms(\cdot)}$ for one round of~\funcSem, to bound
  $\norminf{\linterms(\Theta)}$ it suffices to look at
  inequalities computed via substitutions. Suppose
  that, during the run of the procedure, we reach
  line~\ref{sem:A} of~\funcSem, with respect to some $x
  \in \vec y$ (line~\ref{sem:outer-loop}), define the
  set $A$, and later compute all the relevant
  substitutions for all $\alpha \in A$, according to
  lines~\ref{sem:case-1} to~\ref{sem:case-8}. Let
  $\alpha'$ be one of the inequalities resulting from a
  substitution of some $\alpha$. Then, $\alpha'$ has
  strictly fewer variables from $\vec y$ than $\alpha$.
  Indeed, in all substitutions of
  lines~\ref{sem:case-1} to~\ref{sem:case-8}, $\pow{x}$
  is removed (recall that all variables from $\vec y$ only
  appear in powers), and it is replaced with
  $\lambda(\sigma)$ (and later with $\pow{w_{\sigma}}$
  or $0$), a constant $2^j$, or a term $2^j \cdot
  \pow{v}$ where $v \in \vec y$ is another variable
  appearing in $\alpha$. In the next call to the
  procedure, inequalities obtained form $\alpha'$ via
  substitution will lose at least one other variable from
  $\vec y$, and so on (this is part of our termination
  argument for~\Cref{algo:top-level}). However,
  $\alpha$ only has at most $\maxvars(\phi) \leq v$
  many exponentiated variables and $\maxvars$ does not
  change during the procedure (see bound
  on~\Cref{lemma:sem-one}), hence iterated
  substitutions of this form can be done only $v$ many
  times. It then suffices to iterate $v$ times the
  function $c \mapsto 2^{12} \cdot v^2 \cdot c^3$ corresponding
  to the bound on $\norminf{\linterms(\cdot)}$ obtained
  in~\Cref{lemma:sem-one} for one call to~\funcSem, in
  order to compute an upper bound
  to~$\norminf{\linterms(\Theta)}$. After $i$ iterations
  we obtain a bound of $(2^{12} \cdot v^2)^{3^i-1} \cdot c^{3^i}$, as
  indeed, 
  \[ 
    (2^{12} \cdot v^2)^{3^i-1} \cdot c^{3^i} \ \mapsto \ (2^{12} \cdot v^2)((2^{12} \cdot v^2)^{3^i-1} \cdot c^{3^i})^{3} \ \leq \ (2^{12} \cdot v^2)^{3^{i+1}-1}c^{3^{i+1}},
  \]
  and therefore we conclude that
  $\norminf{\linterms(\Theta)} \leq
  (2^{12} \cdot v^2)^{3^v-1}c^{3^v} \leq c^{2^{2(v+2)}}$.

  We now move to $\card{\homterms(\Theta)}$. The
  arguments are similar to the ones used to
  bound~$\norminf{\linterms(\Theta)}$. For brevity, we
  rely on the notion of \emph{substitute} of a
  homogeneous term introduced in the proof
  of~\Cref{lemma:sem-one}. In that lemma, we have
  already counted the number of substitutes for one
  iteration of~\funcSem, and conclude that each
  homogeneous term~$\rho$ generates $2^6 \cdot v^3 \cdot \log(c)$ many substitutes. As discussed in the case
  of~$\norminf{\linterms(\Theta)}$, these substitutes
  have fewer variables from~$\vec y$ than $\rho$.
  Since~\funcSem does not increase the $\maxvars$ of the
  considered formulae, iterating~\funcSem never
  generates more than $(2^6 \cdot v^3 \cdot
  \log(\norminf{\linterms(\Theta)}))^v$ many
  substitutes per homogeneous term
  in~$\homterms(\phi)$. We then have a bound on the
  number of substitutes overall, meaning the ones that appear at some point in the procedure, of: 
  \begin{align*}
    \sum_{i=0}^v \big(h \cdot (2^6 \cdot v^3 \cdot \log(\norminf{\linterms(\Theta)}))^i\big) 
    & \leq h \cdot (2^6 \cdot v^3 \cdot \log(\norminf{\linterms(\Theta)}))^{v+1}\\
    &\hspace{-1cm} \leq h \cdot (2^6 \cdot v^3 \cdot 2^{2(v+2)} \cdot \log(c))^{v+1}
    \leq h \cdot 2^{6 \cdot (v+1)^2} \cdot \log(c)^{v+1}.
  \end{align*}
  Note that this bound also includes the original homogeneous terms form $\homterms(\phi)$, since the above sum starts at $i=0$.
  To conclude our analysis on
  $\card{\homterms(\Theta)}$, it suffices now to bound
  the number of homogeneous terms generated by calls
  to~\funcSem, outside of substitutions. Let us fix a
  homogeneous term $\rho$ of an inequality $\rho+c'<0$ (with
  $c' \in \Zed$) that at some point during the run
  of~\Cref{algo:top-level} occurs in the set~$I$ of
  line~\ref{sem:I}. Hence, $\rho$ is either in
  $\homterms(\phi)$ or it is one of the substitutes
  considered above. Associated to $\rho$ there is a single
  term $\sigma$, homogeneous and including all variables that
  are not from $\vec y$. We count the number of
  inequalities added in lines~\ref{sem:case-1}
  to~\ref{sem:case-8} and
  lines~\ref{semenov:guard-for-lambda}
  and~\ref{semenov:lambda-sub} for that specific $\rho$
  its $\sigma$. We already performed this computation in~\Cref{lemma:sem-one}, the only difference now being that we need to set $G \coloneqq 13 \cdot \log(v \cdot \norminf{\linterms(\Theta)})$. 
  According to Equation~\eqref{lemma7:counting-hom} in~\Cref{lemma:sem-one}, 
  for every $\rho$ (and its $\sigma$) the procedure adds 
  \begin{align*}
             & 2 \cdot v^2\cdot G + 3 \cdot v^2 + 6v + 6 \\ 
    {}\leq{} & 26 \cdot v^3 \cdot 2^{2(v+2)} \cdot \log(c) + 3 \cdot v^2 + 6v + 6 
    \leq 2^{5(v+2)} \log(c)
  \end{align*}
  many homogeneous terms. As argued in the case of~$\norminf{\linterms(\Theta)}$, 
these homogeneous terms do not play a role in subsequent invocation of~\funcSem, 
as they are all either from $\powcomp$, or they only feature variables not in $\vec y$. Hence, the overall number of homogeneous terms in $\Theta$ can be found by multiplying the above bound by the number of terms $\rho$ we previously computed, and adding 
  the remaining $\card{\vec y}^2$ homogeneous terms from line~\ref{sem:union-gamma}, which are the same across all iterations of \funcSem, and thus needs to be counted only once overall:
  \begin{align*}
    \card{\homterms(\Theta)} &\leq (h \cdot 2^{6 \cdot (v+1)^2} \cdot \log(c)^{v+1}) \cdot (2^{5(v+2)}\log(c)) + \card{\vec y}^2\\
    & \leq h \cdot 2^{20 \cdot v^2} \cdot \log(c)^{v+2}  + \card{\vec y}^2
  \end{align*} 

  We now move to the computation of $\boolnum(\Theta)$, by computing first the bound for $\boolnum(\theta_i)$ with $i \in [1,k]$. We know from~\Cref{l:correct-semenov-nice-var} that, 
after a call to~\funcSem, each formula in the returned set has a variable from $\vec y$ that only occurs in inequalities and divisibility constraints from \powcomp. 
  This implies that a bound on $\boolnum(\theta_i)$ can be obtained by iterating, $\card{\vec y}$ times, the bound on $\boolnum(\theta_i)$ found in~\Cref{lemma:sem-one}, and taking into account the number of homogeneous terms computed above. Notice that this bound is overly pessimistic, as we are using $\card{\homterms(\Theta)}$ instead of $\card{\homterms(\theta_i)}$. But it is good enough to achieve our claimed complexity bounds. We get: 
  \begin{align*}
    \boolnum(\theta_i) &\leq b + \card{\vec y} \cdot (\card{\homterms(\theta_i)} \cdot (v+15) + \card{\vec y})\\
    &\leq b + \card{\vec y} \cdot ((h \cdot 2^{20 \cdot v^2} \cdot \log(c)^{v+2}  + \card{\vec y}^2) \cdot (v+15) + \card{\vec y})\\ 
    &\leq b + h \cdot 2^{20 \cdot v^2+2} \cdot \log(c)^{v+2} \cdot \card{\vec y}^3 \cdot (v+15)\\ 
    &\leq b + h \cdot 2^{26\cdot v^2} \cdot \log(c)^{v+2} \cdot \card{\vec y}^3 -1. 
  \end{align*}
  Then, $\boolnum(\Theta) \leq \sum_{i=1}^k (\boolnum(\theta_i)+1)$, as usual, and so 
  $\boolnum(\Theta) \leq k \cdot (b + h \cdot 2^{26 \cdot v^2} \cdot \log(c)^{v+2} \cdot \card{\vec y}^3)$.

  We now estimate $k$. The situation is similar to the case of $\boolnum$. Note that \funcSem returns a set in which each formula has a variable 
from $\vec y$ that only occurs in inequalities and divisibility constraints from {\powcomp}. Thus it suffices to iterate $\card{y}$  times the bound on $k$ found in~\Cref{lemma:sem-one}, taking into account the growth of $\card{\homterms(\Theta)}$ and $\norminf{\linterms(\Theta)}$. We get: 
  \begin{align*}
    k &\leq \Big((v+1)^{10 \cdot \card{\homterms(\Theta)}} \cdot \log(\norminf{\linterms(\Theta)})^{\card{\homterms(\Theta)}} \cdot \card{\vec y}\Big)^{\card y}\\
    &\leq \Big((v+1)^{10 \cdot (h \cdot 2^{20 \cdot v^2} \cdot \log(c)^{v+2}  + \card{\vec y}^2)} \cdot \log(c^{2^{2(v+2)}})^{(h \cdot 2^{20 \cdot v^2} \cdot \log(c)^{v+2}  + \card{\vec y}^2)} \cdot \card{\vec y}\Big)^{\card{\vec y}}\\
    &\leq (v+1)^{11 \cdot (h \cdot 2^{20 \cdot v^2+1} \cdot \log(c)^{v+2}  \cdot \card{\vec y}^3)} \cdot \log(c^{2^{2(v+2)}})^{(h \cdot 2^{20 \cdot v^2} \cdot \log(c)^{v+2}  \cdot \card{\vec y}^3)}\\
    &\leq 2^{Q},
  \end{align*}
  where $Q \coloneqq {(h \cdot 2^{20 \cdot v^2+1} \cdot \log(c)^{v+2}  \cdot \card{\vec y}^3)(11 \cdot \log(v+1)+2(v+2)+\log \log (c))}$ is such that:
  \begin{align*}
    Q & \leq {(h \cdot 2^{20 \cdot v^2+1} \cdot \log(c)^{v+2}  \cdot \card{\vec y}^3)(9 \cdot (v+1) \cdot \log (c))} \\ 
    & \leq  {h \cdot 2^{21 \cdot v^2+6} \cdot \log(c)^{v+3}  \cdot \card{\vec y}^3}\\ 
    & \leq  {h \cdot \log(c)^{30 \cdot v^2}  \cdot \card{\vec y}^3}
    &\text{recall $v \geq 2$ and $c \geq 4$.}
  \end{align*}

  Lastly, we need to estimate the number of quantifiers added to $\Pi'$. This is straightforward. We have already computed the overall number of homogeneous terms for which terms of the form $\lambda(\sigma)$ need to be considered. These are the original homogeneous terms from $\phi$, together with all the substitutes, and there are  $h \cdot 2^{6 \cdot (v+1)^2} \cdot \log(c)^{v+1}$ many in total. This is also an upper bound to the number of quantifiers added to $\Pi'$.

  By inspection, the sizes of the output formulae are essentially a bound on the runtime of the procedure: again, as 
stressed in~\Cref{lemma:presburger-one-round}, here one has to disregard the ``automatic'' term normalizations discussed in~\Cref{section:preliminaries}, which might decrease the size of the formula. Then, the upper bounds in the parameter table, together with $\len{\Psi}$ to parse the initial formula, are enough to estimate the running time of the procedure.
\end{proof}

The following lemma combines multiple iterations of \funcPA and \funcSem.

\begin{restatable}{proposition}{PASEMOneBlock}
  \label{prop:PA-Sem-OneBlock}
  Consider $\Psi(\vec x, \vec y, \vec z)$ in $\QFPC$ in which variables $\vec x = (x_1,\dots,x_\vL)$ always occur linearly, variables $\vec y = (y_1,\dots,y_\vE)$ always occur in powers, and $\vL,\vE \geq 1$. Let~${\cT = \QFPC}$, and run~\Cref{algo:top-level} from line~\ref{top:init-q-and-d} with $Q = \{(\vec x\vec y, \Psi)\}$. 
  At the end of the {\rm\textbf{while}} loop of line~\ref{top:inner-loop}, $D = \{\theta_1,\dots,\theta_k\}$ satisfies the following table:
  \begin{center}
    \renewcommand\arraystretch{1.4}
    \setlength{\tabcolsep}{4pt}
    \begin{tabular}{|g|c|c|c|c|c|c|}
        \hline
        \rowcolor{light-gray}
        & $\card\homterms$ 
        & $\maxvars$ 
        & $\norminf{\linterms(\cdot)}$ 
        & $\fmod$
        & $\boolnum$\\
        \hline
        \hline
        $\Psi$
        & $h \geq 2$
        & $v \geq 2$ 
        & $c \geq 4$ 
        & $m \geq 4$
        & $b$
        \\
        \hline
        \hline
        $\bigvee_{j=1}^k\theta_j$
        &
          $h^{L+1} \cdot 2^{27 V^2 \vL} \cdot \log( c \cdot m )^{V+2}  + \vE^2$
        & $V$ 
        & $c^{2^{4 (\vL + V)}} \cdot m^{2^{4 V}}$
        & $c^{h^{2\vL+2}} m$
        & $k \cdot b'$\\ 
        \hline
    \end{tabular}
  \end{center}
  where $V \coloneqq \min(2^L v, L + E + \card{\vec z})$, $\boolnum(\theta_j) < b' \coloneqq b + h \cdot 2^{35\cdot V^2 \vL} \cdot \log(c \cdot m)^{V+2} \cdot \vE^3$ and $k \leq {2 \uparrow {(h \cdot 2^{60 \cdot (\vL+2) \cdot V^2}(\log(c \cdot m))^{30 \cdot V^2} \cdot \vE^3)}}$. The number of quantifiers added to 
  $\Pi'$ is at most ${h^{\vL+1} \cdot 2^{2^4 \cdot V^2 \vL} \cdot \log(c \cdot m)^{V+1}}$. 
  The procedure runs in~${{\len{\Psi}} \uparrow (h + \log(c \cdot m)) \uparrow {\poly{\vL,V}}}$.
\end{restatable}

\begin{proof}
  Let $\Theta = \bigvee_{j=1}^k \theta_j$.
  The proof follows by essentially chaining~\Cref{prop:qePA-multiV} and~\Cref{prop:Sem-rounds}. 
  Let us first add an assumption that, w.l.o.g., will simplify our reasoning: 
  we consider a slight variation 
  of~\Cref{algo:top-level} where $\text{pop}(Q)$ in line~\ref{top:inner-loop} always pops a pair $(\vec w,\psi)$ that respects the order of the if-then-else 
  chain in lines~\ref{top:move-to-d} to~\ref{top:call-semenov}. That is:
  \begin{itemize}
    \item if $Q$ contains pairs $(\vec w, \psi)$ with $\vec w$ empty, then $\text{pop}(Q)$ will return one of these pairs, 
    \item else if $Q$ contains pairs $(\vec w, \psi)$ in which there is $x \in \vec w$ not appearing in $\psi$, then $\text{pop}(Q)$ will return one of these pairs, 
    \item else if $Q$ contains pairs $(\vec w, \psi)$ in which some $x \in \vec w$ appears only linearly in $\psi$, then $\text{pop}(Q)$ will return one of these pairs, 
    \item else, it will return an arbitrary pair (if $Q$ is non-empty, otherwise the loop in line~\ref{top:inner-loop} terminates).
  \end{itemize}
  It is clear that this modification does not change the set $D$ we obtain at the end of the while loop of line~\ref{top:inner-loop}, and hence we do not lose generality.
  The simplification that this assumption brings in terms of the proof is that now, since in $\Psi$ all $\vec x$ occurs only linearly and all $\vec y$ occur 
only in powers, 
  there is an iteration of the while loop where $Q$ only contains formulae $\psi(\vec y, \vec z)$ where all variables from $\vec x$ have been eliminated thanks to \funcPA (or by relying on the trivial case where a variable from $\vec x$ does not occur in the formula anymore, i.e., the first of the four cases above). At that iteration,~\funcSem was never called.
  Let $Q' = \{(\vec w_1,\psi_1),\dots,(\vec w_\ell,\psi_\ell)\}$ be the content of $Q$ at that iteration.
  By applying~\Cref{prop:qePA-multiV} (and performing some simplifications), we get the following table of parameters for $Q'$, where $i \in [1,\ell]$.
  \begin{center}
    \renewcommand\arraystretch{1.2}
    \setlength{\tabcolsep}{3pt}
    \begin{tabular}{|g|c|c|c|c|c|c|}
        \hline
        \rowcolor{light-gray}
        & $\card\homterms$ 
        & $\maxvars$ 
        & $\norminf{\linterms(\cdot)}$ 
        & $\fmod$
        & $\boolnum$\\
        \hline
        \hline
        $\Psi$
        & $h \geq 2$
        & $v \geq 2$ 
        & $c \geq 4$ 
        & $m \geq 4$
        & $b$
        \\
        \hline
        \hline
        $\psi_i$
        & $h$
        & $V$ 
        & $c^{2^{2\vL+1}}(c + m)$
        & $c^{2^{\vL+2}}m$
        & $b+4 \cdot \vL \cdot V$
        \\ 
        \hline
        $\bigvee_{j=1}^\ell\psi_j$
        & $h^{\vL+1}$
        & \ditto
        & \ditto
        & $c^{h^{2\vL+2}} m$
        & $k (\ditto\, + 1)$\\ 
    \hline
    \end{tabular}
\end{center}
where $V = \min(2^L v, L + E + \card{\vec z})$ (as in the statement) and $\ell \leq h^{\vL} \cdot c^{2^{5\vL+5}v} \cdot m^{2^{2\vL+2}v}$. Since all variables from $\vec x$ are removed, starting from $Q'$ the function~\funcPA will never be called, but~\funcSem will be called instead (see~\Cref{l:sem-no-accidental-linear}). 

We can now rely on~\Cref{prop:Sem-rounds}, applied to each $\psi_i$, in order to conclude the proof. The growth of the parameters $\fmod(\Theta)$ and $\maxvars(\Theta)$ is easy to estimate: since \funcSem does not change them, they are as in the table above. Similarly, in order to compute $k$ and $\boolnum(\Theta)$ it suffices to apply~\Cref{prop:Sem-rounds} to each $\psi_i$, and multiply the max among the resulting bounds by $\ell$ (i.e.~the number of $\psi_i$). We get:
\allowdisplaybreaks
\begin{align*}
  k 
  & \leq (h^{\vL} \cdot c^{2^{5\vL+5}v} \cdot m^{2^{2\vL+2}v}) \cdot \max_{i=1}^\ell\big(2^{\card{\homterms(\psi_i)} \cdot \log(\norminf{\linterms(\psi_i)})^{30 \cdot (\maxvars(\psi_i))^2} \cdot \vE^3}\big)\\ 
  & \leq (h^{\vL} \cdot c^{2^{5\vL+5}v} \cdot m^{2^{2\vL+2}v}) \cdot 2^{h \cdot \log(c^{2^{2\vL+1}}(c + m))^{30 \cdot V^2} \cdot \vE^3}\\ 
  & \leq 2 \uparrow (h \cdot 2^{60 \cdot (\vL+2) \cdot V^2}(\log(c \cdot m))^{30 \cdot V^2} \cdot \vE^3).
\end{align*}

The reasoning is similar for $\boolnum(\Theta)$. It suffices to apply to $\psi_i$ the bounds from~\Cref{prop:Sem-rounds}, to obtain bounds for $\boolnum(\theta_j)$ ($j \in [1,k])$. Then, as usual, $\boolnum(\Theta) = k(1 + \max_{i=1}^j\boolnum(\theta_j))$. 
\begin{align*}
\boolnum(\theta_j) 
  &\leq \max_{i=1}^\ell(\boolnum(\psi_i) + \card{\homterms(\psi_i)} \cdot 2^{26\cdot (\maxvars(\psi_i))^2} \cdot \log(\norminf{\linterms{(\psi_i)}})^{\maxvars(\psi_i)+2} \cdot \vE^3 -1)\\
  &\leq b + 4 \cdot \vL \cdot V + h \cdot 2^{26\cdot V^2} \cdot \log( c^{2^{2L+1}}(c+m))^{V+2} \cdot \vE^3 -1\\
  &\leq b + h \cdot 2^{35\cdot V^2 \vL} \cdot \log(c \cdot m)^{V+2} \cdot \vE^3 -1
\end{align*}

What is left is to estimate $\norminf{\linterms(\Theta)}$, $\card{\homterms(\Theta)}$, and the number of quantifiers in $\Pi'$.
Below, for clarity, let us set $\Omega \coloneqq \bigvee_{i=1}^\ell \psi_i$.
For $\norminf{\linterms(\Theta)}$ it suffices to compose the results from the table above with the bound on~\Cref{prop:Sem-rounds}:
\begin{align*}
  \norminf{\linterms(\Theta)} 
  &\leq \norminf{\linterms(\Omega)}^{2^{2 \cdot (\maxvars(\Omega)+2)}}
  \leq \Big( c^{2^{2\vL+1}}(c + m) \Big)^{2^{2 \cdot (V+2)}}
  \leq c^{2^{4 \cdot (\vL + V)}} \cdot m^{2^{4 \cdot V}}
\end{align*}

We now move to $\card{\homterms(\Theta)}$. 
Again to simplify the proof, we fictitiously increase each $\psi_i$ with a tautology so that they all have the same set of inequalities. That is, instead of considering $\psi_i$, let us consider the equivalent formula $\psi_i \land (\gamma \lor \lnot \gamma)$ where $\gamma \coloneqq \bigwedge\linterms(\Omega)$. Notice that $\card{\homterms}(\psi_i) \leq h^{L+1}$, whereas $\maxvars(\psi_i)$, $\norminf{\linterms(\psi_i)}$ and $\fmod(\psi_i)$ are as before.
Applying~\Cref{prop:Sem-rounds} on these $\psi_i$ generates formulae that have more homogeneous terms that the original ones, but as we now made uniform the set of $\homterms(\psi_i)$ across all $\psi_i$, we can simply obtain a bound on $\card\homterms(\Theta)$ by applying the bounds in~\Cref{prop:Sem-rounds} directly to the bounds of $\Omega$:
\begin{align*}
  \card\homterms(\Theta) &\leq \card{\homterms(\Omega)} \cdot 2^{20 \cdot \maxvars(\Omega)^2} \cdot \log(\norminf{\linterms(\Omega)})^{\maxvars(\Omega)+1}  + \vE^2\\ 
  &\leq h^{L+1} \cdot 2^{20 V^2} \cdot \log( c^{2^{2\vL+1}}(c + m) )^{V+1}  + \vE^2\\ 
  &\leq h^{L+1} \cdot 2^{27 V^2 \vL} \cdot \log( c \cdot m )^{V+1}  + \vE^2.
\end{align*}
Lastly, the number of quantifiers added to $\Pi'$ is computed in a similar way, from the bound on~\Cref{prop:Sem-rounds}. We conclude that this number is bounded by:
\begin{align*}
  & \card{\homterms(\Omega)} \cdot 2^{6 \cdot (\maxvars(\Omega)+1)^2} \cdot \log(\norminf{\linterms(\Omega)})^{\maxvars(\Omega)+1}\\
  \leq{}& h^{\vL+1} \cdot 2^{6 \cdot (V+1)^2} \cdot \log(c^{2^{2\vL+1}}(c+m))^{V+1}\\
  \leq{}& h^{\vL+1} \cdot 2^{16 \cdot V^2 \vL} \cdot \log(c \cdot m)^{V+1}.
  &&\qedhere
\end{align*}
\end{proof}

\subsection{Proof of Theorem~\ref{theo:pow-3exp}}

\Cref{theo:pow-3exp} follows directly from~\Cref{prop:PA-Sem-All}. 

\PASEMALL* 

\begin{proof}
  Note that, since $\Phi$ does not have divisibility constraints and 
  each quantified variables appear only linearly or in powers, 
  the quantifier-free part of $\Phi$ is by definition in~$\Sem$.

  The proof of the statement is by induction on $\alt(\Phi)$, 
  We prove that after handling $K$ many quantifier blocks, the formula obtained in line~\ref{top:update-phi} of the algorithm --- below referred to as $\Psi_K$ --- satisfies the following bounds:
  \begin{center}
    \renewcommand\arraystretch{1.4}
    \setlength{\tabcolsep}{4pt}
    \begin{tabular}{|g|c|c|c|c|c|c|}
        \hline
        \rowcolor{light-gray}
        & $\card\homterms$ 
        & $\maxvars$ 
        & $\norminf{\linterms(\cdot)}$ 
        & $\fmod$
        & $\boolnum$
        \\
        \hline
        \hline
        $\Phi$
        & $h \geq 2$
        & $v \geq 2$ 
        & $c \geq 4$ 
        & \cellcolor{light-gray}
        & $b$
        \\
        \hline
        \hline
        $\Psi_K$
        &
          $\vH_K \coloneqq (E \cdot h \cdot \log c) 
          \uparrow {(2 \cdot v)^{2^4 \cdot L \cdot K^2}}$
        & $2^{K \cdot \vL} v$
        & $2^{\vH_K}$
        & $2^{\vH_K}$
        & $b \cdot 2^{\vH_K}$
        \\ 
        \hline
    \end{tabular}
  \end{center}
  and the number of quantified variables added to $\Pi'$ is at most $\vH_K$. Recall that all these variables are added to the next block, and they all only occur in powers in $\Psi_K$ (this increases the number ``$E$'' of powers occurring variables in the next block, while keeping the number of variables that only occur linearly bounded by~$L$). 
  Below, we write $E_K$ for the number of quantified variables occurring in power terms within the innermost block of $\Psi_K$. This quantity
  is bounded by $\vH_K+E$. 

  The induction is relatively straightforward, and simply relies on the bound obtained for a single quantifier block in~\Cref{prop:PA-Sem-OneBlock}. Note that the base case for $K = 1$ follows directly from that proposition. 
  In the induction step, suppose that the bounds above are satisfied for $\Psi_K$ with $K \geq 1$. We show that they are also satisfied for $\Psi_{K+1}$.

  \allowdisplaybreaks
  \begin{description}
    \item[bound on $\maxvars$:] $\maxvars(\Psi_{K+1}) \leq 2^L \maxvars(\Psi_K) \leq 2^L(2^{KL}v) = 2^{(K+1)L}v$.
    \item[bound on $\boolnum$:] See bounds on $k$ and $b'$ in~\Cref{prop:PA-Sem-OneBlock}. Then,
      \begin{align*}
        &\boolnum(\Psi_{K+1})\\
          \leq{}& 
            \Big( 
              2^{(\card{\homterms(\Psi_K)} \cdot 2^{60 (L+2) (2^{L} \cdot \maxvars(\Psi_K))^2} \log(\norminf{\linterms{(\Psi_K)}} \cdot \fmod(\Psi_K)))^{30 \cdot (2^L \maxvars(\Psi_K))^2} \cdot \vE_K^3}  
            \Big) \cdot{}\\
            & 
            \Big(\boolnum(\Psi_{K})+\homterms(\Psi_{K}) \cdot 
            2^{35(2^L \maxvars(\Psi_K))^2 L}
             \log(\norminf{\linterms{(\Psi_K)}} \cdot \fmod(\Psi_K))^{2^L \cdot \maxvars(\Psi_K)+2} \vE_K^3
            \Big)
          \\
          \leq{}& 
          2^{H_K \cdot 2^{60(L+2)(2^{(K+1)L}v)^2} (2 \cdot H_K)^{30(2^{(K+1)L}v)^2} \vE_k^3}
          \cdot\\
          &
          (b \cdot 2^{H_K} + H_K \cdot 2^{35(2^{(K+1)L}v)^2L} (2 \cdot H_K)^{2^{(K+1)L}v+2}\vE_K^3)\\
          \leq{}& 
          2^{2^{(60(L+2)+30)(2^{(K+1)L}v)^2} (H_K)^{30(2^{(K+1)L}v)^2+1} \vE_K^3}
          \cdot\\
          &
          b \cdot 2^{H_K + 35(2^{(K+1)L}v)^2L+2^{(K+1)L}v+1 + \log((H_K)^{2^{(K+1)L}v+3}) + \log(\vE_K^3)}\\ 
          \leq{}& 
          b \cdot 2^{P}
      \end{align*}
      where $P \coloneqq 2 \cdot 2^{(60(L+2)+30)(2^{(K+1)L}v)^2} (H_K)^{30(2^{(K+1)L}v)^2+1} E_K^3$ is further bounded as:
      \begin{align*}
        P 
        & = 2^{(60(L+2)+30)(2^{(K+1)L}v)^2+1} (H_K)^{30(2^{(K+1)L}v)^2+1} E_K^3\\
        & \leq (H_K)^{(60(L+2)+30)(2^{(K + 1)L}v)^2
        + 30(2^{(K+1)L}v)^2+7}
        &\text{note: $2H_K \geq E_K$}\\
        & \leq (H_K)^{61 \cdot 2^{3(K+1)L}v^2}\\
        & \leq \left( 
          (E \cdot h \cdot \log c)^{(2 \cdot v)^{2^4 \cdot L \cdot K^2}}
        \right)^{2^6 \cdot 2^{3(K+1)L}v^2}\\ 
        & \leq 
        (E \cdot h \cdot \log c)^{(2 \cdot v)^{2^4 \cdot L \cdot K^2 + 3(K+1)L+6}}\\
        & \leq 
        (E \cdot h \cdot \log c)^{(2 \cdot v)^{2^4 \cdot L \cdot (K+1)^2}} 
        = H_{K+1}
      \end{align*}
    \item[bound on $\card{\homterms}$:] 
      \begin{align*}
        \card{\homterms(\Psi_{K+1})} 
        &\leq \homterms(\Psi_{K})^{L+1}
        2^{27(2^{K\cdot L}v)^2 L} \log(\norminf{\linterms(\Psi_K)} \cdot \fmod(\Psi_K))^{2^{K \cdot L}v+2} + \vE_K^2\\
        &\leq (H_K)^{L+1}2^{27(2^{K\cdot L}v)^2 L}
        (2 H_K)^{2^{K \cdot L}v+2}
        + \vE_K^2\\
        &\leq 
         2^{29 \cdot L \cdot (2^{K \cdot L}v)^2}
         (H_K)^{3 \cdot 2^{K \cdot L}v} + E_K^2\\ 
        &\leq P \leq H_{K+1} 
        \hspace{5cm}\text{$P$ defined as above.}
      \end{align*}
    \item[bound on $\norminf{\linterms{(\cdot)}}$:]
      \begin{align*}
        \norminf{\linterms(\Psi_{K+1})} 
        &\leq \norminf{\linterms(\Psi_{K})}^{2^{4(L+2^{K \cdot L}v)}} \cdot \fmod(\Psi_{K})^{2^{4(2^{K \cdot L}v)}}\\
        &\leq 2^{2 \cdot H_K \cdot 2^{4(L+2^{K \cdot L}v)}}\\
        & \leq 2^P \leq 2^{H_{K+1}}.
      \end{align*}
    \item[bound on $\fmod$:]
      \begin{align*}
        \fmod(\Psi_{K+1}) 
        & \leq 
          \norminf{\linterms(\Psi_{K})}^{(\card{\homterms(\Psi_{K})})^{2\vL+2}} \fmod(\Psi_K)\\
        & \leq 
          2^{H_K \cdot (H_K)^{2\vL+2}} \cdot 2^{H_K}\\
        & \leq 2^{(H_K)^{2\vL+4}}\\
        &\leq 2^P \leq 2^{H_{K+1}}.
      \end{align*}
  \end{description}
  The running time follows as usual. Again the sizes of the output formulae are essentially a bound on the runtime of the procedure.
  Once more, as stressed in the proof of~\Cref{lemma:presburger-one-round}, one disregards the ``automatic'' term normalizations discussed in~\Cref{section:preliminaries}, which might decrease the size of the formula. Then, the upper bounds in the parameter table, together with $\len{\Phi}$ to parse the initial formula, are enough to estimate the running time of the procedure.
\end{proof}

%% file: appendix-differences.tex
\section{Summary of changes}

We have made the following changes compared to the previous
version (in conference proceedings) of this paper.

\begin{itemize}

\item
We have clarified the definitions of sets
$\linterms(\aformula)$ and $\homterms(\aformula)$.
These sets may contain terms with powers.
(The mnemonic for the choice of notation is that,
 \emph{if} no powers are involved, these sets
 will contain linear polynomials and homogeneous linear polynomials,
 respectively.)

\item
We have enhanced the wording of the Master procedure,
using ``push pair'' and ``bulk push'' instead of
``add pair'' and ``add'' to refer to operations on $Q$.

\item
We have corrected the pseudocode of function \funcLinearise.
The previous version contained a typo (incorrect dummy variable
in set comprehension) and omitted the inequality $\abs{x} \ge r'$.

\item
We have narrowed down the range of the quantity $k$ in function \funcPA.
In the proceedings, this quantity ranged over~$[-r,r]$ with $r \coloneqq a \cdot (2 \cdot \norminf{\linterms(\phi)}+ g \cdot \fmod(\phi))$, where $g \coloneqq \Pi\{ b : (b,t) \in T \text{ for some } t \}$ and $a$, $T$ and $\phi$ are as defined in \funcPA. 
This improvement does not affect the correctness of the procedure,
and there is no added difficulty in the analysis.

\item
We have clarified the input/output specification
of function \funcSem.

\item 
The original complexity analysis had a few minor counting errors (e.g.,~the bound $24 \cdot b$ from~\Cref{lemma:lin-bounds} was stated as $22 \cdot b$ in the proceedings),
which we have corrected.
None of these errors affected the results of the paper, 
and in particular all bounds in \Cref{prop:PA-Sem-All} remain asymptotically the same.

\end{itemize}

%% file: bibliography.bib
@inproceedings{phstrongcobham,
author = {Hieronymi, Philipp and Schulz, Christian},
title = {A Strong Version of {Cobham}'s Theorem},
year = {2022},
booktitle = {{STOC}}
}

@article{Berman80,
  author    = {Leonard Berman},
  title     = {The complexity of logical theories},
  journal   = {Theor.~Comput.~Sci.},
  volume    = {11},
  number    = {1},
  pages     = {71-77},
  year      = {1980}
}

@article{bes,
  author    = {Alexis B\`es},
  title     = {A Survey of Arithmetic Definability},
  journal   = {Soc.~Math.~Belgique}, 
  pages     = {1-54},
  year      = {2002}
}

@book{classicaldecision,
  author    = {Egon B{\"{o}}rger and
               Erich Gr{\"{a}}del and
               Yuri Gurevich},
  title     = {The Classical Decision Problem},
  year      = {1997},
 publisher = {Springer}
}

@article{pointexpansion,
  author    = {Point, Fran{\c{c}}oise}, 
  title     = {On decidable extensions of {Presburger} arithmetic: from {A.\ Bertrand} numeration systems to {Pisot} numbers}, 
  journal   = {{JSL}}, 
  volume    = {65}, 
  number    = {3}, 
  pages     = {1347--1374},
  year      = {2000}
}

@incollection{Pre29,
  author    = {Moj\.{z}esz Presburger},
  title     = {{\"Uber die Vollst\"andigkeit eines gewissen Systems der Arithmetik ganzer Zahlen, in welchem die Addition als einzige Operation hervortritt}},
  booktitle = {Comptes Rendus du I~Congr\`es des Math\'ematiciens des Pays Slaves},
  pages     = {92-101},
  year      = {1929}
}

@article{semenovprespower,
  author    = {Aleksei L. Semenov},
  title     = {On certain extensions of the arithmetic of addition of natural numbers},
  journal   = {Math.~USSR Izv.},
  volume    = {15},
  number    = {2},
  pages     = {401-418},
  year      = {1980}
}

@article{semenovpresexp,
  author    = {Aleksei L. Semenov},
  title     = {Logical theories of one-place functions on the set of natural numbers},
  journal   = {Math.~USSR Izv.},
  volume    = {22}, 
  number    = {3},
  pages     = {587--618},
  year      = {1984}
}

@article{Weispfenning90,
  author    = {Volker Weispfenning},
  title     = {{The Complexity of Almost Linear Diophantine Problems}},
  journal   = {J. Symb. Comput.},
  volume    = {10},
  number    = {5},
  pages     = {395--404},
  year      = {1990}
}

@inproceedings{0001HM22,
  author       = {Dmitry Chistikov and
                  Christoph Haase and
                  Alessio Mansutti},
  title        = {Geometric decision procedures and the {VC} dimension of linear arithmetic
                  theories},
  booktitle    = {{LICS}},
  year         = {2022},
}

@inproceedings{DurandH10,
  author       = {Antoine Durand{-}Gasselin and
                  Peter Habermehl},
  title        = {On the Use of Non-deterministic Automata for {Presburger} Arithmetic},
  booktitle    = {{CONCUR}},
  year         = {2010}
}

@article{Klaedtke08,
  author       = {Felix Klaedtke},
  title        = {Bounds on the automata size for {Presburger} arithmetic},
  journal      = {{ACM} Trans. Comput. Log.},
  volume       = {9},
  number       = {2},
  pages        = {11:1--11:34},
  year         = {2008}
}

@article{GIVAN2002105,
  author = {Robert Givan and David McAllester and Carl Witty and Dexter Kozen},
  title       = {Tarskian Set Constraints},
  journal     = {Information and Computation},
  volume      = {174},
  number      = {2},
  pages       = {105-131},
  year        = {2002}
}

@inproceedings{RayaHK23,
  author       = {Rodrigo Raya and
                  Jad Hamza and
                  Viktor Kun{\v{c}}ak},
  title        = {On the complexity of convex and reverse convex prequadratic constraints},
  booktitle      = {{LPAR}},
  year         = {2023}
}

@inproceedings{LechnerOW15,
  author       = {Antonia Lechner and
                  Jo{\"{e}}l Ouaknine and
                  James Worrell},
  title        = {On the Complexity of Linear Arithmetic with Divisibility},
  booktitle    = {{LICS}},
  year         = {2015}
}

@article{comptonhenson,
title = {A uniform method for proving lower bounds on the computational complexity of logical theories},
journal = {APAL},
volume = {48},
number = {1},
pages = {1-79},
year = {1990},
author = {Kevin J. Compton and C. {Ward Henson}},
}

@inproceedings{pointcherlin,
title={On extensions of {Presburger} Arithmetic},
author = {Gregory Cherlin and Fran{\c{c}}oise Point},
year = {1986},
url = {https://webusers.imj-prg.fr/~francoise.point/papiers/cherlin_point86.pdf},
  booktitle = {4th Easter Conference on Model Theory},
  series = {Humboldt-Univ. Berlin Seminarberichte},
  volume = {86},
  pages = {17--34},
}

@inproceedings{christophnparith,
  author    = {Florent Gu{\'{e}}pin and
               Christoph Haase and
               James Worrell},
  title     = {On the Existential Theories of {B{\"{u}}chi} Arithmetic and Linear
               {$p$}-adic Fields},
  booktitle = {{LICS}},
  year      = {2019}
}

@article{GinsburgS66,
  author = {Seymour Ginsburg and Edwin H. Spanier},
  title = {{Semigroups, {Presburger} formulas, and languages.}},
  volume = {16},
  journal = {Pacific Journal of Mathematics},
  number = {2},
  pages = {285--296},
  year = {1966}
}

@article{Opp78,
  title={A $2^{2^{2^{pn}}}$ upper bound on the complexity of {P}resburger arithmetic},
  author={Oppen, Derek C.},
  journal={{JCSS}},
  volume={16},
  number={3},
  pages={323--332},
  year={1978},
  publisher={Elsevier}
}

@inproceedings{ReddyLoveland78,
author = {Reddy, C. R. and Loveland, D. W.},
title = {{Presburger} Arithmetic with Bounded Quantifier Alternation},
year = {1978},
booktitle = {{STOC}},
}

@article{FerranteR75,
  author    = {Jeanne Ferrante and
               Charles Rackoff},
  title     = {A Decision Procedure for the First Order Theory of Real Addition with
               Order},
  journal   = {{SIAM} J. Comput.},
  volume    = {4},
  number    = {1},
  pages     = {69--76},
  year      = {1975}
}

@article{Mine06,
  author    = {Antoine Min{\'{e}}},
  title     = {The octagon abstract domain},
  journal   = {High. Order Symb. Comput.},
  volume    = {19},
  number    = {1},
  pages     = {31--100},
  year      = {2006}
}

@inproceedings{KapurZHZLN13,
  author    = {Deepak Kapur and
               Zhihai Zhang and
               Matthias Horbach and
               Hengjun Zhao and
               Qi Lu and
               ThanhVu Nguyen},
  title     = {Geometric Quantifier Elimination Heuristics for Automatically Generating
               Octagonal and Max-plus Invariants},
  booktitle = {Automated Reasoning and Mathematics - Essays in Memory of William
               W. McCune},
  year      = {2013},
  series    = {LNCS},
  volume    = {7788},
  pages     = {189--228},
  publisher = {Springer},
}

@book{NT,
    author = { {Gareth A.} Jones and {J. Mary} Jones },
    title = { Elementary Number Theory },
    year = { 1998 },
    publisher = { Springer}
}

@book{FreudGyarmati,
    title = { Number Theory },
    author = {  R{\'o}bert Freud and Edit Gyarmati },
    year = { 2021 },
    publisher = {AMS},
    series = { Pure and Applied Undergraduate Texts },
    volume = {48},
}

@article{Par66,
  author    = {Rohit J. Parikh},
  title     = {On Context-Free Languages},
  journal   = {J. {ACM}},
  volume    = {13},
  number    = {4},
  pages     = {570--581},
  year      = {1966},
}
